\documentclass[12pt]{iopart}

\expandafter\let\csname equation*\endcsname\relax
\expandafter\let\csname endequation*\endcsname\relax

\usepackage{amsfonts}
\usepackage{amsmath}
\usepackage{amssymb}
\usepackage{bm}
\usepackage{dcolumn}
\usepackage{graphicx}
\usepackage{graphics}
\usepackage[T1]{fontenc}
\usepackage[utf8]{inputenc}
\usepackage{latexsym}
\usepackage{rotating}
\usepackage{cite}
\usepackage[perpage,marginal]{footmisc}
\usepackage[colorlinks=true]{hyperref}
\usepackage[all]{hypcap}
\usepackage{xspace} 
\usepackage[usenames]{color}
\usepackage{mathrsfs}
\usepackage{microtype}

\usepackage{ulem}
\makeatletter
\renewcommand\footnoterule{%
  \kern-3\p@
  \hrule\@width2.5cm
  \kern2.6\p@}
\makeatother

\definecolor {darkgreen}{rgb}{0.2,0.7,0.2}

\newcommand\be{\begin{equation}}
\newcommand\ba{\begin{eqnarray}}
\newcommand\ee{\end{equation}}
\newcommand\ea{\end{eqnarray}}

\newcommand\bw{\begin{widetext}}
\newcommand\ew{\end{widetext}}

\newcommand{\nn}{\nonumber}

\newcommand{\ISCO}{{\mbox{\tiny ISCO}}}
\newcommand{\cont}{{\mbox{\tiny cont}}}

\newcommand{\N}{{\mbox{\tiny N}}}

\newcommand{\tid}{{\mbox{\tiny (tid)}}}

\newcommand{\rot}{{\mbox{\tiny (rot)}}}

\newcommand{\sys}{{\mbox{\tiny sys}}}

\newcommand{\temp}{{\mbox{\tiny temp}}}

\newcommand{\inj}{{\mbox{\tiny i}}}

\newcommand{\mrm}{\mathrm}

\begin{document}
\title[Approximate Universal Relations among Tidal Parameters for NS Binaries]{Approximate Universal Relations among Tidal Parameters for Neutron Star Binaries} 

\author{Kent Yagi}
\address{Department of Physics, Princeton University, Princeton, New Jersey 08544, USA}
\address{eXtreme Gravity Institute, Department of Physics, Montana State University, Bozeman, Montana 59717, USA}

\author{Nicol\'as Yunes}
\address{eXtreme Gravity Institute, Department of Physics, Montana State University, Bozeman, Montana 59717, USA}

\date{\today}

\begin{abstract} 

One of largest uncertainties in nuclear physics is the relation between the pressure and density of supranuclear matter: the equation of state. Some of this uncertainty may be removed through future gravitational wave observations of neutron star binaries by extracting the tidal deformabilities (or Love numbers) of neutron stars, a novel way to probe nuclear physics in the high-density regime. Previous studies have shown that only a certain combination of the individual (quadrupolar) deformabilities of each body (the so-called \emph{chirp} tidal deformability) can be measured with second-generation, gravitational wave interferometers, such as Adv.~LIGO, due to correlations between the individual deformabilities. To overcome this, we search for approximately universal (i.e.~approximately equation-of-state independent) relations between two combinations of the individual tidal deformabilities, such that once one of them has been measured, the other can be automatically obtained and the individual ones decoupled through these relations. We find an approximately universal relation between the symmetric and the anti-symmetric combination of the individual tidal deformabilities that is equation-of-state-insensitive to $20\%$ for binaries with masses less than $1.7M_\odot$. We show that these relations can be used to eliminate a combination of the tidal parameters from the list of model parameters, thus breaking degeneracies and improving the accuracy in parameter estimation. A simple (Fisher) study shows that the universal binary Love relations can improve the accuracy in the extraction of the symmetric combination of tidal parameters by as much as an order of magnitude, making the overall accuracy in the extraction of this parameter slightly better than that of the chirp tidal deformability. These new universal relations and the improved measurement accuracy on tidal parameters not only are important to astrophysics and nuclear physics, but also impact our ability to probe extreme gravity with gravitational waves and cosmology.

\end{abstract}

\maketitle

\section{Introduction}

The equation of state (EoS), the thermodynamic relation between state variables, is key in the description of fluids and solids, but in particular, it is critical in the description of stars. The barotropic equation of state of matter, the relation between pressure and density, at nuclear saturation density  ($\sim 2.5 \times 10^{14}\mrm{g/cm}^3$) has been well-constrained by terrestrial experiments~\cite{Li:2008gp}. For example, heavy-ion collisions~\cite{Tsang:2008fd,Chen:2004si,Li:2005jy} and measurements of the neutron skins of nuclei~\cite{Centelles:2008vu} have constrained the linear density dependence of the nuclear symmetry energy. However, terrestrial experiments cannot constrain the EoS beyond saturation density, rendering the supranuclear EoS one of the largest uncertainties in nuclear physics. 

Neutron stars (NSs) are a perfect testbed to probe nuclear physics in this high-density regime. For example, one can use independent measurements of the NS mass and radius to constrain the EoS~\cite{lattimer_prakash2001,lattimer-prakash-review,Ozel:2012wu,Miller:2013tca}. Current observations of X-ray bursters and quiescent low-mass X-ray binaries have already placed some constraints on the EoS~\cite{steiner-lattimer-brown,ozel-baym-guver,Steiner:2012xt,guver,Lattimer:2013hma,Lattimer:2014sga,Ozel:2015fia}, though these may suffer from large systematic errors due to uncertainties in the astrophysical modeling of the NS sources. Future X-ray pulse profile observations from a hot spot on the NS surface using NICER~\cite{2012SPIE.8443E..13G} and LOFT~\cite{2012AAS...21924906R,2012SPIE.8443E..2DF} may be able to place stronger constraints with less systematics~\cite{Morsink:2007tv,Psaltis:2013zja,Lo:2013ava,Psaltis:2013fha,Miller:2014mca}. Future radio observations of the double binary pulsar J0737-3039~\cite{burgay,lyne,kramer-double-pulsar} are expected to measure the moment of inertia of the primary pulsar~\cite{lattimer-schutz,kramer-wex}, which would also allow for constraints on the EoS.
 
Given the recent direct detection of gravitational waves (GWs) from black hole binaries~\cite{Abbott:2016blz,Abbott:2016nmj}, one expects GWs from NS binaries to also be detected soon. The latter can be a novel probe of nuclear physics because they encode information about the EoS through finite-size effects. During the early inspiral, when the orbital separation is large, the NSs barely feel the gravitational field of their companion, and thus, are not really deformed. As the inspiral proceeds and the orbital separation decreases, the gravitational tidal field of one star at the location of its companion increases in magnitude, creating a deformation on the latter (and vice versa). Such a deformation forces the gravitational field of the deformed stars to not be spherically symmetric any longer, which affects their orbital trajectory. Moreover, since the quadrupolar tidal deformation is time dependent, it also modifies the amount of GW energy carried away from the binary, and thus, the orbital decay rate. Such changes in the trajectory, in turn, imprint directly onto the gravitational waveform. The largest finite-size effect in the waveform is characterized by the electric-type, quadrupolar tidal deformability~\cite{hinderer-love,damour-nagar,binnington-poisson}, which we here refer to as the \emph{tidal deformability} $\lambda$, and which is related to the tidal \emph{Love} number~\cite{love}. The tidal deformability is defined as the linear response of the tidally induced quadrupole moment of an object due to an external tidal field.

Measurability of the NS tidal effects and constraints on the NS EoS with GW observations have been studied in~\cite{read-markakis-shibata,flanagan-hinderer-love,hinderer-lackey-lang-read,damour-nagar-villain,lackey,lackey-kyutoku-spin-BHNS,delpozzo,read-matter,Favata:2013rwa,Yagi:2013baa,Wade:2014vqa,Lackey:2014fwa,Agathos:2015uaa,Hotokezaka:2016bzh}. For example, Ref.~\cite{damour-nagar-villain} carried out an approximate Fisher analysis and found that a certain combination of the individual tidal deformabilities of the binary components can be measured to better than $\mathcal{O}(20\%)$ with second-generation GW interferometers, such as Adv.~LIGO. Bayesian studies that followed~\cite{Wade:2014vqa} showed that the certain combination of the tidal deformabilities may be measurable to the accuracy mentioned above, for sufficiently high signal-to-noise ratio events, although systematic error may also add to the error budget (see also~\cite{Favata:2013rwa,Yagi:2013baa} for studies on systematic errors). References~\cite{delpozzo,Agathos:2015uaa} also showed that the detection of multiple NS binary signals could enhance even further the accuracy to which certain combinations of the tidal deformabilities can be measured. 

The measurement of the tidal deformabilities can also be useful in probing cosmology~\cite{messenger-read,Li:2013via,DelPozzo:2015bna} by using GW sources as \emph{standard sirens}~\cite{Schutz:1986gp,Holz:2005df,Dalal:2006qt,Sathyaprakash:2009xt,Nissanke:2009kt,Cutler:2009qv,Zhao:2010sz,Nishizawa:2010xx,Camera:2013xfa,Nissanke:2013fka}. The main idea here is to measure the luminosity distance (from the GW amplitude) and the redshift of the source independently from a set of GW and electromagnetic-wave observations; since these two quantities are related by cosmological evolution equations, they can be used to estimate cosmological parameters. The original idea was to obtain the redshift from host galaxy identification, provided the GW sky-localization is accurate enough. Typically, however, this is not expected to be the case for a large number of galaxies with second-generation detectors. Another idea is to obtain the redshift from precise measurements of the tidal deformabilities~\cite{messenger-read,Li:2013via,DelPozzo:2015bna}. These measurements allow us to construct the intrinsic mass of the source, provided one knows the correct EoS \emph{a priori}; comparing this mass  to the observed (redshifted) mass, one can then estimate the redshift~\cite{messenger-read,Li:2013via,DelPozzo:2015bna}\footnote{See e.g.~\cite{MacLeod:2007jd,DelPozzo:2011yh,Petiteau:2011we,Nishizawa:2011eq,Taylor:2011fs,Taylor:2012db,Namikawa:2015prh,Oguri:2016dgk} for other possibilities of probing cosmology with GW observations alone.}.

Can one independently extract the individual dimensionless tidal deformabilities $(\bar \lambda_{1}, \bar \lambda_{2})$ of each binary component with an Adv.~LIGO observation? Previous literature claims this is not possible with the sources expected to be detected with Adv.~LIGO because $\bar \lambda_{1}$ and $\bar \lambda_{2}$ are strongly correlated. One can, however, reparameterize the templates with two new tidal parameters, constructed from independent linear combinations of $\bar \lambda_{1}$ and $\bar \lambda_{2}$, in such a way as to diminish the correlations. For example, a commonly-used set of tidal parameters is~\cite{flanagan-hinderer-love,hinderer-lackey-lang-read,Favata:2013rwa,Wade:2014vqa,Lackey:2014fwa}\footnote{The signs of the terms proportional to $\bar \lambda_a$ are opposite those in~\cite{Wade:2014vqa} because we use the convention $m_1 \leq m_2$ (so that $\bar \lambda_a \geq 0$), while Ref.~\cite{Wade:2014vqa} used $m_1 \geq m_2$.}
\begin{align}
\label{eq:Lambda-def}
\bar \Lambda &= f(\eta) \bar \lambda_s + g(\eta)  \bar \lambda_a \,, \\
\label{eq:delta-Lambda-def}
\delta \bar \Lambda &= \delta f(\eta) \bar \lambda_s  + \delta g(\eta) \bar \lambda_a \,,
\end{align}
where $\lambda_{s,a} \equiv {(\bar \lambda_1 \pm \bar \lambda_2)}/{2}$ are symmetric and antisymmetric combinations of the tidal deformabilities, while $\eta \equiv m_{1} m_{2}/(m_{1} + m_{2})^{2} = q/(1 + q)^{2}$ is the symmetric mass ratio, with $m_A$ representing the mass of the $A$th body and $q \equiv m_{1}/m_{2}$ (with $m_1 \leq m_2$) the mass ratio; the functions $[f(\eta),g(\eta),\delta f(\eta), \delta g(\eta)]$ are given in Sec.~\ref{sec:Lambda-delta-Lambda}. These tidal parameters $(\bar \Lambda, \delta \bar \Lambda)$ partially break the degeneracies between $\bar \lambda_{1}$ and $\bar \lambda_{2}$ because they enter at different post-Newtonian (PN) orders in the waveform\footnote{The PN expansion is a series in powers of the ratio of the orbital velocity $v$ of the binary to the speed of light $c$. A term of $n$PN order in a PN series corresponds to one that is of $\mathcal{O}[(v/c)^{2n}]$ or equivalently $\mathcal{O}(x^n)$, relative to the leading order term in the series.}~\cite{flanagan-hinderer-love,hinderer-lackey-lang-read,Favata:2013rwa,Wade:2014vqa,Lackey:2014fwa}, i.e.~the tidal part of the gravitational waveform phase in the Fourier domain is given by~\cite{Wade:2014vqa}
\be
\Psi_\mrm{tidal} = -\frac{3}{128 \eta\, x^{5/2}} \left[ \frac{39}{2} \bar \Lambda \, x^5 + \left(\frac{3115}{64} \bar \Lambda - \frac{6595}{364} \sqrt{1 - 4 \eta}\, \delta \bar \Lambda  \right) x^6 + \mathcal{O}(x^7) \right]\,,
\ee
where $x \equiv [\pi (m_1+m_2) f]^{2/3}$ with $f$ the GW frequency. Nevertheless, one can only measure $\bar \Lambda$ in this parametrization, because $\delta \bar \Lambda$ enters at too high a PN order to affect the template's phase sufficiently~\cite{Wade:2014vqa}.  Since $\bar \Lambda$ is the dominant tidal parameter entering the waveform, we shall call it the dimensionless \emph{chirp} tidal deformability, in analogy with the chirp mass being the dominant mass parameter in the waveform.

In this paper, we search for a way to overcome this problem, i.e.~a way to extract \emph{both} tidal deformabilities of the binary from an Adv.~LIGO GW observation. We accomplish this by finding approximately universal relations, i.e.~relations that are approximately EOS-insensitive, among independent tidal parameters. In general, these relations serve two important purposes:
\begin{enumerate}
\item {\bf{To Improve Parameter Estimation.}} The universal relations can be used to eliminate some of the tidal parameters from the template parameter vector in a data analysis study. This breaks degeneracies between the parameters removed and those left in the template and improves the latter's parameter estimation accuracy. 
\item {\bf{To Extract Both Tidal Deformabilities.}}  Given the measurement of a given combination of tidal parameters (e.g.~$\bar \Lambda$), these relations allow us to automatically obtain the other combination (e.g.~$\delta \bar \Lambda$) to the accuracy of the approximate universality. These two independent combinations can then be easily decoupled to find $\bar \lambda_{1}$ and $\bar \lambda_{2}$. 
\end{enumerate}
 
Universal relations in NSs are not really new (see the recent review~\cite{Yagi:2016bkt} and references therein for various NS universal relations). In particular, universal relations among the NS moment of inertia, the tidal deformability and the quadrupole moment (the so-called \emph{I-Love-Q} relations)~\cite{I-Love-Q-Science,I-Love-Q-PRD} can be used to eliminate the individual quadrupole moments from the template parameter vector. This breaks degeneracies between the quadrupole moment and the individual NS spins, allowing us to extract the latter more accurately. Such relations also allow us to probe extreme gravity without having to know what the correct EoS is. For example, by combining future measurements of the moment of inertia with the double binary pulsar and the tidal deformability with GW observations, one can use the I-Love relation to place constraints on the parity-violating sector of gravity that are six orders of magnitude stronger than the current bound~\cite{I-Love-Q-Science,I-Love-Q-PRD}. Universal relations among the various tidal deformabilities of an isolated NS (multipole Love relations) were found in~\cite{Yagi:2013sva}, while relations among these deformabilities and certain binary parameters were studied in~\cite{Bernuzzi:2014kca}.

\subsection{Executive Summary}

In this paper, we find three different, approximately universal, \emph{binary Love} relations, among the following quantities: 
\begin{itemize}
\item[(i)] the symmetric tidal parameter $\bar \lambda_{s}$ (the average sum of $\bar \lambda_{1}$ and $\bar \lambda_{2}$) and the antisymmetric tidal parameter $\bar \lambda_{a}$ (the average difference of $\bar \lambda_{1}$ and $\bar \lambda_{2}$), 
\item[(ii)] The $\bar \Lambda$ and $\delta \bar \Lambda$ tidal parameters mentioned earlier,
\item[(iii)] The coefficients $\bar \lambda_0^{(k)}$ in the Taylor expansion of the dimensionless tidal deformabilities around a fiducial mass $m_0$~\cite{delpozzo}. 
\end{itemize}
One can implement the first and second parameterizations for any NS binary systems, while the third parameterization can only be applied, in practice, to systems whose NS masses are close to $m_0$. Otherwise, the systematic error on the leading tidal coefficient due to mismodeling the tidal deformability can dominate the statistical one. If one insists on retaining many terms in the Taylor expansion to reduce such systematic error, then correlations among the many $\bar \lambda_0^{(k)}$ parameters may increase the statistical error. If enough systems with masses that cluster together are observed, however, the third parameterization has an advantage that it allows one to combine such observations to increase the measurement accuracy of $\bar \lambda_0^{(k)}$ that is common to all systems~\cite{delpozzo,Agathos:2015uaa}.

We find these new universal relations as follows. We first calculate the tidal deformability of an isolated NS with various realistic EoSs for a set of different NS masses. In doing this, we follow~\cite{hinderer-love} and extract the tidal deformability from the asymptotic behavior of the gravitational potential of tidally-deformed NS solutions, treating the tidal deformations perturbatively. When considering the first and second relations, we then calculate the two independent tidal parameters in each relation, such as $\bar \lambda_{s}[\bar \lambda_{1}(m_{1}),\bar \lambda_{2}(m_{2})]$ and $\bar \lambda_{a}[\bar \lambda_{1}(m_{1}),\bar \lambda_{2}(m_{2})]$, or $\bar \Lambda[\bar \lambda_{1}(m_{1}),\bar \lambda_{2}(m_{2})]$ and $\delta \bar \Lambda[\bar \lambda_{1}(m_{1}),\bar \lambda_{2}(m_{2})]$. Given these, we finally rewrite one of the tidal parameters in terms of the other and in terms of the mass ratio $q$, e.g.~$\bar \lambda_a(\bar \lambda_s,q)$ or $\bar \Lambda(\delta \bar \Lambda,q)$, and study its EoS variation. When considering the third relation, we calculate the $k$th coefficients $\bar \lambda_0^{(k)}(m_0)$ in the Taylor expansion of $\bar \lambda(m)$ about $m=m_0$ for the dimensionless tidal deformability of an isolated NS $\bar \lambda(m)$. We then eliminate $m_0$ using the zeroth-coefficient in the expansion, $\bar \lambda_0^{(0)}$, so as to find $\bar \lambda_0^{(k)} (\bar \lambda_0^{(0)})$, and then we study its EoS variation.

\begin{figure*}[htb]
\begin{center}
\includegraphics[width=7.5cm,clip=true]{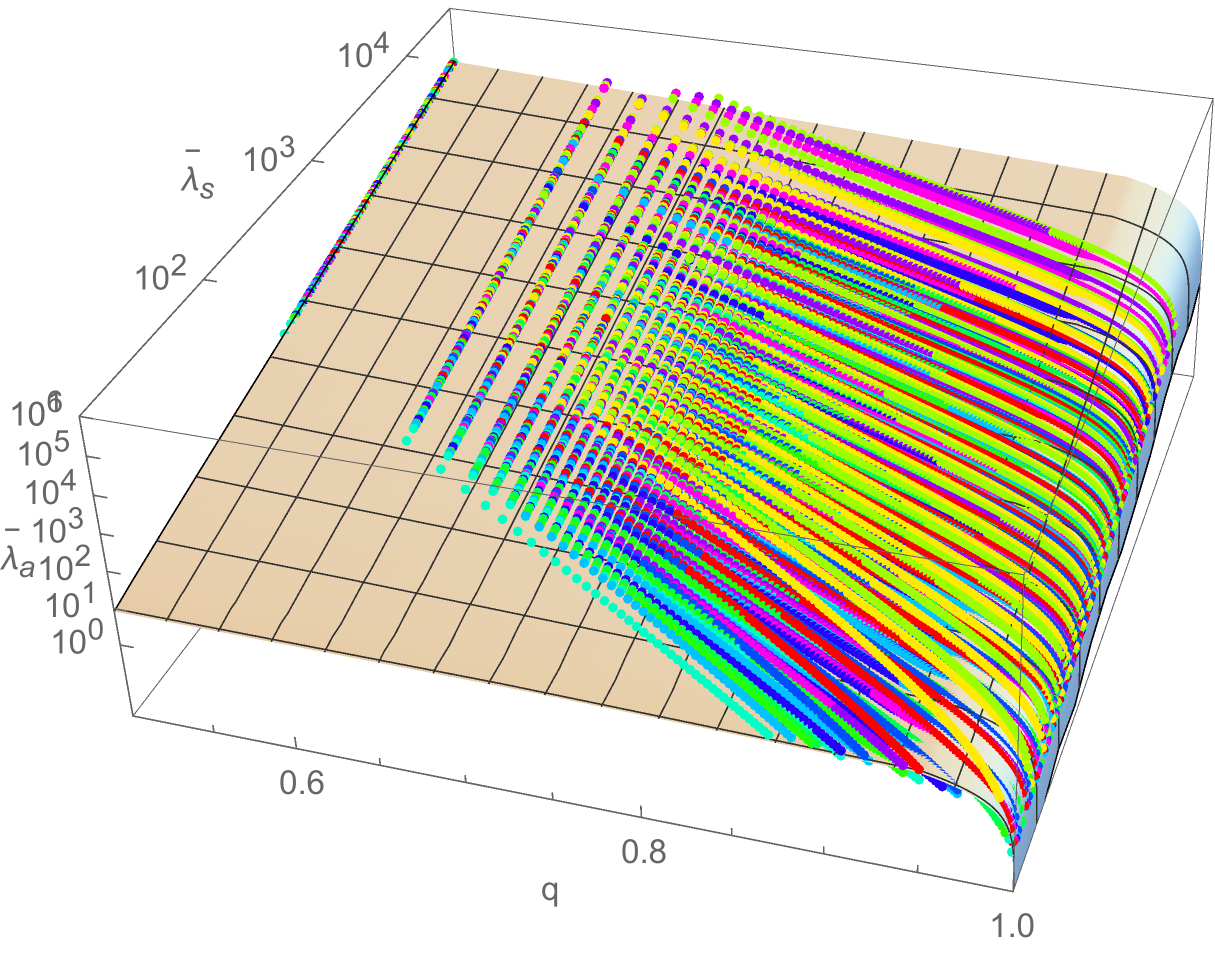}  
\includegraphics[width=7.5cm,clip=true]{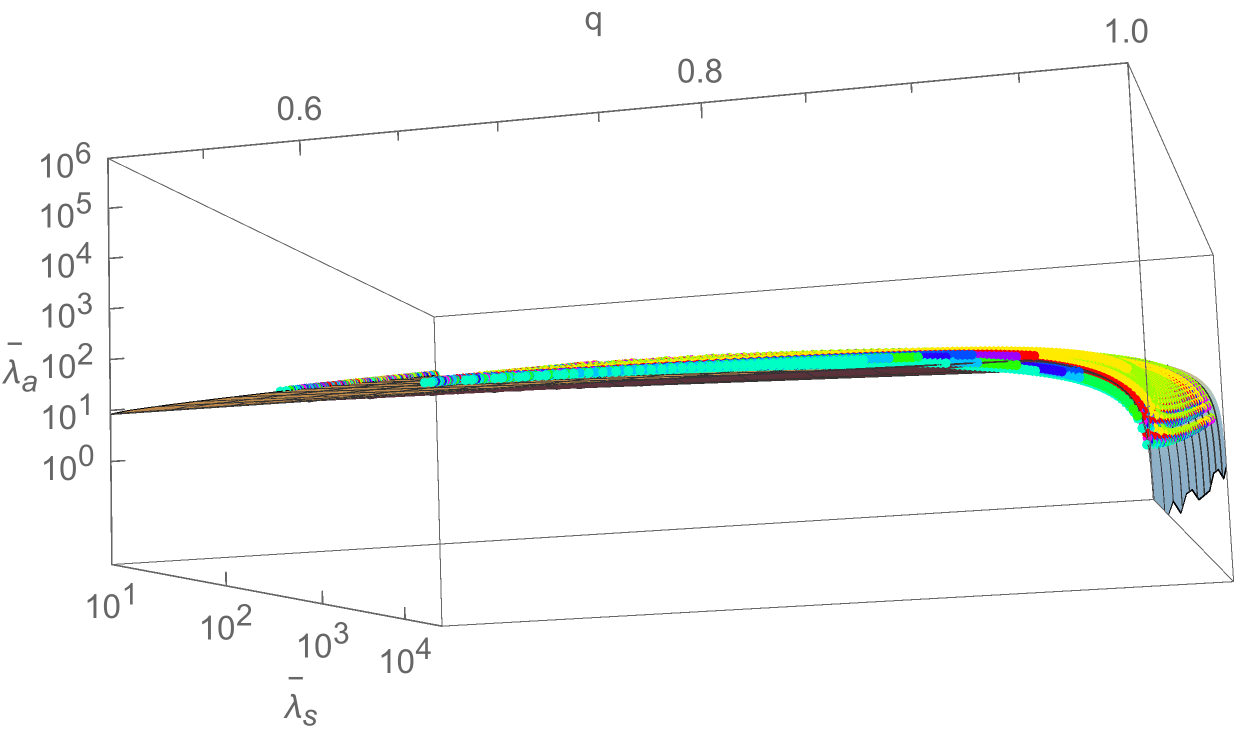}  
\caption{\label{fig:lambdas-lambdaa-3D} Universal relation between $\bar \lambda_s$ and $\bar \lambda_a$, the averaged sum and difference of the tidal deformabilities of two NSs in a binary, in the mass ratio $q$ plane from two different viewpoints. Points represent numerical results with eleven different realistic EoSs, while the approximately universal plane is constructed from a fit to all the numerical data. Observe that the data lies approximately on the fitted plane, irrespective of the EoS used to construct each data point. 
}
\end{center}
\end{figure*}

Figure~\ref{fig:lambdas-lambdaa-3D} shows the $\bar \lambda_s$--$\bar\lambda_a$ relation, namely $\bar \lambda_a$ as a function of $\bar \lambda_s$ and $q$, for various EoSs. The plane in the figure is simply a fit to all the numerical data. Observe that all the data points lie approximately on this \emph{approximately universal plane} irrespective of the EoSs. The relative fractional difference between any data point and the plane never exceeds $20\%$ for NS masses less than $1.7M_\odot$. Similar figures can be constructed for the $\bar \Lambda$--$\delta \bar \Lambda$ and the $\bar \lambda_0^{(0)}$--$\bar \lambda_0^{(k)}$ relations, although the universality for the former is weaker compared to the $\bar \lambda_s$--$\bar\lambda_a$ universality while the latter does not depend on $q$. The $\bar \lambda_0^{(0)}$--$\bar \lambda_0^{(k)}$ relation is particularly useful when studying the possibility of using tidal deformabilities to probe cosmology with GWs, as the knowledge of the correct EoS, or the mass dependence of the tidal deformability, is crucial in such an analysis. These results show that once one measures one of the tidal parameters in the relations, one can automatically obtain the other. This in turn allows one to determine the tidal deformability of each body independently, allowing us to extract more astrophysical information beyond just the masses and spins of the binary.

The universality seems to become stronger as one decreases the mass ratio for fixed total mass, but it deteriorates as one increases the total mass for a fixed mass ratio. The former can be easily understood by considering the $\bar \lambda_s$--$\bar\lambda_a$ relation in the $q \to 0$ limit, since then $\bar \lambda_a = \bar\lambda_s$ and this is exactly EoS universal. The latter, however, is somewhat more difficult to explain, because one intuitively expects the universal relations to improve as one approaches the black hole (BH) limit, i.e.~as the NS mass, and thus the NS compactness, increase. Although it is true that $\bar \lambda_{1}(m)$ and $\bar \lambda_{2}(m)$ approach their BH values in this limit, their averaged difference $\bar \lambda_{a}$ is sensitive to \emph{how} this limit is approached, i.e.~to the slope of the $\bar \lambda$--$m$ relation. This slope decreases with increasing mass, essentially because $\bar \lambda(m)$ decreases very fast with stellar compactness, which is simply because  massive stars deform less than light stars. Recalling that the relative fractional difference is a function of the difference in $\bar \lambda_{a}$ with different EoSs \emph{divided} by $\bar \lambda_{a}$ for a reference EoS, the fact that the universality deteriorates in the BH limit is then simply a consequence of the slope of the $\bar \lambda$--$m$ relation decreasing faster than the EoS variability in $\bar \lambda_{a}$.

\begin{figure}[htb]
\begin{center}
\includegraphics[width=8.cm,clip=true]{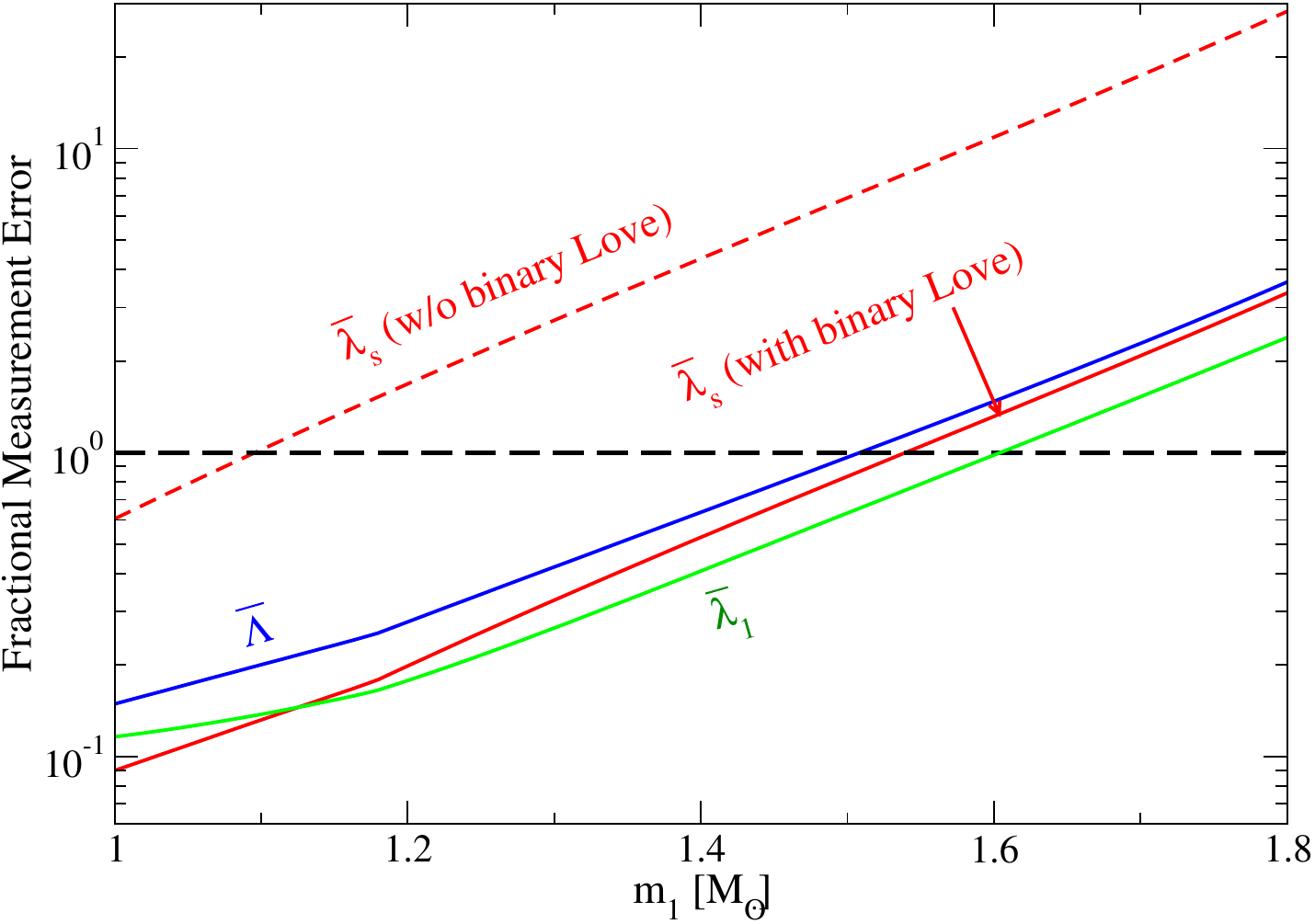}  
\caption{\label{fig:Fisher-prior} Estimated fractional measurement accuracy of the symmetric tidal deformability $\bar \lambda_s$, given GW observations of NS binaries with Adv.~LIGO, as a function of the smaller NS mass. The fractional measurement error is defined by $\Delta \bar \lambda_s/\bar \lambda_s$ with $\Delta \bar \lambda_s$ representing the measurement error of $\bar \lambda_s$. The deformability is measurable with an error less than $100\%$ (horizontal black dashed line). The figure was generated through a Fisher analysis that assumed a signal-to-noise ratio (SNR) of 30, a mass ratio of 0.9, the AP4 EoS~\cite{APR}, and the zero-detuned Adv.~LIGO noise curve~\cite{ajith-zero-detuned}. The red solid (dashed) curve shows the accuracy with (without) using the universal binary Love relation of Fig.~\ref{fig:lambdas-lambdaa-3D}. Observe that such a relation improves the measurement accuracy by approximately an order of magnitude. For reference, we also show the measurement accuracy of the chirp tidal deformability $\bar \Lambda$ used in~\cite{flanagan-hinderer-love,hinderer-lackey-lang-read,Favata:2013rwa,Wade:2014vqa,Lackey:2014fwa}, which is not improved much by using the universal relation. We further show the fractional measurement accuracy of the tidal deformability ($\bar \lambda_1$) of one of the NSs (the one with the smaller mass) in the binary. Observe that such an accuracy is better than that of $\bar \Lambda$ by 50\%.
}
\end{center}
\end{figure}

Figure~\ref{fig:Fisher-prior} shows a measure of the improvement in parameter accuracy of $\bar \lambda_s$ when one uses the $\bar \lambda_s$--$\bar\lambda_a$ relation. More precisely, we carried out two Fisher analyses: one with a template family that includes both tidal parameters (red dashed curve), and one with a template family that uses the approximately universal relations to eliminate one of the tidal parameters from the parameter vector (red solid curve), in both cases assuming a fiducial realistic NS EoS. Observe that parameter estimation accuracy improves by roughly an order of magnitude when using the universal relations. This is because the relation allows us to eliminate $\bar \lambda_a$ from the search parameters, which breaks the degeneracy between $\bar \lambda_s$ and $\bar \lambda_a$. We have checked that the fractional systematic error on $\bar \lambda_s$, due to the EoS variation in the $\bar \lambda_s$--$\bar\lambda_a$ relation, is smaller than $10^{-2}$, and thus negligible in this figure. This is mainly because the term in the GW phase that depends on $\bar \lambda_a$ is proportional to the difference in the masses of the two NSs, and hence, the systematic error is suppressed in the near equal-mass regime, precisely where the binary Love relations are less EoS-universal. On the other hand, we find that the $\bar \Lambda$--$\delta \bar \Lambda$ relation does not really affect the measurement accuracy of the chirp tidal deformability $\bar \Lambda$ (blue dot-dashed curve). This is because the correlation between $\bar \Lambda$ and $\delta \bar \Lambda$ is weaker than that between $\bar \lambda_s$ and $\bar\lambda_a$, since $\bar \Lambda$ and $\delta \bar \Lambda$ enter at different PN order in the waveform phase. We also found that the measurement accuracy of $\bar \lambda_0^{(0)}$ improves by a few times when one uses the $\bar \lambda_0^{(0)}$--$\bar \lambda_0^{(k)}$ relations. 

These findings have an impact on various branches of physics. On the astrophysics front, the individual tidal deformability of each NS brings important additional information, in addition to the mass and spin of NSs. On the nuclear physics front, the improved measurement accuracy of tidal parameters increases our ability to constrain the EoS with future GW observations.  On the experimental relativity front, such an improvement in the measurement accuracy of the tidal deformability strengthens projected constraints on alternative theories of gravity through universal relations between e.g.~the tidal deformability and the moment of inertia. On the cosmology front, the $\bar \lambda_0^{(0)}$--$\bar \lambda_0^{(k)}$ relation improves our knowledge of how the tidal deformability depends on the intrinsic NS mass, which helps breaking the degeneracy between the mass and redshift, allowing us to probe cosmology with GW observations alone.

Some of the results discussed above were already presented in the letter~\cite{Yagi:2015pkc}, which we here explain in much more detail and extend by finding new results. For example, the binary Love relation between $\bar \Lambda$ and $\delta \bar \Lambda$ was not shown in~\cite{Yagi:2015pkc}. We also carry out analytic calculations in the Newtonian limit to obtain a better understanding of the universality and to create a base function for fits. Furthermore, we estimate not only statistical errors but also systematic errors on tidal parameters due to (i) the EoS variation in the binary Love relations and (ii) not including one of the tidal parameters in the search parameter set that can be eliminated with the binary Love relations. We also extend the binary Love relations for NSs to quark stars (QSs) and to I-Love-Q relations for a binary system (binary I-Love-Q relations).

\subsection{Organization}

The remainder of this paper presents the details of the results discussed above and it is organized as follows. In Sec.~\ref{sec:binary-Love}, we derive universal relations among two (or more) independent tidal parameters in the gravitational waveform of NS binaries. In Sec.~\ref{sec:Fisher}, we carry out a parameter estimation study using the Fisher matrix and estimate the measurement accuracy of tidal parameters with Adv.~LIGO. In Sec.~\ref{sec:applications}, we discuss the possibility of applying the universal relations to probe astrophysics, nuclear physics, experimental relativity and cosmology. We end in Sec.~\ref{sec:conclusion} by presenting possible avenues for future work. We use geometric unit $c=1=G$ throughout.

\section{Universal Binary Love Relations}
\label{sec:binary-Love}

In this section, we focus on relations among various tidal parameters that enter the gravitational waveform of a non-spinning NS/NS binary with masses $m_A$, where the subscript $A$ denotes the $A$th body. We assume $m_1 \leq m_2$ throughout the paper. We adopt eleven realistic EoSs that can support a NS with a mass above $2M_\odot$: AP3~\cite{APR}, AP4~\cite{APR}, SLy~\cite{SLy}, WFF1~\cite{Wiringa:1988tp}, WFF2~\cite{Wiringa:1988tp}, ENG~\cite{1996ApJ...469..794E}, MPA1~\cite{1987PhLB..199..469M}, MS1~\cite{Mueller:1996pm}, MS1b~\cite{Mueller:1996pm}, LS~\cite{LS} with nuclear incompressibility of 220MeV (LS220) and Shen~\cite{Shen1,Shen2}, where for the latter two, we assume a temperature of 0.1MeV with a neutrino-less and beta-equilibrium condition. These eleven EoSs can be classified by their stiffness, namely how the pressure increases given an increase in energy density. Such a classification is summarized in Table~\ref{table:EoS}.

\Table{\label{table:EoS} Three different classes of EoSs and their selected members.
\vspace{3mm}
}
\br
EoS class & selected members  \\
\mr
soft & WFF1, WFF2, SLy, AP4   \\
intermediate & AP3, MPA1, LS220, ENG   \\
stiff & Shen, MS1, MS1b  \\
\br
\endTable

The tidal parameter that impacts the waveform the most is the electric-type, $\ell = 2$ tidal deformability (or just tidal deformability for short), which we denote\footnote{For rotating NSs, tidal effects enter first through their quadrupole moment, which appear at lower PN order than $\lambda$ in the gravitational waveform. The effect of the quadrupole, however, is suppressed for slowly-rotating NSs because it is proportional to the NS spin squared.} as $\lambda$. Let us consider a NS of mass $m$ that is immersed in the field of an external companion. The tidal deformability is then the linear response of the (symmetric trace-free) tidally induced quadrupole moment tensor $Q_{ij}^{\tid}$ of the NS due to the external quadrupolar tidal tensor $\mathcal{E}_{ij}$ induced by its companion~\cite{hinderer-love,damour-nagar,binnington-poisson}:
\be
\label{eq:lambda-def}
Q_{ij}^{\tid} = - \lambda \, \mathcal{E}_{ij}\,.
\ee
The tidal deformability $\lambda$ is extracted from the asymptotic behavior of the gravitational potential, or the $(t,t)$ component of the metric of  the NS, via
\begin{align}
\frac{1 - g_{tt}}{2} &= - \frac{m}{r} +  \frac{1}{2}  \mathcal{E}_{ij} x^i x^j \left[1 + \frac{\alpha_1}{r} + \frac{\alpha_2}{r^2}  -3 \frac{\lambda}{r^5} + \mathcal{O}\left( \frac{m^6}{r^6} \right)  \right] + \mathcal{O} \left(\mathcal{E}_{L} x^L \right) \quad (L \geq 3)\,,
\end{align}
where the NS is located at the origin, $n^i = x^i/r$ is a unit vector pointing from the NS to a field point, $\alpha_1$, $\alpha_2$ are constants and $\mathcal{O} \left(\mathcal{E}_{L} x^L \right)$ represents terms at the $\ell=3$ order and higher.

One can calculate $g_{tt}$, and thus extract $\lambda$, by constructing tidally deformed NS solutions. One first constructs a background solution for a non-spinning, spherically symmetric NS and then adds a tidal deformation as a perturbation. One solves the background and perturbed Einstein equations in the interior region with a given EoS, imposing regularity at the stellar center. One then matches the interior solution to an analytic, exterior solution at the stellar surface to determine integration constants (see e.g.~\cite{hinderer-love,damour-nagar,I-Love-Q-PRD} for a detailed explanation of this procedure), which then determines $g_{tt}$ modulo an overall constant. With this metric component at hand, one can then extract $Q_{ij}^{\tid}$ and ${\cal{E}}_{ij}$ by expanding $g_{tt}$ far from the NS and reading off the coefficient of $n^{\langle ij \rangle}/r^{3}$ for the quadrupole moment and $n^{ij} r^2$ for the external tidal field, where $\langle \rangle$ stands for the symmetric and trace free operator. From these, one can easily extract $\lambda$ from Eq.~\eqref{eq:lambda-def}.

Finite-size effects encoded in the GWs emitted by NS binaries are then dominantly controlled by two tidal parameters, $\lambda_1$ and $\lambda_2$. Two parameters enter the waveform because $\lambda_{A}$ depends on the NS mass and NSs in binaries typically have different masses. For later convenience, we define the dimensionless tidal deformability $\bar \lambda_A \equiv \lambda_A/m_A^5$~\cite{I-Love-Q-Science,I-Love-Q-PRD}. Higher (multipole) order tidal deformabilities also enter the waveform at higher PN order but they can be expressed in terms of $\lambda_{A}$ through the multipole Love relations~\cite{Yagi:2013sva}.

When constructing a GW template family for NS binaries, one must choose two independent tidal parameters to represent finite-size effects, but this choice is not unique as any function of $\bar \lambda_{A}$ would in principle be acceptable. We first consider two different sets [($\bar \lambda_s$,$\bar \lambda_a$)~\cite{I-Love-Q-Science,I-Love-Q-PRD} in Sec.~\ref{sec:lambdas-lambdaa} and ($\bar \Lambda$,$\delta \bar \Lambda$)~\cite{flanagan-hinderer-love,hinderer-lackey-lang-read,Favata:2013rwa,Wade:2014vqa,Lackey:2014fwa} in Sec.~\ref{sec:Lambda-delta-Lambda}] and look for universal relations, which as argued in the Introduction, will also depend on the mass ratio $q := {m_1}/{m_2}$. We then consider a Taylor expansion of $\bar \lambda$ around a fiducial mass~\cite{messenger-read,damour-nagar-villain,delpozzo,Agathos:2015uaa} and look for EoS-insensitive relations among the coefficients in the expansion (Sec.~\ref{sec:prime}).

\subsection{$\bar \lambda_s$--$\bar \lambda_a$ Relation}
\label{sec:lambdas-lambdaa}

Let us first consider relations between two dimensionless tidal parameters, $\bar \lambda_s$ and $\bar \lambda_a$, defined below Eq.~\eqref{eq:delta-Lambda-def}~\cite{I-Love-Q-Science,I-Love-Q-PRD}. Notice that the symmetric tidal parameter $\bar \lambda_s$ corresponds to the mean dimensionless tidal deformability, while the antisymmetric tidal parameter $\bar \lambda_a$ is the mean difference between the two tidal deformabilities. For an equal-mass system, they reduce to $\bar \lambda_s = \bar \lambda_1 = \bar\lambda_{2} $ and $\bar \lambda_a = 0$ respectively.  

\subsubsection{Newtonian Limit}

Before studying the relativistic relation between $\bar \lambda_s$ and $\bar \lambda_a$, let us investigate the relations in the so-called \emph{Newtonian limit}. The latter, also known as the non-relativistic limit, is nothing but the leading-order expansion in stellar compactness $C_A \equiv m_A/R_A$, where $m_A$ and $R_A$ are the NS mass and radius respectively of the $A$th star. One can carry out calculations analytically in such a limit for a polytropic EoS,
\be
p = K \rho^{1 + 1/n}\,,
\ee
where $p$ and $\rho$ correspond to the pressure and energy density respectively, $K$ is a constant and $n$ is the polytropic index. For such a polytrope, $m_A$ and $\bar \lambda_A$ are related to the compactness via~\cite{I-Love-Q-PRD}
\be
\label{eq:love-Newton}
m_A \propto C_A^{(3-n)/2}\,, \quad \bar \lambda_A = \frac{\alpha_n^{(\bar \lambda)}}{C_A^{5}}\,,
\ee
respectively, where the coefficient $\alpha_n^{(\bar \lambda)}$ depends on the polytropic index $n$. For an $n=0$ and $n=1$ polytrope, one finds $\alpha_0^{(\bar \lambda)} = {1}/{2}$  and $\alpha_1^{(\bar \lambda)} = 5/\pi^2 - 1/3$~\cite{I-Love-Q-PRD}. From Eq.~\eqref{eq:love-Newton} and the definition of $\bar \lambda_s$ and $\bar \lambda_a$, one finds
\be
\label{eq:lambdas-lambdaa-C-Newton}
\bar \lambda_{s,a} = \frac{\alpha_n^{(\bar \lambda)}}{2} \left( \frac{1}{C_1^5} \pm \frac{1}{C_2^5} \right) = \frac{\alpha_n^{(\bar \lambda)}}{2 C_1^5} \left[ 1 \pm q^{10/(3-n)} \right]\,, 
\ee
where the $+$ and $-$ correspond to  $\bar \lambda_{s}$ and $\bar \lambda_{a}$ respectively. Since in the Newtonian limit $C_{A} \ll 1$, Eqs.~\eqref{eq:love-Newton} and~\eqref{eq:lambdas-lambdaa-C-Newton} tell us that large values of $\lambda_{A}$ or large values of $\lambda_{s}$ also correspond to this limit. Solving for $q$ in terms of $\bar\lambda_{s}$ and then inserting this in Eq.~\eqref{eq:lambdas-lambdaa-C-Newton} for $\bar \lambda_{a}$, one finds the $\bar\lambda_{s}$--$\bar\lambda_{a}$ relation in the Newtonian limit
\be
\label{eq:lambdas-lambdaa-Newton}
\bar \lambda_a = F_n^{(\bar \lambda_a)}(q) \ \bar \lambda_s\,, \quad F_n^{(\bar \lambda_a)} (q) \equiv \frac{1-q^{10/(3-n)}}{1+q^{10/(3-n)}}
\,.
\ee
Observe that $F_n^{(\bar \lambda_a)}(1) = 0$ as $\bar \lambda_a$ vanishes in this case, but $F_n^{(\bar \lambda_a)}(0) = 1$ as then $\bar \lambda_a = \bar \lambda_s$. 

\fulltable{\label{table:coeff-n} Approximate polytropic index $n$ in the Newtonian regime obtained by fitting the $\bar \lambda_s$--$\bar \lambda_a$ relation (when $C_A < 0.05$ and for different realistic EoSs) to Eq.~\eqref{eq:lambdas-lambdaa-Newton}.
\vspace{3mm}}
\br
 &  AP3 & AP4 & SLy & WFF1 & WFF2 & ENG & MPA1 & MS1 & MS1b  & LS220 & Shen & mean \\
\mr
$n$ &  0.795 & 0.875 & 0.832 & 0.977 & 0.869 & 0.873 & 0.714 & 0.411 & 0.581 & 0.613 & 0.632 & 0.743 \\
\br
\endfulltable

\begin{figure}[htb]
\begin{center}
\includegraphics[width=8.cm,clip=true]{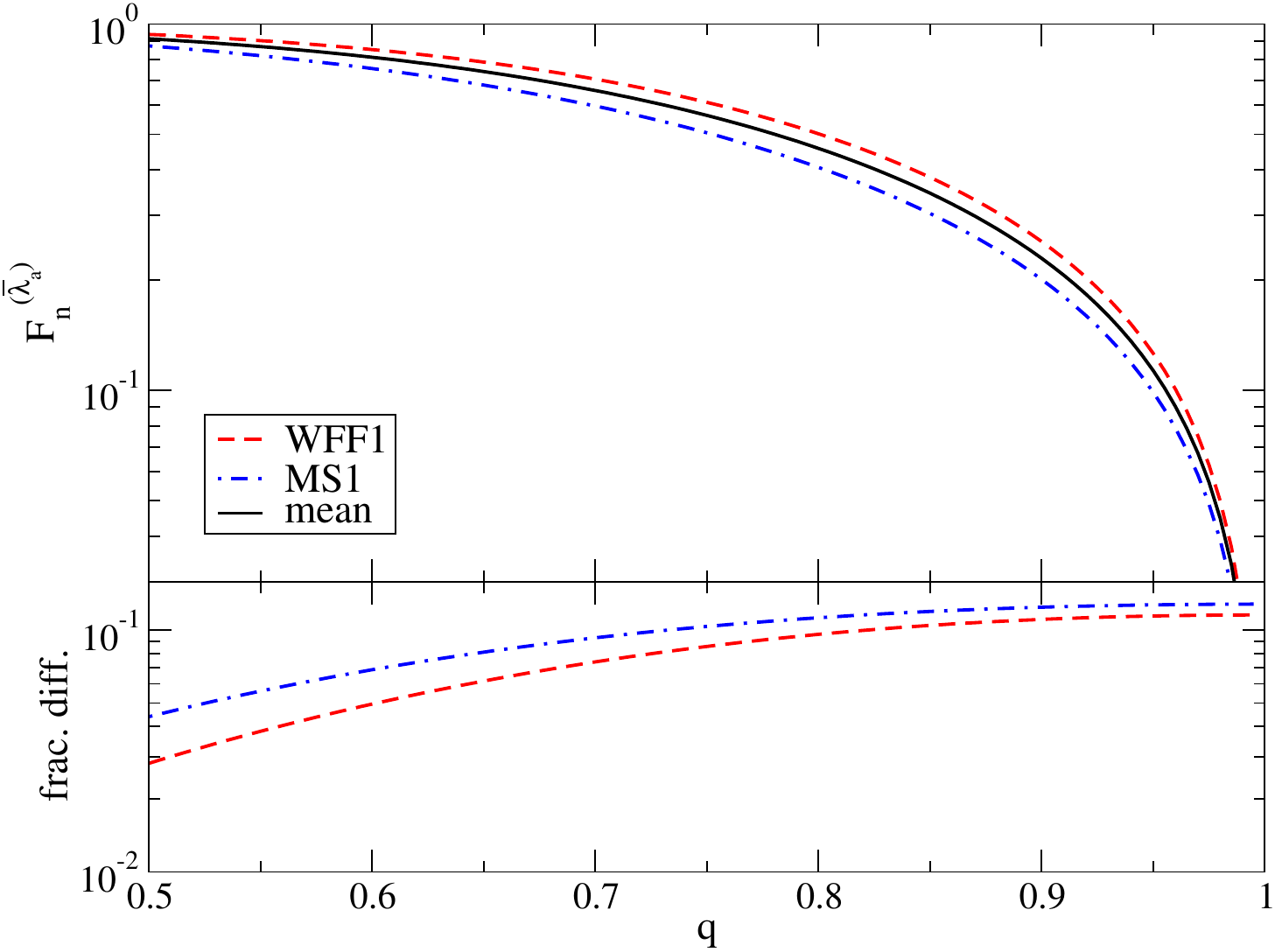}  
\caption{\label{fig:lambdas-lambdaa-Newton} (Top) Coefficient $F_n^{(\bar \lambda_a)}$ of the $\bar \lambda_s$--$\bar \lambda_a$ relation in the Newtonian limit [see Eq.~\eqref{eq:lambdas-lambdaa-Newton}] against $q$ for a polytropic EoS with $n=0.977$, $n=0.411$ and $n=0.743$, which correspond to the fitted polytropic indices for WFF1, MS1 and the mean value respectively. (Bottom) Absolute value of the fractional difference of the WFF1 or MS1 relations relative to the mean $n$ case of Table~\ref{table:coeff-n}.
}
\end{center}
\end{figure}

Let us now investigate $F_n^{(\bar \lambda_a)}(q)$ for values of $n$ that approximate more realistic EoSs in the Newtonian limit. The latter can be estimated  by first constructing a sequence of NS solutions using a set of realistic EoSs with $C_A < 0.05$, and then computing the realistic $\bar \lambda_s$--$\bar\lambda_a$ relation and fitting Eq.~\eqref{eq:lambdas-lambdaa-Newton} to it. We show the best fit value of $n$ in Table~\ref{table:coeff-n}, where notice that the mean value is $n \sim 0.743$. The top panel of Fig.~\ref{fig:lambdas-lambdaa-Newton} shows $F_n^{(\bar \lambda_a)}$ against $q$ with $n$'s that approximate the WFF1 (the softest EoS considered) and MS1 EoSs (the stiffest EoS considered), as well as the mean $n$. Observe that $F_n^{(\bar \lambda_a)}$ is close to unity for $q \sim 0.5$,  while it drops quickly to zero in the limit $q \to 1$ as explained earlier. The bottom panel shows the relative fractional difference between the WFF1 or the MS1 relations and the mean $n$ relation. Notice that the $n$-dependence is smaller in the $q \sim 0.5$ region compared to that in the $q \sim 1$ region. But even in the latter case, the Newtonian relation is approximately EoS-insensitive to an accuracy of $\sim 13\%$.

One may think that the EoS-universality should be exact in the $q \to 1$ limit, since then $F_n^{(\bar \lambda_a)} \to 0$, but this contradicts the bottom panel of Fig.~\ref{fig:lambdas-lambdaa-Newton}. Such a naive argument does not hold because Fig.~\ref{fig:lambdas-lambdaa-Newton} considers the relative fractional difference, $(F_n^{(\bar \lambda_a)}-F_{\bar n}^{(\bar \lambda_a)})/F_{\bar n}^{(\bar \lambda_a)}$ (where $\bar n$ is the mean $n$ of Table~\ref{table:coeff-n}), and not the absolute difference. Although the numerator of this fraction vanishes in the $q \to 1$ limit, the denominator does as well. Thus, the fact that the fractional difference in $F_n^{(\bar \lambda_a)}$ does not vanish in the $q\to 1$ limit is a reflection of the ratio $(F_n^{(\bar \lambda_a)}-F_{\bar n}^{(\bar \lambda_a)})/F_{\bar n}^{(\bar \lambda_a)}$ being finite in this limit. We can see this explicitly by expanding the relative fractional difference around $q = 1$:
\begin{align}
\label{eq:lambdas-lambdaa-q1}
\frac{F_n^{(\bar \lambda_a)}-F_{\bar n}^{(\bar \lambda_a)}}{F_{\bar n}^{(\bar \lambda_a)}} &= \frac{n-\bar n}{3 - \bar n} + \mathcal{O}\left[ (1-q)^2 \right] \nn \\
& =  - 0.128  + \mathcal{O}\left[ (1-q)^2 \right] \ (\mathrm{MS1})\,, 
\end{align}
where we evaluated the second line with the data of Table~\ref{table:coeff-n} for the MS1 EoS. Notice that the correction is of $\mathcal{O}\left[ (1-q)^2 \right]$ and not of $\mathcal{O}\left[ (1-q) \right]$. Equation~\eqref{eq:lambdas-lambdaa-q1} agrees with the $13\%$ variation in the $\lambda_s$--$\lambda_a$ relation in the Newtonian, $q \to 1$ limit shown in the bottom panel of Fig.~\ref{fig:lambdas-lambdaa-Newton}.

\Table{\label{table:coeff} Coefficients of the fit in Eq.~\eqref{eq:fit} for various EoS-insensitive relations. The last row shows the r-squared value of the fit.
\vspace{3mm}
}
\br
$x^{-1}$ & $\bar \lambda_s$ & $\bar \Lambda$ & $\bar \lambda_s$ & $\bar \lambda_s$   \\
$y$ & $\bar{\lambda}_a$ & $\delta \bar \Lambda$ & $\bar{Q}_s$ & $\bar{Q}_a$   \\
\mr
$n$ & 0.743 & 0.743 & 1 & 1 \\
$\alpha$ & -1 & -1 & 1/5 & 1/5 \\ 
$a$ & 0.07550 & 0.07319 & 0.1697 & $-0.01298$ \\
$b_{11}$ & $-2.235$ &$-3.598$ & $-11.22$ & $0.8593$  \\
$b_{12}$ & 0.8474 & $1.773$ & $18.90$ & $-0.6626$ \\
$b_{21}$ & 10.45 & $18.50$  & $35.95$ & $4.529$  \\
$b_{22}$ & $-3.251$ & $-10.84$  & $-44.35$ & $4.862$  \\
$b_{31}$ & $-15.70$ & $-30.45$ & $-31.23$ & $6.398$  \\
$b_{32}$ & 13.61 & $42.60$ & $57.49$ & $-6.911$ \\
$c_{11}$ & $-2.048$ & $-1.925$ & $-9.112$ & $0.8971$  \\
$c_{12}$ & $0.5976$ & $-0.2747$ & $18.21$ & $-0.6850$ \\
$c_{21}$ & 7.941 & $4.788$  & $-0.04962$ & $-4.967$  \\
$c_{22}$ & 0.5658 & $9.178$  & $4.085$ & $5.373$  \\
$c_{31}$ & $-7.360$ & $-1.240$ & 73.60 & $7.730$  \\
$c_{32}$ & $-1.320 $ & $-13.55$ & $-3.177$ & $-8.524$ \\
\mr
$r^2$ & 0.9996 & 0.9939 & 0.9998 & 0.9971  \\
\br
\endTable

\subsubsection{Relativistic Relations}
\label{sec:lambdas-lambdaa-relativistic}

We now turn our attention to the relativistic relation. We construct this as explained at the beginning of Sec.~\ref{sec:binary-Love}, without expanding in powers of compactness. Figure~\ref{fig:lambdas-lambdaa-3D} presents $\bar \lambda_a$ against $\bar \lambda_s$ and $q$ from two different viewpoints for eleven different realistic EoSs with $m_A > 1 M_\odot$. Inspired by the Newtonian relation in Eq.~\eqref{eq:lambdas-lambdaa-Newton}, we created a fit for the relation given by
\be
\label{eq:fit}
y = F_{n}^{(y)} (q) \; x^\alpha \; \frac{a + \sum_{i=1}^3 \sum_{j=1}^2 b_{ij} q^j x^{i/5} }{a + \sum_{i=1}^3 \sum_{j=1}^2 c_{ij} q^j x^{i/5} }\,,
\ee
with $x=1/\bar \lambda_s$, $y=\bar \lambda_a$ and $n = \bar n = 0.743$. The fitted coefficients $\alpha$, $a$, $b_{ij}$ and $c_{ij}$ are given in the second column of Table~\ref{table:coeff}. Using the Newtonian relation of Eq.~\eqref{eq:lambdas-lambdaa-Newton} as a controlling factor in the fit has one big advantage: by construction, the relation reduces to (i) this equation in the Newtonian limit when $\bar \lambda_s \to \infty$, (ii) $\bar \lambda_a \to \bar \lambda_s$ with $q \to 0$ and (iii) $\bar \lambda_a \to 0$ with $q \to 1$ (as $F_{n}^{(\bar \lambda_a)} (1) = 0$). The fit corresponds to a double (or \emph{bivariate}) expansion in $q$ and $\bar \lambda_s^{-1/5}$, where we expand in this power of $\bar \lambda_{s}$ because $\bar \lambda_s^{-1/5} \propto C_A$ in the Newtonian limit. Namely, we simultaneously expand asymptotically in small mass ratios and in the Newtonian region of the 3D plot in Fig.~\ref{fig:lambdas-lambdaa-3D}. Although the fit becomes less accurate as $\bar \lambda_s \to 0$ and $q \to 1$, the fit is reasonably good as $q \leq 1$ and $1/\bar \lambda_s \leq 0.5$. This fit is depicted as the blue approximately invariant plane of Fig.~\ref{fig:lambdas-lambdaa-3D}. Observe that the binary Love relation for each EoS lies nicely on this plane, which shows that the relation is quite EoS-insensitive. Let us quantify this statement further. 

\begin{figure}[htb]
\begin{center}
\includegraphics[width=7.5cm,clip=true]{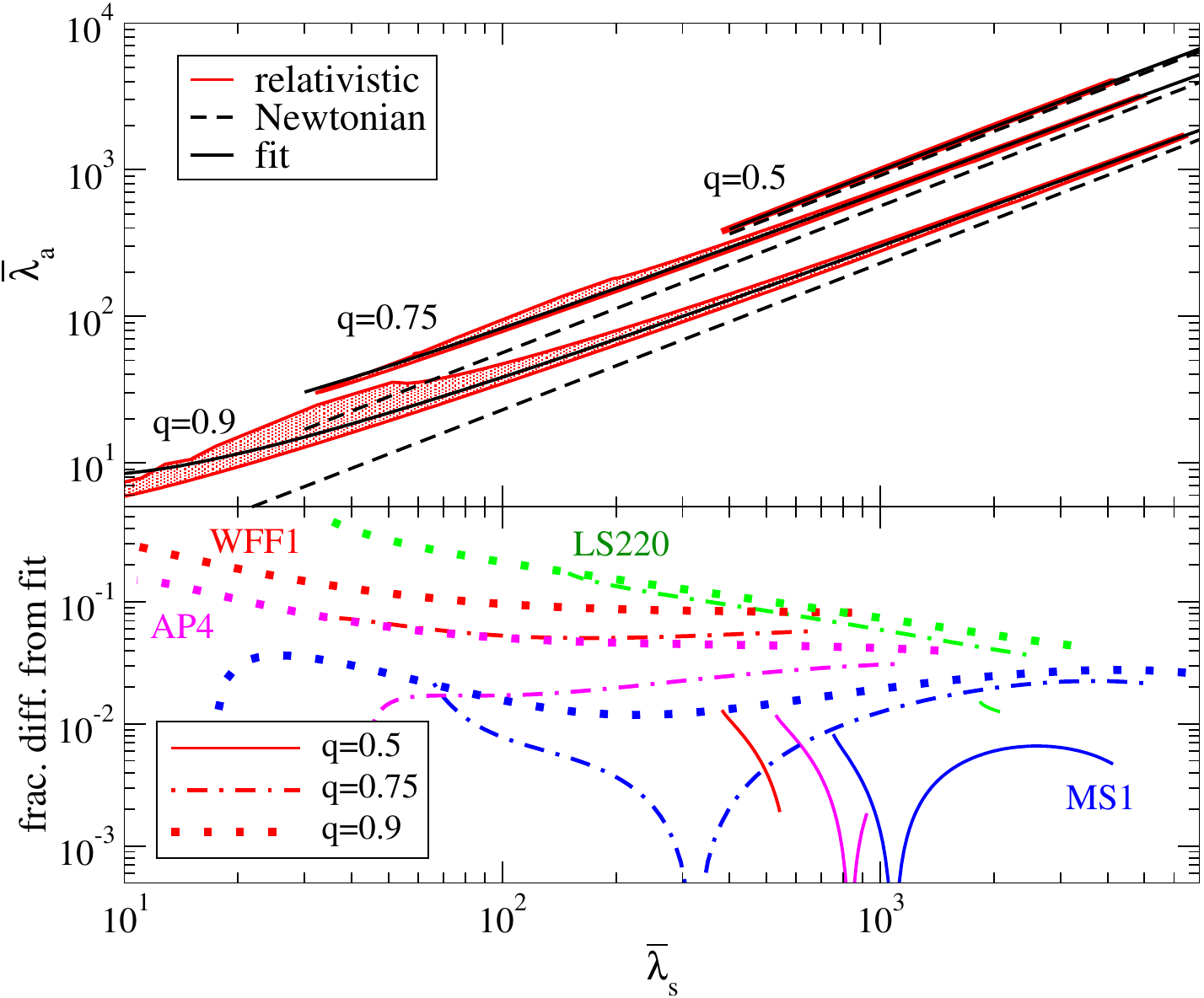}  
\includegraphics[width=6.2cm,clip=true]{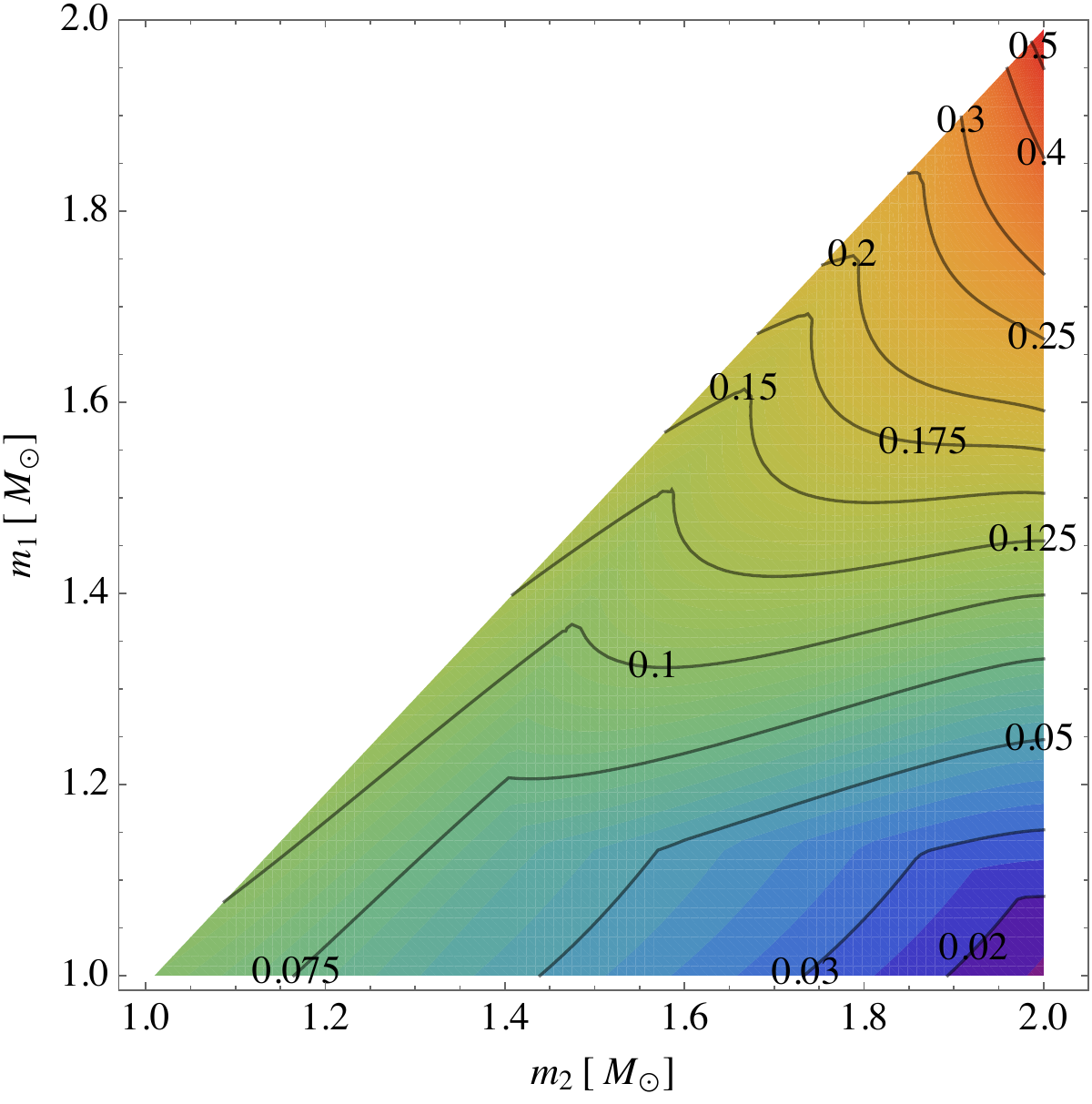}  
\caption{\label{fig:lambdas-lambdaa} (Top left) The  $\bar \lambda_s$--$\bar \lambda_a$ relation for $q = 0.5$, 0.75 and 0.9, together with the fit in Eq.~\eqref{eq:fit} and the Newtonian relation with $n=0.743$ in Eq.~\eqref{eq:lambdas-lambdaa-Newton}. The red shaded regions correspond to the parameter space spanned by the relation with different EoSs. (Bottom left) Absolute value of the fractional difference between the numerical data and the fit, with WFF1 (red), MS1 (blue), LS220 (green) and AP4 (magenta) EoSs. 
(Right) Maximum absolute fractional difference among eleven realistic EoSs between the numerically obtained $\bar \lambda_s$--$\bar \lambda_a$ relation and the fit for various NS masses with $m_1 \leq m_2$. Observe that the relation is EoS-insensitive to $\sim 15\%$ for NS binaries with the NS mass smaller than $1.6M_\odot$.
} 
\end{center}
\end{figure}

The top left panel of Fig.~\ref{fig:lambdas-lambdaa} shows the $\bar \lambda_s$--$\bar \lambda_a$ relation for NSs with $q = 0.5$, 0.75 and 0.9 (constant $q$ hypersurfaces of Fig.~\ref{fig:lambdas-lambdaa-3D}), where $q=0.75$ correspond to the mass ratio of the currently-known double neutron star binary J0453+1559~\cite{Martinez:2015mya}, while $q=0.9$ roughly corresponds to that of J0737-3039 (the double binary pulsar)~\cite{burgay,lyne,kramer-double-pulsar} and J1756-2251~\cite{Faulkner:2004ha}. The relation is given by a single curve for a fixed $q$ and a given EoS (see Fig.~1 of~\cite{Yagi:2015pkc}), with the single parameter along the curve being the mass, or equivalently, the compactness. The red shaded region for each $q$ corresponds to the parameter space spanned by the relation with different EoSs. A smaller area corresponds to a stronger universality. We also show the fit of Eq.~\eqref{eq:fit}, as well as the Newtonian relation of Eq.~\eqref{eq:lambdas-lambdaa-Newton} with the mean $n=\bar{n}$. The bottom panel of this figure shows the relative fractional difference between the numerical data and the fit. We selected four representative realistic EoSs: WFF1, AP4, LS220 and MS1, which lead to stars with a radius of 10.4km, 11.4km, 12.6km and 14.9km respectively when the NS mass is 1.4$M_\odot$. We only considered cases in which the smallest NS has a mass larger than $1 M_\odot$ and the largest NS has a mass smaller than the maximum allowed to have a stable configuration given the EoS. For example, the LS220 EoS predicts a maximum NS mass of $m_{\rm{LS220}}^{\max} \sim 2.1 M_{\odot}$, and thus, when $q=0.5$ there is only a very small subset of binary masses $(m_{1},m_{2})$ that satisfy $ 1 M_{\odot} \leq m_{A} \leq m_{\rm{LS220}}^{\max}$ and with which one can compute the $\bar \lambda_s$--$\bar \lambda_a$ relation. 

The left panel of Fig.~\ref{fig:lambdas-lambdaa} contains several features that ought to be discussed further. First, observe that the numerical data approaches the analytic Newtonian relation in the large $\bar \lambda_s$ limit. This makes sense since $\bar{\lambda}_{s} \propto C_{A}^{-5}$, and thus, as $C_{A} \to 0$ then $\bar{\lambda}_{s} \to \infty$ and one approaches the Newtonian limit. Second, observe that when $q = 0.5$, the difference between the Newtonian and numerical relation is small, while it becomes larger as one increases $q$. This is because the relation reduces to $\bar \lambda_a = \bar \lambda_s$ in the $q \to 0$ limit both in the Newtonian and relativistic regimes, while the relativistic correction to the Newtonian relation becomes more important as one increases $q$. This is one of the reasons that we chose the Newtonian relations as the controlling factor of the fit in Eq.~\eqref{eq:fit}.

Let us now consider how universal the relation is when exploring a large set of EoSs, a fact that cannot be inferred by considering only the four EoSs of the bottom left panel of Fig.~\ref{fig:lambdas-lambdaa}. The $(m_{1},m_{2})$ contour plot in the right panel shows the \emph{maximum} fractional difference between the fit and the numerical data obtained using all eleven realistic EoSs we considered in this paper. Observe that the approximate universality holds to $\sim 20\%$ if $m_1 \lesssim 1.6M_\odot$ and for all $m_{2}$. For example, the maximum fractional difference for $(m_1,m_2) = (1.3,1.4)M_\odot$ is roughly $11\%$. On the other hand, observe that the maximum fractional difference can reach a maximum of $\sim 50\%$ when $m_{1} \sim 2 M_{\odot} \sim m_{2}$. In fact, notice that the maximum fractional difference increases from $\sim 10\%$ to $\sim 50\%$ as one considers stars on the $q = 1$ line of ever larger mass. 

\begin{figure}[htb]
\begin{center}
\includegraphics[width=8.cm,clip=true]{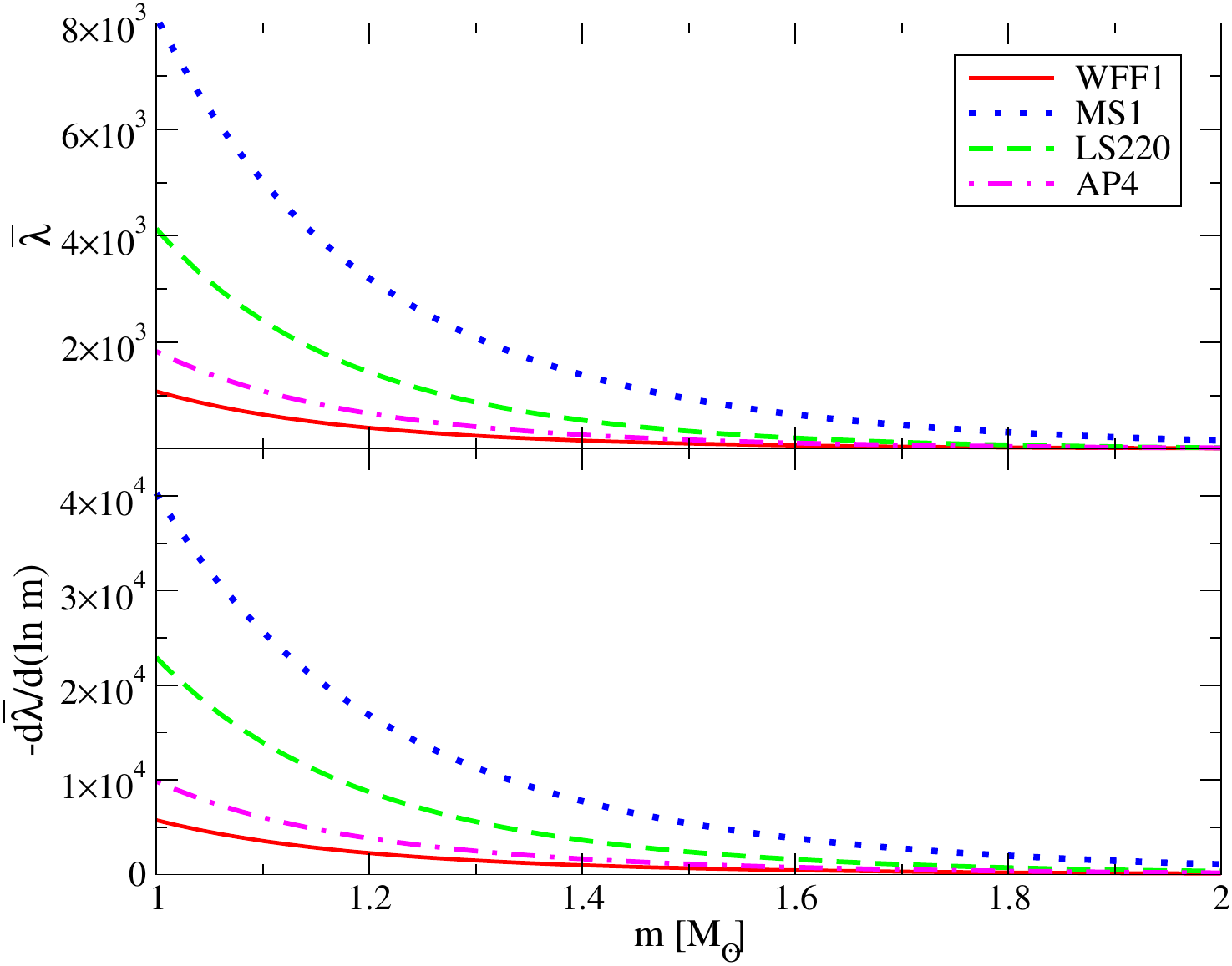}  
\caption{\label{fig:lambda-M-diff} $\bar \lambda$ (top) and $d \bar \lambda / d \ln m$ (bottom) as a function of $m$ for an isolated NS with four representative EoSs. Observe how $\bar \lambda$ becomes more insensitive to $m$ as one increases $m$.
}
\end{center}
\end{figure}

Why does the universality become worse as one increases the NS mass? One might find this puzzling since the stellar sequence approaches the BH limit as one increases the mass (or equivalently, as one increases the stellar compactness), and in this limit, the universality should become exact (because of the no-hair theorems of General Relativity (GR)). Indeed, $\bar \lambda_{1}(m)$ and $\bar \lambda_{2}(m)$ do approach their BH values in the BH limit, but their \emph{averaged difference}, $\bar \lambda_{a}$, is sensitive to \emph{how} this limit is approached, or in other words, to the slope of the $\bar \lambda$--$m$ relation. Figure~\ref{fig:lambda-M-diff} shows $\bar \lambda$ (top) and $-d \bar \lambda / d \ln m$ (bottom) against $m$ for four different EoSs. Observe that $\bar \lambda$ becomes less sensitive to $m$ as one increases $m$, since $-d \bar \lambda / d \ln m$ decreases. Physically, the slope of the $\bar \lambda$--$m$ relation decreases with increasing mass because $\bar \lambda(m)$ decreases very fast with stellar compactness, which, in turn, is simply due to the fact that more dense stars deform less than less dense stars [see Eq.~\eqref{eq:love-Newton}]. Therefore, when one calculates the relative fractional difference in $\bar \lambda_a$, both the numerator and the denominator of this fraction decrease as $m$ increases. That the universality deteriorates in the BH limit is then simply a consequence of the slope of the $\bar \lambda$--$m$ relation decreasing faster than the EoS variability in $\bar \lambda_{a}$.

\begin{figure}[htb]
\begin{center}
\includegraphics[width=8.cm,clip=true]{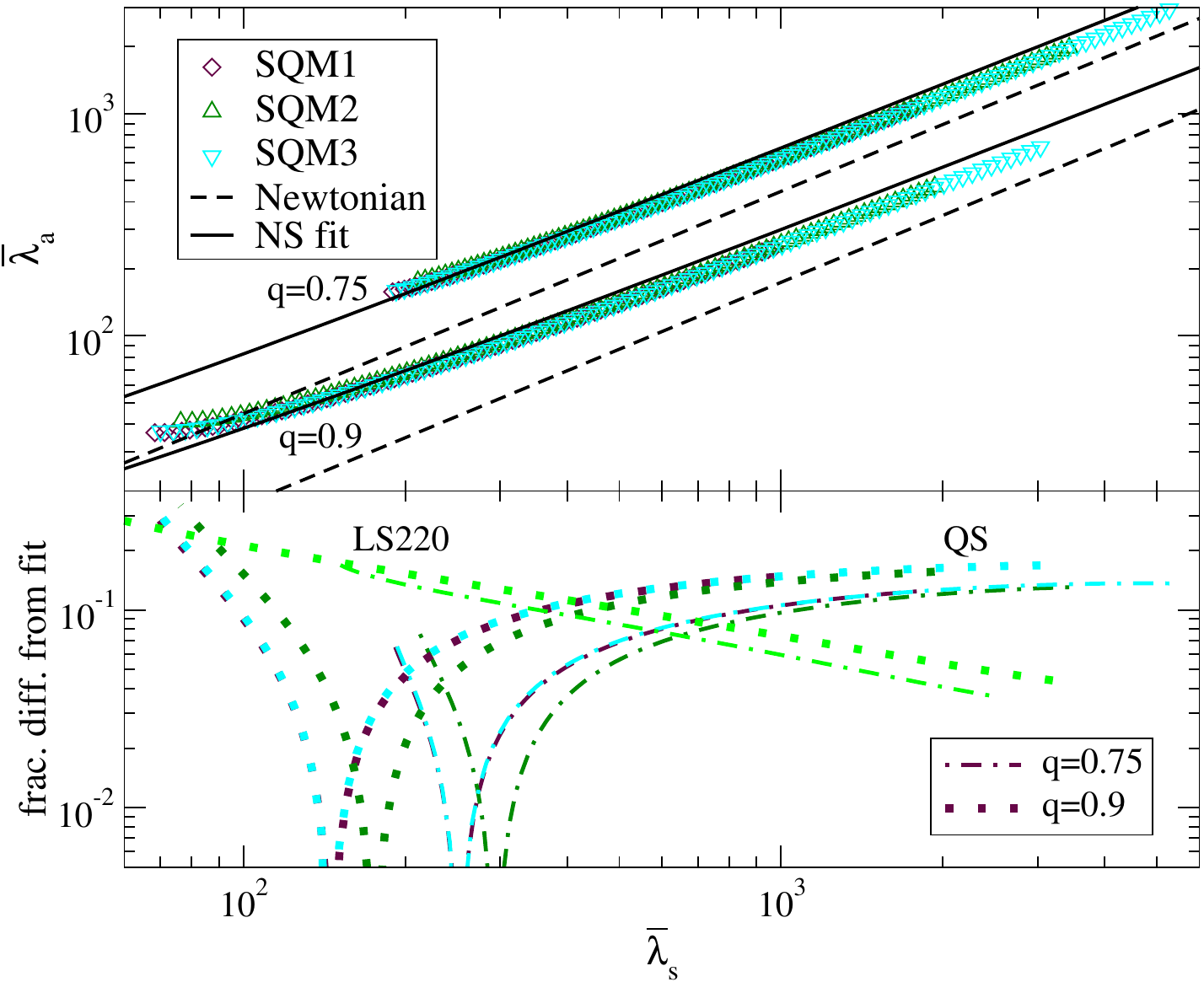}  
\caption{\label{fig:lambdas-lambdaa-QS} Same as Fig.~\ref{fig:lambdas-lambdaa} but for QSs. The fit is the one created for the NS relations and the Newtonian relation is with $n=0$. In the bottom panel, we also show the absolute fractional difference between the NS fit and the LS220 relation for reference.
}
\end{center}
\end{figure}

Let us end this section by studying the binary Love relations for QSs. The top panel of Fig.~\ref{fig:lambdas-lambdaa-QS} presents the $\bar \lambda_s$--$\bar \lambda_a$ relations using three different QS EoSs, SQM1, SQM2 and SQM3~\cite{SQM}. We also show the fit created for the NS relations and the Newtonian relation with $n=0$. The latter choice is made because the QS EoSs are similar to a constant density EoS (i.e.~an $n=0$ polytropic EoS) in the low-pressure regime~\cite{lattimer_prakash2001}. One sees that the QS relations are slightly different from the NS ones, in particular when $q$ is close to 1. 
The bottom panel of Fig.~\ref{fig:lambdas-lambdaa-QS} shows the absolute fractional difference between the QS relations and the NS fit. 
When $q = 0.75$ or 0.9, the difference between the two relations can be as large as $\mathcal{O}(10)\%$ in the large $\bar \lambda_s$ regime, which is comparable to the intrinsic EoS-variation in the NS relations. This shows that one can still apply the relations for NSs to QSs. That is, a separate set of QS universal relations is not needed at this stage.

\subsection{$\bar \Lambda$--$\delta \bar \Lambda$ Relation}
\label{sec:Lambda-delta-Lambda}

Another way to choose two independent tidal parameters in the gravitational waveform is through the coefficients in the PN expansion of the GW phase. For non-spinning compact binaries, finite-size effects enter the GW phase first at $5$PN order relative to the leading-order (Newtonian) term~\cite{flanagan-hinderer-love}, with 1PN corrections to this entering at 6PN order. We can then choose the two independent tidal parameters to be the combination of tidal deformabilities that enter first as a coefficient of the 5PN (the chirp tidal deformability $\bar \Lambda$) and the 6PN ($\delta \bar \Lambda$) terms in the phase defined in Eqs.~\eqref{eq:Lambda-def} and~\eqref{eq:delta-Lambda-def}, where 
\begin{align}
f(\eta) &\equiv  \frac{16}{13} (1 + 7 \eta - 31 \eta^2)\,, 
\\
g(\eta) &\equiv - \frac{16}{13} \sqrt{1- 4 \eta} (1 + 9 \eta - 11 \eta^2)\,,
\\
\delta f(\eta) &\equiv \sqrt{1-4\eta} \left( 1-\frac{13272}{1319} \eta + \frac{8944}{1319} \eta^2 \right)\,, 
\\
\delta g(\eta) &\equiv 
-  \left( 1 - \frac{15910}{1319} \eta + \frac{32850}{1319} \eta^2 + \frac{3380}{1319} \eta^3 \right)\,.
\end{align}
Such a choice was made e.g. in~\cite{flanagan-hinderer-love,hinderer-lackey-lang-read,Favata:2013rwa,Wade:2014vqa,Lackey:2014fwa}\footnote{$\bar \Lambda$ and $\delta\bar \Lambda$ in Eqs.~\eqref{eq:Lambda-def} and~\eqref{eq:delta-Lambda-def} are equivalent to $\tilde \Lambda$ and $\delta \tilde \Lambda$ in Eqs.~(5) and~(6) of~\cite{Wade:2014vqa}. The signs in front of terms proportional to $\bar \lambda_a$ are different because we assume $m_1 \leq m_2$ (so that $\bar \lambda_a$ is positive definite) in this paper, while Ref.~\cite{Wade:2014vqa} assumed $m_1 > m_2$.}. Observe that for an equal mass binary, these parameters reduce to $\bar \Lambda = \bar \lambda_s = \bar \lambda_A$ and $\delta \bar \Lambda = 0$. Although such parametrization is less intuitive than $(\bar \lambda_s,\bar \lambda_a)$, 
the former has the data analysis advantage that the correlation between $\bar \Lambda$ and $\delta \bar \Lambda$ is smaller relative to that between $\bar \lambda_s$ and $\bar \lambda_a$. 

\subsubsection{Newtonian Limit}

As in Sec.~\ref{sec:lambdas-lambdaa}, let us first investigate the relation in the Newtonian limit, i.e.~to leading order in an expansion about small compactness. Such a relation can easily be obtained by substituting the Newtonian $\bar \lambda_s$--$\bar \lambda_a$ relation of Eq.~\eqref{eq:lambdas-lambdaa-Newton} into the definition of $\bar \Lambda$ and $\delta \bar \Lambda$ in Eqs.~\eqref{eq:Lambda-def} and~\eqref{eq:delta-Lambda-def} so as to obtain $\bar \Lambda (\bar \lambda_s, \eta)$ and $\delta \bar \Lambda (\bar \lambda_s, \eta)$. Next, we solve $\bar \Lambda (\bar \lambda_s, \eta)$ for $\bar \lambda_s$ to obtain $\bar \lambda_s (\bar \Lambda, \eta)$. Finally, we substitute this into $\delta \bar \Lambda (\bar \lambda_s, \eta)$ to obtain $\delta \bar \Lambda (\bar \Lambda, \eta)$, which leads to the Newtonian relation
\be
\label{eq:Lambda-delta-Lambda-Newton}
\delta \bar \Lambda = F_n^{(\delta \bar \Lambda)} (q) \, \bar \Lambda\,, 
\ee
where
\be
\label{eq:Fn_delta_Lambda}
F_n^{(\delta \bar \Lambda)} \equiv -\frac{13}{21104} \frac{1}{1 + q}   \frac{\left[1-F_n^{(\bar \lambda_a)} (q)\right] N_1(q) - \left[1+F_n^{(\bar \lambda_a)} (q)\right] q^6 N_1(1/q)}{\left[1-F_n^{(\bar \lambda_a)} (q)\right] N_2(q) + \left[1+F_n^{(\bar \lambda_a)} (q)\right] q^5 N_2(1/q)} \,,
\ee
with 
\be
N_1(q) := 11005 q^2+7996 q-1319\,, 
\qquad
N_2(q) := 12q+1\,. 
\ee
In the small mass-ratio limit, $F_n^{(\delta \bar \Lambda)}$ becomes $F_n^{(\delta \bar \Lambda)} = 13/16 + \mathcal{O}(q^2)$, and hence the relation is exactly EoS-universal. On the other hand, in the equal-mass limit, the fractional difference between the relation with a polytropic index of $n$ and $\bar n$ becomes 
\begin{align}
\label{eq:Lambda-delta-Lambda-q1}
\frac{F_n^{(\delta \bar \Lambda)}-F_{\bar n}^{(\delta \bar \Lambda)}}{F_{\bar n}^{(\delta \bar \Lambda)}} &= \frac{2947}{643+768 \bar n} \frac{(n-\bar n)}{3-n} + \mathcal{O}\left[ (1-q)^2 \right] \nn \\
& =  - 0.311  + \mathcal{O}\left[ (1-q)^2 \right] \ (\mathrm{MS1})\,, 
\end{align}
where, in the last equality, we set $n = 0.411$ and $\bar n = 0.743$ (see Table~\ref{table:coeff-n}). Observe that the fractional difference is larger than the one for the Newtonian $\bar \lambda_s$--$\bar \lambda_a$ relation by $2947/(643+768 \bar n) \sim 2.43$ for $\bar n = 0.743$. We thus expect the relativistic $\bar \Lambda$--$\delta \bar \Lambda$ relations to approach the Newtonian relations more slowly than the $\bar{\lambda}_{s}$--$\bar{\lambda}_{a}$ ones do as the compactness increases. 

\begin{figure}[htb]
\begin{center}
\includegraphics[width=7.5cm,clip=true]{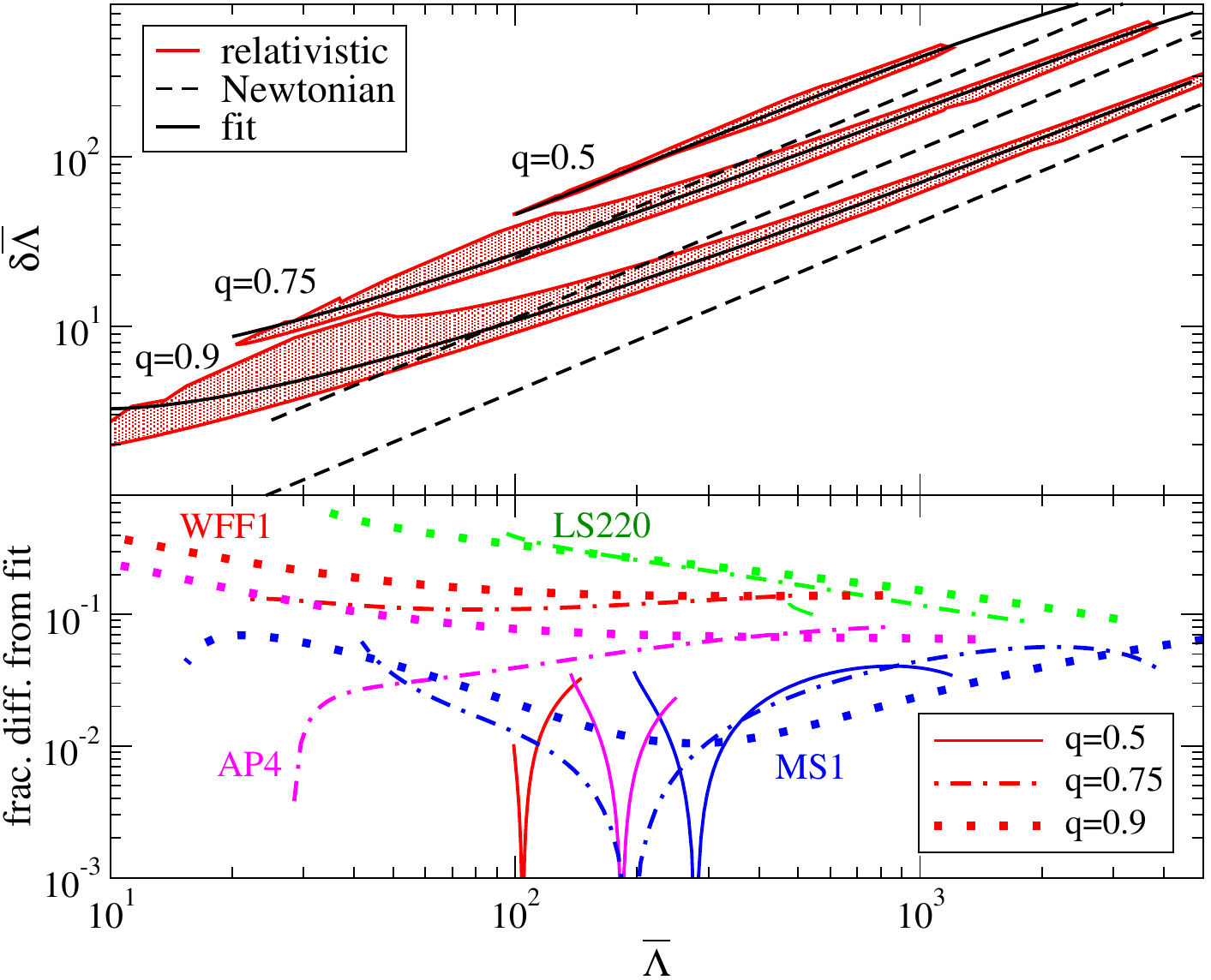}  
\includegraphics[width=6.2cm,clip=true]{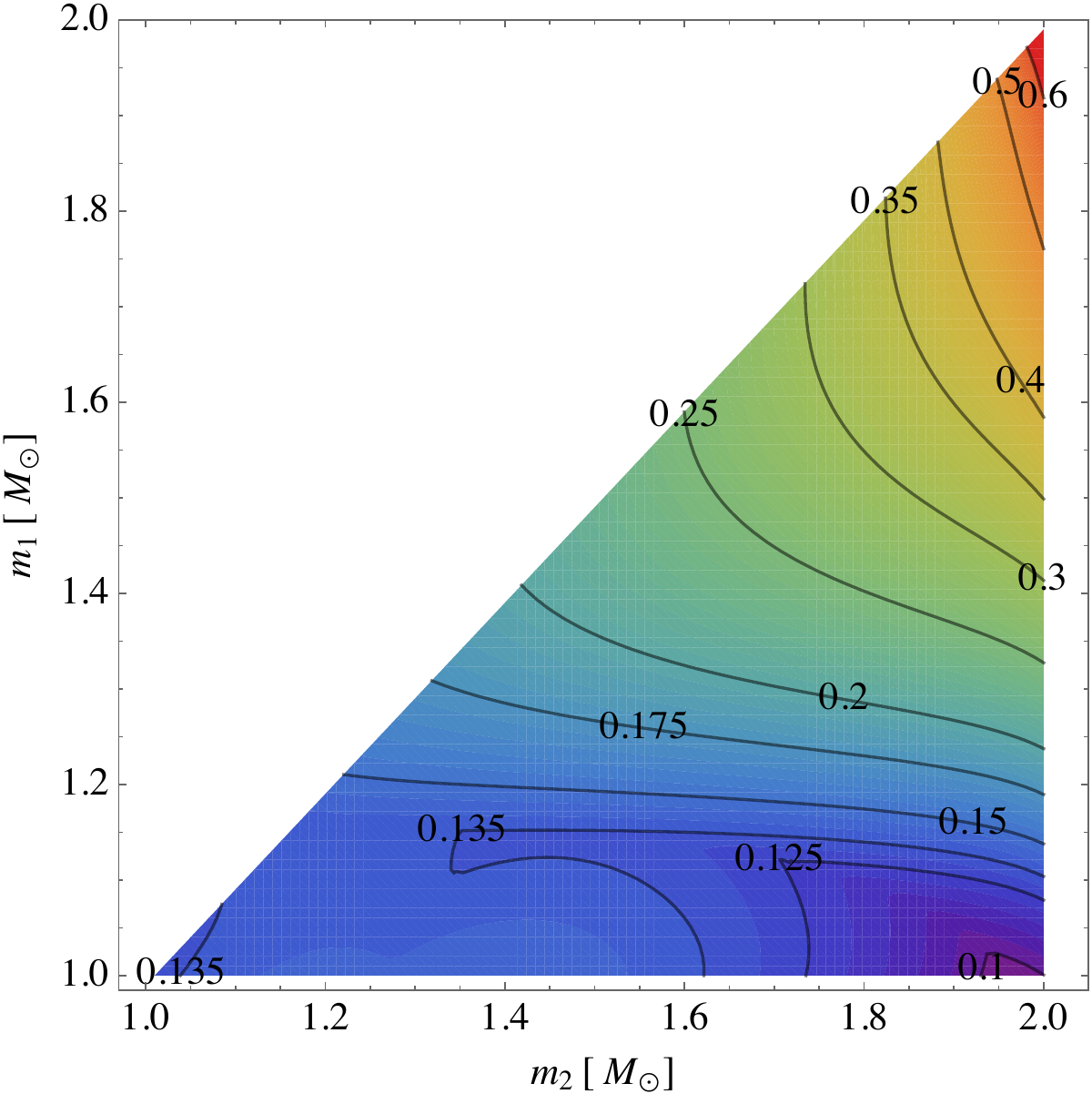}  
\caption{\label{fig:Lambda-delta-Lambda}  (Left) Same as Fig.~\ref{fig:lambdas-lambdaa} but for the $\bar \Lambda$--$\delta \bar \Lambda$ relation. Observe that the absolute fractional difference is larger than that for the $\bar \lambda_s$--$\bar \lambda_a$ relation in Fig.~\ref{fig:lambdas-lambdaa}.
}
\end{center}
\end{figure}

\subsubsection{Relativistic Relations}

Let us now return to the relativistic relations. Once relativistic NSs have been constructed with a given EoS, one can easily obtain the relativistic $\bar \Lambda$--$\delta \bar \Lambda$ relations, just as we did in Sec.~\ref{sec:lambdas-lambdaa-relativistic} for the $\bar \lambda_s$--$\bar \lambda_a$ relation. The top left panel of Fig.~\ref{fig:Lambda-delta-Lambda} is the same as that of Fig.~\ref{fig:lambdas-lambdaa} but for the $\bar \Lambda$--$\delta \bar \Lambda$ relation. As in the latter figure, we focus on fixed $q=0.5$, 0.75 and 0.9, and also plot the Newtonian relation with $n=0.743$ and a fit to the numerical data. The latter is created from the fitting function of Eq.~\eqref{eq:fit}, with $x=1/\bar \Lambda$, $y=\delta \bar \Lambda$, $n = 0.743$, $F_n^{(\delta \bar \Lambda)}$ given in Eq.~\eqref{eq:Fn_delta_Lambda} and the fitting coefficients of Table~\ref{table:coeff}. As in the $\bar \lambda_s$--$\bar \lambda_a$ case of Fig.~\ref{fig:lambdas-lambdaa}, the numerical results approach the Newtonian relation in the large $\bar \Lambda$ limit. The bottom left panel of Fig.~\ref{fig:Lambda-delta-Lambda} shows the absolute fractional difference between the numerical data and the fit. As in the $\bar \lambda_s$--$\bar \lambda_a$ relation, the difference becomes larger as one increases $q$. 

Let us now study the maximum absolute fractional difference of the $\bar \Lambda$--$\delta \bar \Lambda$ relation. The right panel of Fig.~\ref{fig:Lambda-delta-Lambda} shows an $(m_1,m_2)$ contour plot of the maximum absolute fractional difference between the fit and all eleven realistic EoSs considered in this paper. Comparing this to the right panel of Fig.~\ref{fig:lambdas-lambdaa}, one sees that the fractional error is larger in the $\bar \Lambda$--$\delta \bar \Lambda$ relation than in the $\bar \lambda_s$--$\bar \lambda_a$ relation. Even then, the EoS-variation of the relation is less than $25\%$ for $m_1 \leq m_2 \lesssim 1.6 M_\odot$. For a system of $(m_1,m_2) = (1.3,1.4)M_\odot$, the maximum absolute fractional difference is roughly $16\%$. On the other hand, as one increases the mass for a fixed $q$, the EoS variation becomes larger for the same reason as it does in the $\bar \lambda_s$--$\bar \lambda_a$ case.

\subsection{$\bar \lambda_0^{(0)}$--$\bar \lambda_0^{(k)}$ Relations}
\label{sec:prime}

Yet another parametrization of finite size effects in the GW phase of inspiraling NS binaries is through a Taylor expansion of the dimensionless tidal deformability $\bar \lambda$ around a reference mass $m_0$, taking the coefficients in the expansion as tidal parameters:
\be
\label{eq:Taylor}
\bar \lambda_{(\N)} (m) \equiv  \sum_{k=0}^N \frac{\bar \lambda_0^{(k)}}{k!} \left(1 - \frac{m}{m_0}  \right)^k\,, 
\qquad
\bar \lambda_0^{(k)}  \equiv  (-1)^k m_0^k  \frac{d^k \bar \lambda}{d m^k} \bigg|_{m=m_0}\,.
\ee
Of course, what enters the GW phase is $\bar{\lambda}_{1} = \bar{\lambda}_{(\infty)}(m_{1})$ and $\bar{\lambda}_{2} = \bar{\lambda}_{(\infty)}(m_{2})$. A similar parametrization was suggested e.g.~in Refs.~\cite{messenger-read,damour-nagar-villain,delpozzo,Agathos:2015uaa}. The advantage of choosing this kind of Taylor-expanded parametrization is that the coefficients $\bar \lambda_0^{(k)}$ are constant, and thus, they are the same irrespective of the NS mass and one can increase the measurement accuracy of such tidal parameters with multiple GW detections. The disadvantage is that one might have to include not only $\bar \lambda_0^{(0)}$ but also higher order tidal parameters if the NS mass is not close to $m_0$; this introduces correlations between the tidal parameters that could deteriorate the accuracy to which $\bar{\lambda}_{0}^{(0)}$ can be measured.

\begin{figure}[htb]
\begin{center}
\includegraphics[width=8.cm,clip=true]{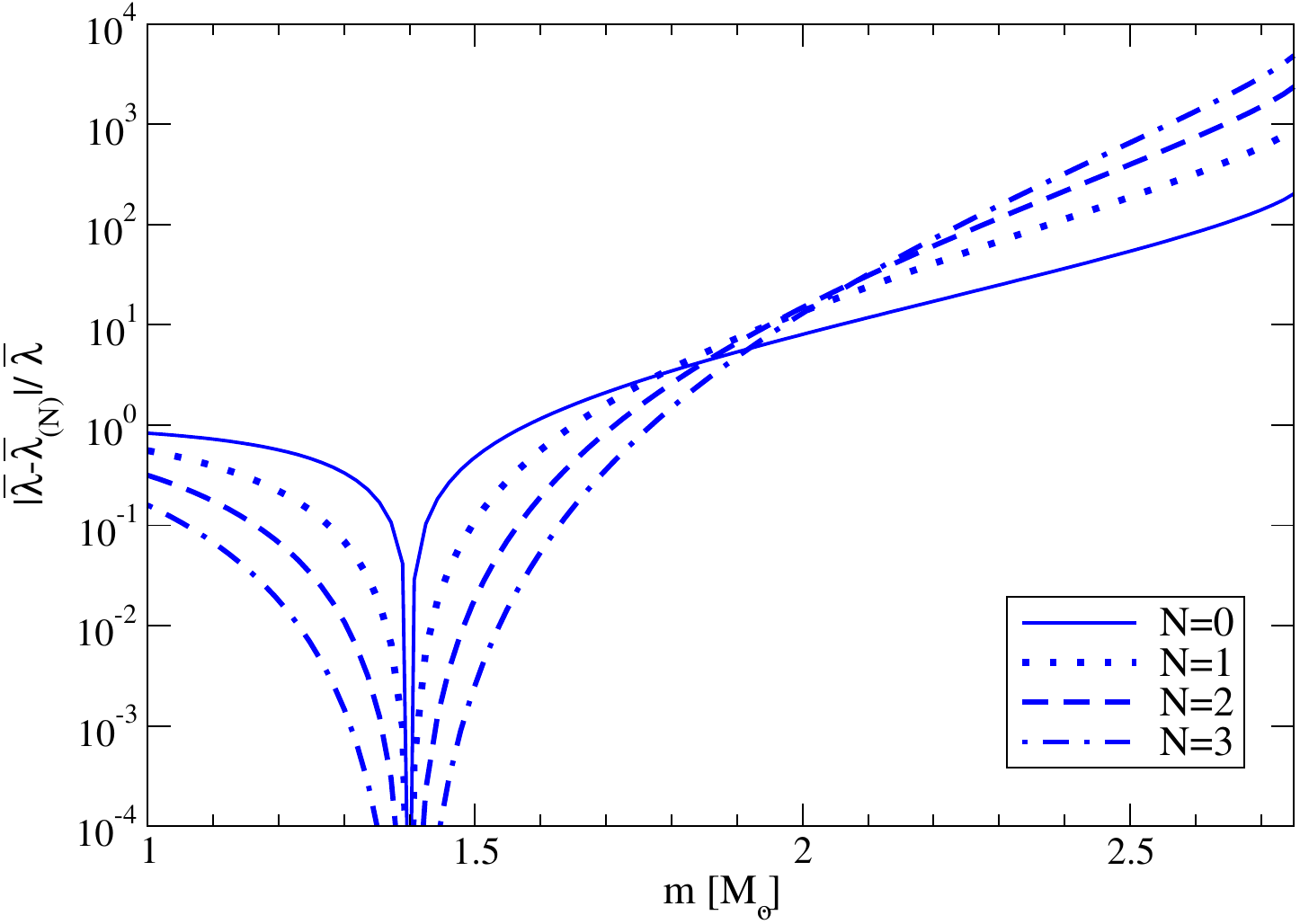}  
\includegraphics[width=6.cm,clip=true]{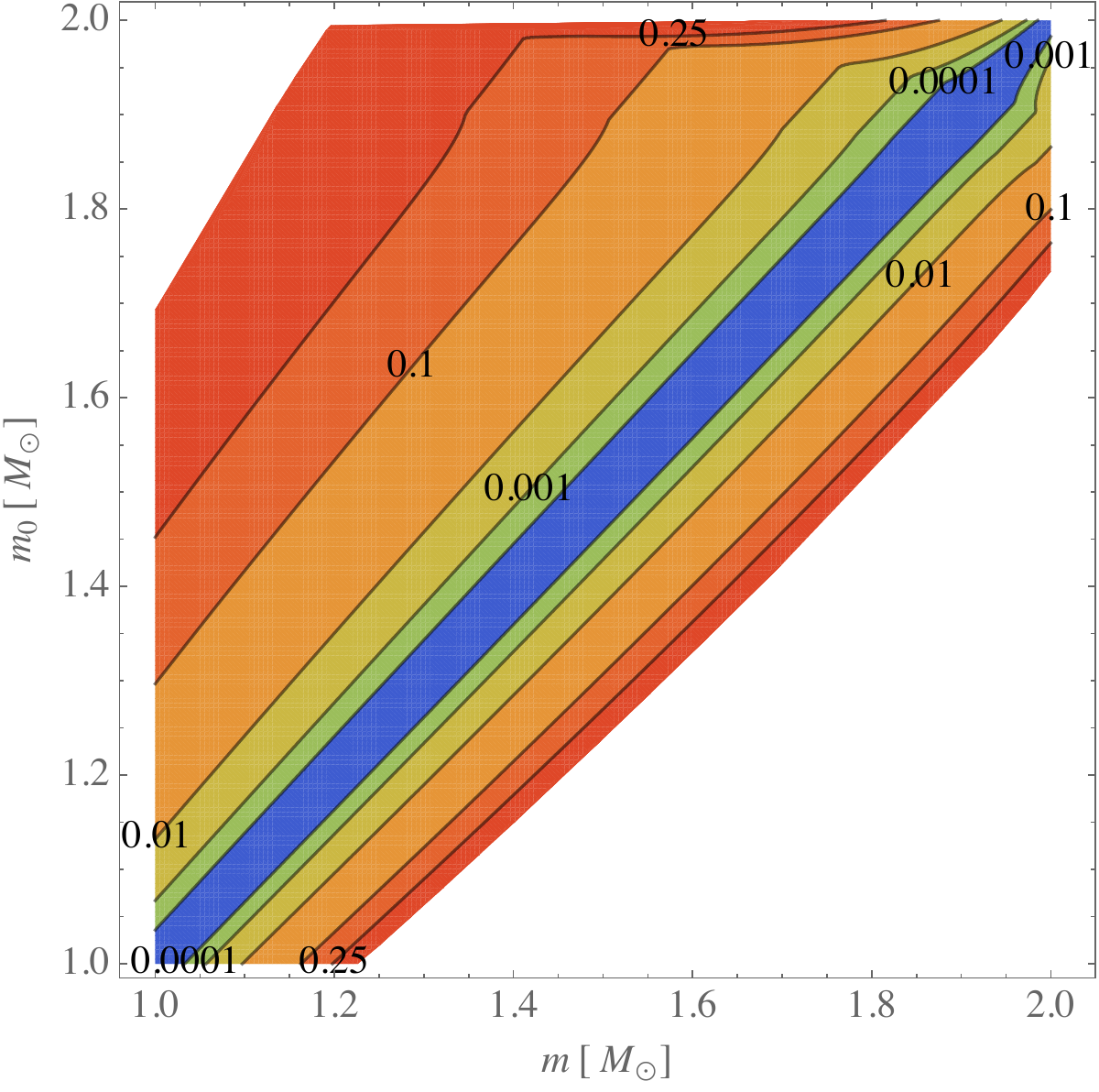}  
\caption{\label{fig:lambdan-diff} (Left) Absolute fractional difference between $\bar \lambda$ and $\bar \lambda_{(\N)}$ (given by Eq.~\eqref{eq:Taylor}) for various $N$. We choose the MS1 EoS and $m_0 = 1.4 M_\odot$. Observe that the series converge when $m < 1.9M_\odot$.
(Right) Maximum absolute fractional difference between $\bar \lambda$ and $\bar \lambda_{(3)}$ among the eleven realistic EoSs with various $(m,m_0)$. 
}
\end{center}
\end{figure}

Whether Eq.~\eqref{eq:Taylor} is a good representation of $\bar \lambda$ depends on whether the Taylor expansion converges and the rate at which it does so. The left panel of Fig.~\ref{fig:lambdan-diff} shows the relative fractional error between this series representation and the correct $\bar \lambda$, for different truncations $N$, with the MS1 EoS and an expansion mass of $m_{0} = 1.4 M_{\odot}$. One sees that the more terms that are added to the series, the smaller the error only in the region $m < 1.9M_\odot$. Thus, the series converges in this region, but it diverges in the high mass region $m > 1.9M_\odot$. This shows that the $\bar \lambda_0^{(k)}$ parameterization is not suitable for NS binaries with masses that are sufficiently different from $m_0$. The right panel of Fig.~\ref{fig:lambdan-diff} shows the maximum absolute fractional error between $\bar \lambda$ and $\bar \lambda_{(3)}$ using the eleven realistic EoSs in the $(m,m_0)$ plane. This contour plot shows the range of $m$ for a given $m_0$ with which one can use up to 3 terms in the series to approximate $\bar \lambda$ to a given accuracy. For example, the fractional difference is always smaller than $10\%$ for $m_0 = 1.4M_\odot$ when $1.1M_\odot<m < 1.6M_\odot$.

\subsubsection{Newtonian Limit}

Let us now consider the $\bar \lambda_0^{(0)}$--$\bar \lambda_0^{(k)}$ relation in the Newtonian limit. First, from Eq.~\eqref{eq:love-Newton}, one finds that $\bar \lambda_A \propto m_A^{-10/(3-n)}$. Thus, by taking derivatives of this relation with respect to $m_A$, one finds that
\be
\label{eq:Gnk}
\bar \lambda_0^{(k)} = G_{n,k} \ \bar \lambda_0^{(0)}\,, \quad G_{n,k} \equiv  \frac{\Gamma \left( k + \frac{10}{3-n} \right)}{\Gamma \left(\frac{10}{3-n} \right)}\,,
\ee
where $\Gamma (x)$ is the Gamma function. $G_{n,k}$ for representative values of $n$ (corresponding to WFF1, MS1 and mean EoSs in Table~\ref{table:coeff-n}) and $k$ are shown in Table~\ref{table:Gnk}, together with the fractional difference from the mean EoS shown in brackets. Observe how the EoS variation increases as one increases $k$.

\Table{\label{table:Gnk} $G_{n,k}$ in Eq.~\eqref{eq:Gnk} for $k=1,2,3$ using polytropic EoSs with the polytropic indices corresponding to WFF1, MS1 and mean EoSs shown in Table~\ref{table:coeff-n}. Absolute fractional difference from the mean EoS values is given in brackets.
\vspace{3mm}
}
\br
k & 1 & 2 & 3 \\
\mr
WFF1 & $4.943 \, (0.116)$ & $29.38 \, (0.221)$ & $204.0 \, (0.318)$  \\
MS1 & $3.862 \, (0.128)$ & $18.78 \, (0.219)$ & $110.1 \, (0.288)$ \\ 
\mr
mean & $4.431$ & $24.06$ & $154.7$  \\
\br
\endTable

\Table{\label{table:coeff2} Coefficients of the fit in Eq.~\eqref{eq:fit2} for the $\bar \lambda_0^{(0)}$--$\bar \lambda_0^{(k)}$ relations. The last row represents the r-squared value of the fit.
\vspace{3mm}}
\br
k & 1 & 2 & 3 \\
\mr
$a_{1,k}$ & $0.4443$ & $0.3344$ & $-0.1334$  \\
$a_{2,k}$ & $2.726$ & $6.568$ & $11.35$ \\
$a_{3,k}$ & $-0.6350$ & $-0.4671$ & $-3.928$  \\
\mr
$r^2$ & 0.9984 & 0.9955 & 0.9923  \\
\br
\endTable

\subsubsection{Relativistic Relations}

We now look at the EoS variation in the $\bar \lambda_0^{(0)}$--$\bar \lambda_0^{(k)}$ relations in the relativistic regime. We first calculate these relations numerically and then fit them to 
\be
\label{eq:fit2}
\bar \lambda_0^{(k)} = G_{\bar n,k} \ \bar \lambda_0^{(0)} \left( 1 + \sum_{i=1}^3 a_{i,k} \ x^{i} \right)\,, 
\ee
where $x \equiv (\bar \lambda_0^{(0)} )^{-1/5}$ and the coefficients for each $k$ are given in Table~\ref{table:coeff2}. This fit will then allow us to estimate the degree of EoS variability of the numerical results. 

\begin{figure}[htb]
\begin{center}
\includegraphics[width=7.5cm,clip=true]{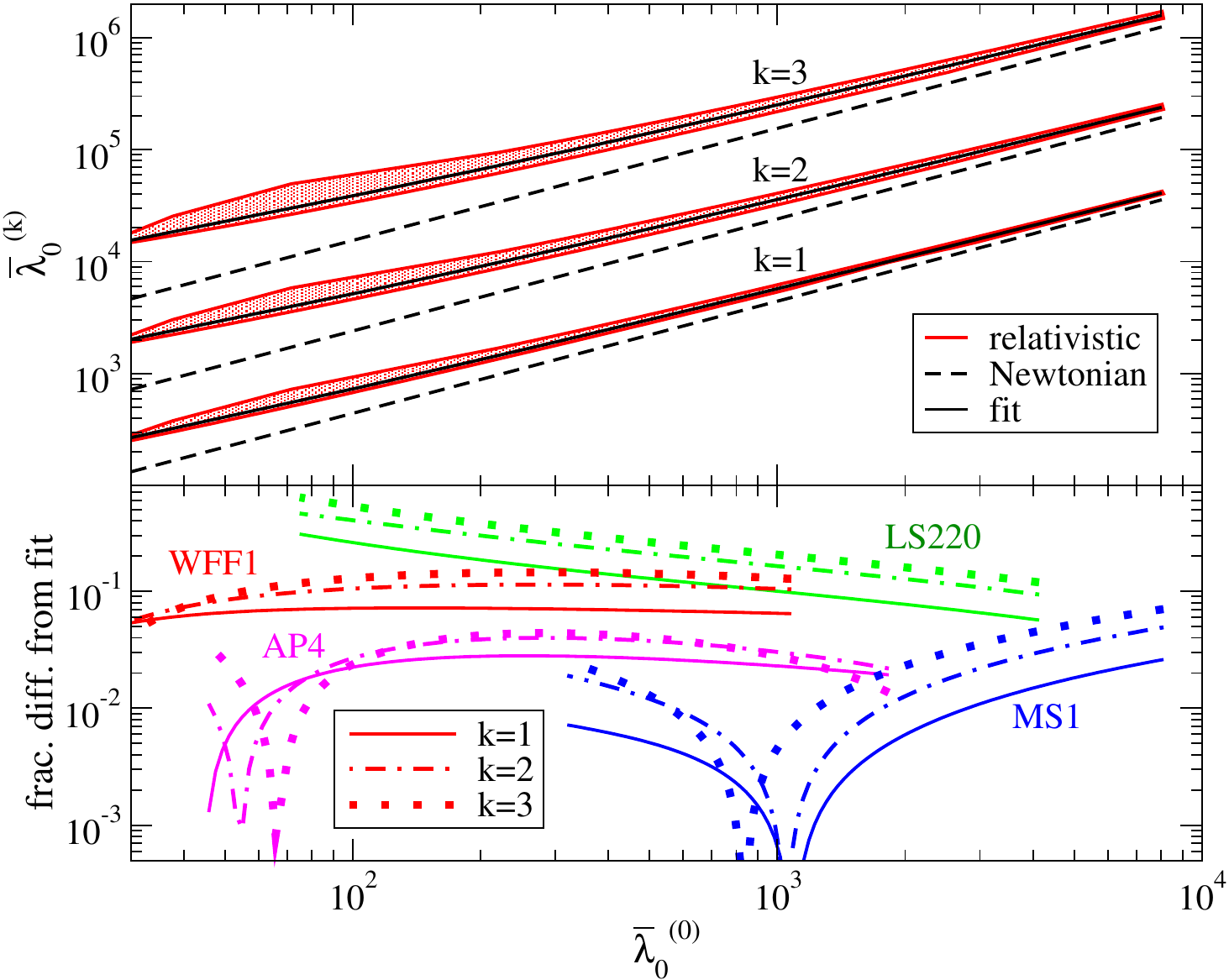}  
\includegraphics[width=7.9cm,clip=true]{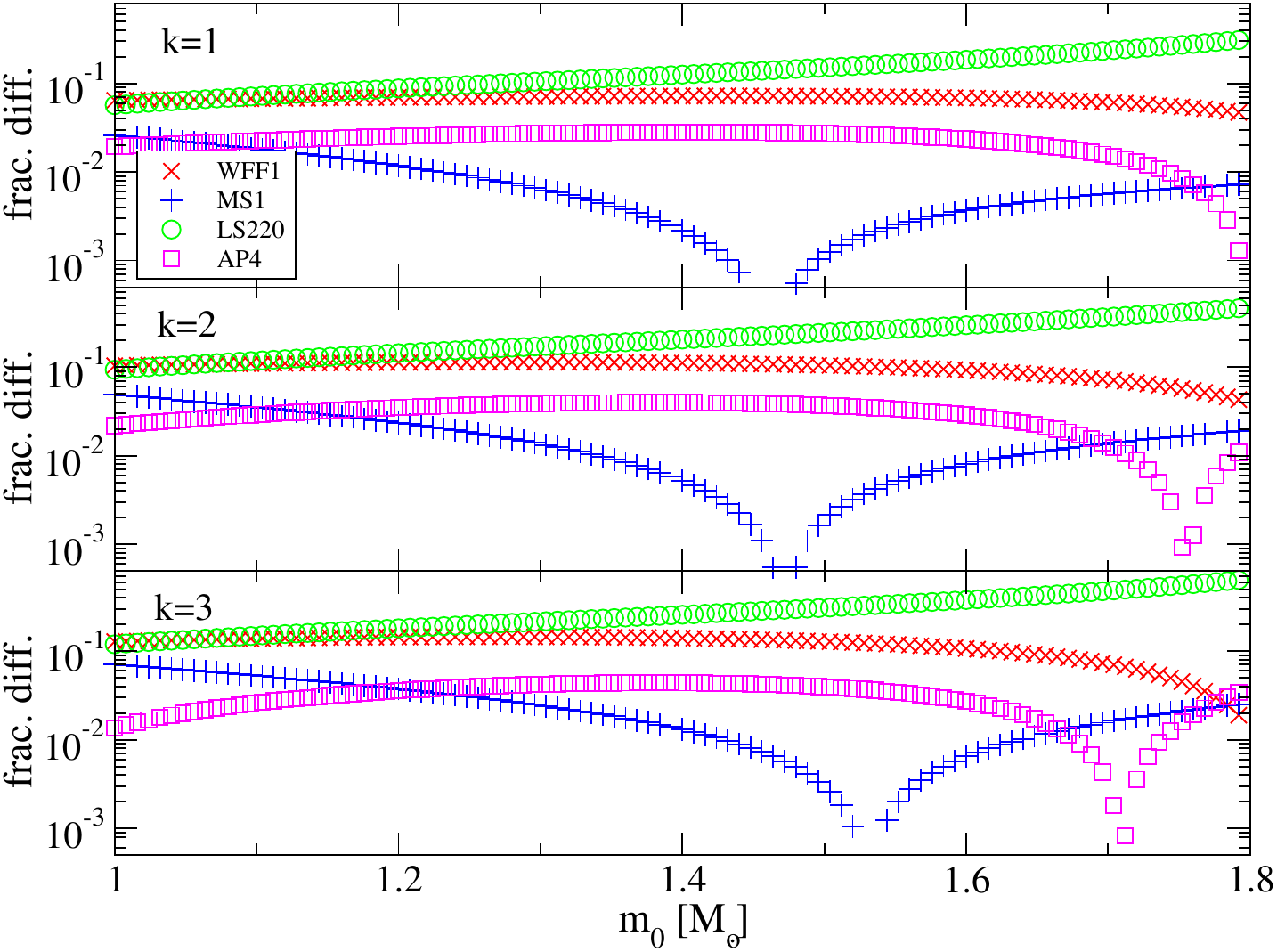}  
\caption{\label{fig:lambda-lambdak}  (Left) Same as the left panel of Fig.~\ref{fig:lambdas-lambdaa} but for the $\bar \lambda_0^{(0)}$--$\bar \lambda_0^{(k)}$ relations.
(Right) Absolute fractional difference of the $\bar \lambda_0^{(0)}$--$\bar \lambda_0^{(k)}$ relations from the fit against the fiducial mass $m_0$ for $k=1$, 2 and 3. Observe that with e.g. $m_0=1.4M_\odot$, the universality holds to $\sim 10\%$, $\sim 20\%$ and $\sim 25\%$ for $k=1,2,3$ respectively.
}
\end{center}
\end{figure}

The top left panel of Fig.~\ref{fig:lambda-lambdak} presents the regions spanned by the relations with varying EoSs obtained numerically as a function of $\bar \lambda_0^{(0)}$. One finds a single curve for a fixed $k$ and EoS (see Fig.~1 of~\cite{Yagi:2015pkc}) with $m_{0}$ the single parameter that varies along each curve. The top panel of the figure also shows the fit of Eq.~\eqref{eq:fit2}, while the bottom panel shows absolute values of the relative fractional difference between the numerical results and the fit. Observe that the difference becomes larger as one increases $k$. The top panel also shows the Newtonian relation of Eq.~\eqref{eq:Gnk} with $n=0.743$ (dashed lines). Again, observe that the relations approach the Newtonian ones in the large $\bar \lambda_0^{(0)}$ limit. The right panel of Fig.~\ref{fig:lambda-lambdak} shows the EoS variation in the $\bar \lambda_0^{(0)}$--$\bar \lambda_0^{(k)}$ relations for each fixed $m_0$ (analogous to the bottom left panel). For example, when $m_0 = 1.4M_\odot$ as chosen in~\cite{messenger-read,delpozzo,Agathos:2015uaa}, the universality holds to $\sim 10\%$ for $k=1$ but to $\sim 20\%$ and $\sim 25\%$ for $k=2$ and 3 respectively.

\section{Parameter Estimation}
\label{sec:Fisher}

In this section, we estimate the impact of the universal binary Love relations on EoS constraints with GW observations of NS binaries. We obtain this estimate through a Fisher analysis~\cite{finn,cutlerflanagan}, which should provide a rough measure of the accuracy with which one may be able to extract best-fit parameters. For a more robust estimate, in particular for low SNR signals, one needs to carry out a Bayesian analysis~\cite{delpozzo,Wade:2014vqa,Lackey:2014fwa,Agathos:2015uaa}.

\subsection{Preliminary}

We begin by reviewing the main ideas behind a Fisher analysis. For stationary and Gaussian noise, and in the large SNR limit, the posterior probability distribution of the template parameter vector ${\theta}^{a}$ with a given signal $s$ is approximately given by~\cite{Poisson:1995ef,berti-buonanno}
\be
\label{eq:posterior}
p({\theta}^{a}|s) \propto p^{(0)}({\theta}^{a}) \exp\left[ - \frac{1}{2} \Gamma_{ab} \left( \theta^a - \hat \theta^a \right) \left( \theta^b - \hat \theta^b \right) \right]\,,
\ee
with $\hat \theta^a$ the best-fit parameter that maximizes the probability distribution and $p^{(0)}$ the prior distribution. The Fisher matrix $\Gamma_{ab}$ is defined by
\be
\Gamma_{ab} \equiv \left( \frac{\partial h}{\partial \theta^a} \bigg| \frac{\partial h}{\partial \theta^b} \right)\,,
\ee
where $h$ is the waveform template (a model for the response of the instrument due to an impinging GW) 
and the inner product is defined by
\be
\label{eq:inner-product}
(a|b) \equiv 2 \int^\infty_0 \frac{\tilde a^* \tilde b + \tilde b^* \tilde a}{S_n(f)}\, df\,,
\ee
with a tilde and * denoting the Fourier transform and complex conjugate operations respectively, and $S_n(f)$ the noise spectral density. We follow~\cite{cutlerflanagan,Poisson:1995ef,berti-buonanno} and consider a crude prior distribution on certain parameters, given by a Gaussian centered around $\bar{{\theta}}^{a}$:
\be
\label{eq:prior}
p^{(0)} \propto \exp\left[ - \frac{1}{2} \sum_a \left( \frac{\theta^a - \bar{\theta}^a}{\sigma_{\theta^a}}\right)^2 \right]\,.
\ee
The explicit choice of $\bar \theta^a$ is irrelevant for calculating the statistical error on $\theta^a$. Since a product of two Gaussian distributions gives a new Gaussian,
one obtains the root-mean-square of $\theta^a$ as
\be
\Delta \theta^a = \sqrt{\left(\tilde \Gamma^{-1} \right)^{aa}}\,, 
\qquad 
\tilde \Gamma_{ab} \equiv \Gamma_{ab} + \frac{1}{\sigma_{\theta^a}^2} \delta_{ab}\,. 
\ee
We note that one is likely to use a uniform prior in actual data analysis, but here we use a Gaussian prior for simplicity. A more detailed, Bayesian analysis with a more natural prior is beyond the scope of this paper and we leave it for future work.

Any results from a GW parameter estimation study depends strongly on the template family used to search for the signal. We here use a sky-averaged, restricted PN waveform~\cite{berti-buonanno}. This means that we only keep leading PN order terms in the amplitude, but in the phase we retain terms up to 3.5PN order in the point-particle contribution~\cite{arun35PN} and up to 7.5PN order in the tidal contribution~\cite{damour-nagar-villain}. We do not include electric-type tidal deformabilities beyond quadrupole order, or magnetic-type tidal deformabilities, since their contributions are negligible to GW observations with second-generation GW interferometers~\cite{hinderer-lackey-lang-read,damour-nagar-villain,Yagi:2013sva}. A more detailed data analysis study could use more sophisticated templates, like the effective-one-body (EoB) ones of~\cite{baiotti,damour-nagar-villain,Bernuzzi:2014owa}. Such sophisticated waveforms would allow us to reduce systematic errors on tidal parameters that would be introduced by the restricted PN model due to neglecting~\cite{Favata:2013rwa,Yagi:2013baa} or incorrectly modeling~\cite{Wade:2014vqa} point-particle higher PN terms in the waveform. Nonetheless, the conclusions derived of this paper about the \emph{relative increase} in accuracy to which parameters can be measured, should not be affected. 

Regardless of the particular waveform model used, one can always decompose the template parameter vector ${\theta}^{a}$ into ${\theta}^{a} = {\theta}_\mrm{pp}^{a} + {\theta}_\mrm{tid}^{a}$, where ${\theta}_\mrm{tid}^{a}$ contains tidal parameters only and ${\theta}_\mrm{pp}^{a}$ contains all other parameters, present for example when modeling the objects in the point-particle approximation. For the particular waveform model we use in this paper, 
\be
{\theta}_\mrm{pp}^{a} = \left( \ln \mathcal{M}, \ln \eta, t_c, \phi_c, \ln D_L \right)\,,
\ee
where $\mathcal{M} \equiv M \eta^{3/5}$ the chirp mass with $M = m_1 + m_2$ representing the total mass, $t_c$ and $\phi_c$ are the time and phase at coalescence respectively and $D_L$ is the luminosity distance to the source. The tidal contribution ${\theta}_\mrm{tid}^{a}$ can be one of the following:
\begin{align}
\label{eq:tid-par-1}
{\theta}_\mrm{tid}^{a} &= (\bar \lambda_s,\bar \lambda_a)\,,
\\
\label{eq:tid-par-2}
{\theta}_\mrm{tid}^{a} &= (\bar \Lambda, \delta \bar \Lambda)\,,
\\
\label{eq:tid-par-3}
{\theta}_\mrm{tid}^{a} &= (\bar \lambda_0^{(0)},\bar \lambda_0^{(1)},\bar \lambda_0^{(2)},...)\,.
\end{align}
The choice of ${\theta}_\mrm{tid}^{a}$ defines three distinct parameterizations of the same waveform family. 

Another key ingredient of a GW parameter estimation study is the representation of the noise. 
We model it through the spectral noise density of the detector, which we approximate with the zero-detuned Adv.~LIGO curve, presented e.g.~in~\cite{ajith-zero-detuned}. Since Adv.~LIGO is expected to have very large seismic noise below $10$ Hz, we start all integrations at $f_{\min} = 10$Hz. We stop integrations at $f_{\max} = \min(f_\ISCO,f_\mrm{cont})$, where $f_\ISCO = 1/(6^{3/2} \pi M)$ is the frequency at the innermost stable circular orbit for a point-particle in a Schwarzschild spacetime, while $f_\mrm{cont} = \sqrt{M/(R_1+R_2)^3}/\pi$ is the frequency of contact. Pushing the analysis above these frequencies would require waveforms that can model the merger phase accurately, and these are not yet available analytically. 

We consider the following injected signals so that the approximations we made above are valid for observations that second-generation detectors may soon make. When considering the tidal parameterization of Eqs.~\eqref{eq:tid-par-1} and~\eqref{eq:tid-par-2}, we choose a mass ratio of $q = 0.9$ and vary $m_1$. For the parameterization of Eq.~\eqref{eq:tid-par-3}, we set the mass difference to be $0.1M_\odot$ and vary $m_0$, centered around $m_1$ and $m_2$. The luminosity distance is chosen such that the signal-to-noise ratio is 30; a lower value of signal-to-noise ratio would make the Fisher approximation questionable. For the masses we considered, this implies distances between 84--140Mpc. We do not vary the NS EoS, keeping it fixed at AP4. The phase and time of coalescence are nuisance parameters, so they are set to zero in the injections.  We consider a GW detection from a single source with Adv.~LIGO. We leave possible studies of multiple detections and/or detections with third-generation, ground-based interferometers, such as the Einstein Telescope (ET), for future work (though we will consider using ET for probing cosmology with GWs in Sec.~\ref{sec:cosmology}).

\subsection{Results}
\label{sec:results}

\subsubsection{Measurement Accuracy on Tidal Parameters}
\label{sec:accuracy}

Let us first look at the tidal parameterization of Eq.~\eqref{eq:tid-par-1}. We choose the standard deviations $\sigma_{\bar \lambda_s} = 10^4$ and $\sigma_{\bar \lambda_a} = 5 \times 10^3$ in Eq.~\eqref{eq:prior} for the prior on these tidal parameters (see~\ref{app:prior} for a detailed discussion on the impact of such priors on the measurement accuracy of $\bar \lambda_s$). This is a conservative prior, which covers all $\bar \lambda_s$ and $\bar \lambda_a$ with realistic EoSs considered in this paper. Figure~\ref{fig:Fisher-prior} shows the measurement accuracy of $\bar \lambda_s$ as a function of injected $m_1$, both with and without using the $\bar \lambda_s$--$\bar \lambda_a$ relation. Observe that the $\bar \lambda_s$--$\bar \lambda_a$ relation improves the measurement accuracy by approximately an order of magnitude, and renders $\bar \lambda_s$ measurable for $m_1 < 1.5M_\odot$ with the chosen EoS. This increase in accuracy is because the  $\bar \lambda_s$--$\bar \lambda_a$ relation allows us to eliminate $\bar \lambda_a$ from ${\theta}_\mrm{tid}^{a}$, thus breaking the degeneracy between $\bar \lambda_s$ and $\bar \lambda_a$. We find a similar improvement in measurement accuracy on $\bar \lambda_s$ for injections with different EoSs. 

We next consider the ${\theta}_\mrm{tid}^{a} = (\bar \Lambda, \delta \bar \Lambda)$ parametrization in Eq.~\eqref{eq:tid-par-2} with the standard deviations $\sigma_{\bar \Lambda} = 3 \times 10^3$ and $\sigma_{\delta \bar \Lambda} = 10^3$ for their Gaussian priors. In this case, we find that the $\bar \Lambda$--$\delta \bar \Lambda$ relation does not help improve the measurement accuracy of $\bar \Lambda$. This is because $\bar \Lambda$ and $\delta \bar \Lambda$ are essentially uncorrelated. Such a result is consistent with those of~\cite{Wade:2014vqa} in which the authors found that the inclusion of $\delta \bar \Lambda$ in the search parameters does not affect the measurement accuracy $\bar \Lambda$ in a Bayesian analysis. Figure~\ref{fig:Fisher-prior} shows the measurement accuracy of $\bar \Lambda$ as a function of the injected mass $m_{1}$. Observe that the measurement accuracy of $\bar \lambda_s$ with the $\bar \lambda_s$--$\bar \lambda_a$ relation is comparable to, or even slightly better than that of $\bar \Lambda$.  

\begin{figure}[htb]
\begin{center}
\includegraphics[width=8.cm,clip=true]{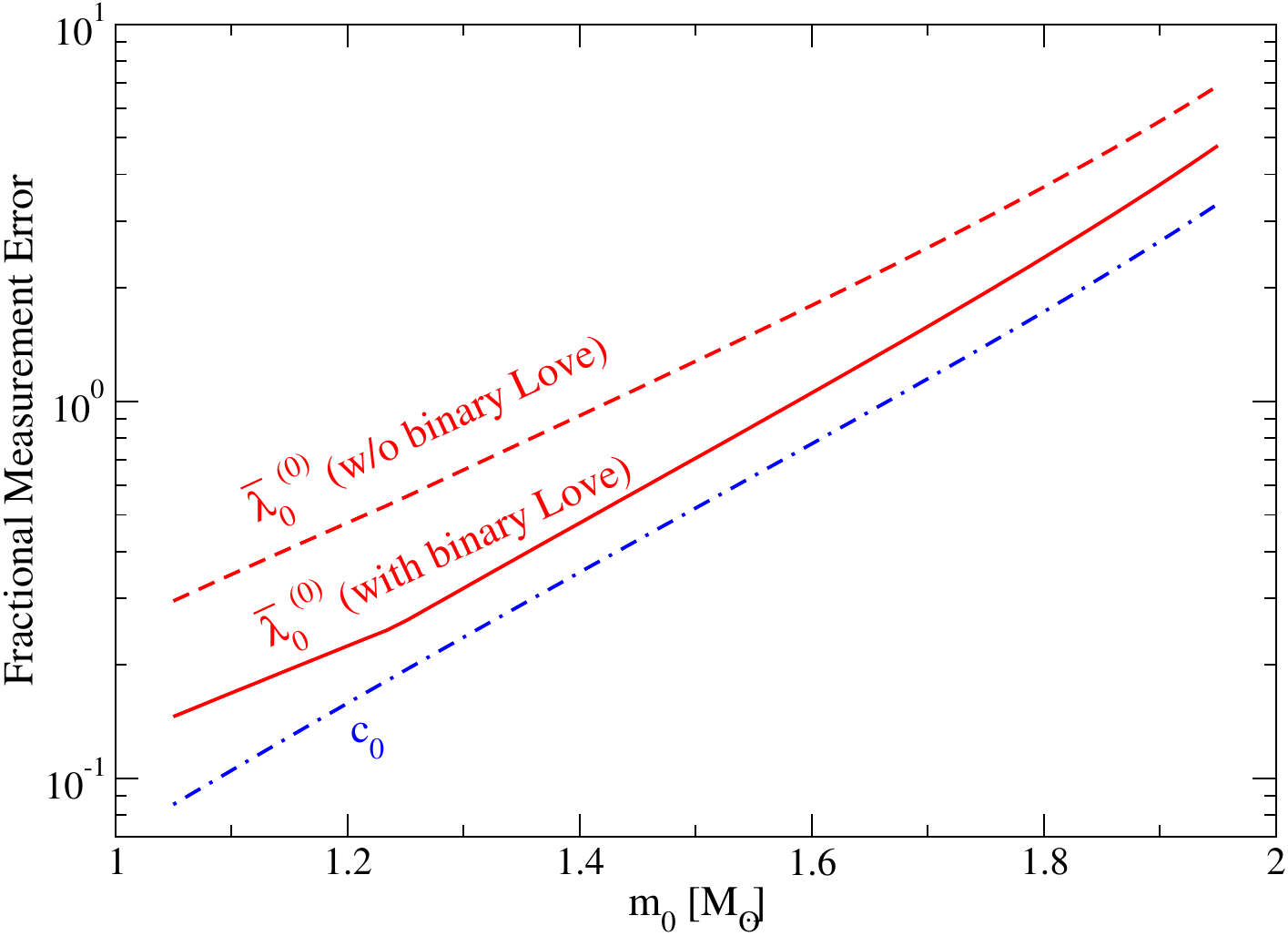}  
\caption{\label{fig:Fisher-prime} Fractional measurement accuracy of $\bar \lambda_0^{(0)}$ as a function of $m_0$ in Eq.~\eqref{eq:Taylor} with and without using the $\bar \lambda_0^{(0)}$--$\bar \lambda_0^{(1)}$ relation. We assume that we detect GWs from a non-spinning NS binaries with Adv.~LIGO with SNR=30. We also assume the AP4 EoS as the fiducial one and $(m_1,m_2) = (m_0 - 0.05M_\odot,m_0 + 0.05M_\odot)$. We include $\bar \lambda_0^{(0)}$ and $\bar \lambda_0^{(1)}$ into search parameters. Observe that the binary Love relation allows us to measure $\bar \lambda_0^{(0)}$ a few times better than the case without using the relation. For comparison, we also show the fractional measurement accuracy of $c_0$, another Taylor-expanded tidal parametrization given by Eq.~\eqref{eq:Taylor-del-pozzo}~\cite{delpozzo,Agathos:2015uaa}.
}
\end{center}
\end{figure}

Finally, let us consider the ${\theta}_\mrm{tid}^{a} = (\bar \lambda_0^{(0)},\bar \lambda_0^{(1)},\bar \lambda_0^{(2)},...)$ parameterization in Eq.~\eqref{eq:tid-par-3}. For simplicity, we truncate the series at $\bar \lambda_0^{(1)}$.  We choose the standard deviations $\sigma_{\bar \lambda_0^{(0)}} = 10^4$ and $\sigma_{\bar \lambda_0^{(1)}} = 5 \times 10^4$ for the Gaussian priors. Figure~\ref{fig:Fisher-prime} shows the measurement accuracy of $\bar \lambda_0^{(0)}$ as a function of $m_0$ with and without using the $\bar \lambda_0^{(0)}$--$\bar \lambda_0^{(1)}$ relation. Observe that such a relation improves the measurement accuracy by 40--100\%. Again, this is because the relation breaks the degeneracy between the tidal parameters. Such an improvement in the measurement accuracy of $\bar \lambda_0^{(0)}$ should improve future constraints on the NS EoS from multiple GW sources~\cite{delpozzo,Agathos:2015uaa}.

Let us now compare the measurement accuracy $\bar \lambda_0^{(0)}$ to a similar parameterization chosen in~\cite{delpozzo,Agathos:2015uaa}, in which the authors Taylor expand the \emph{dimensional} tidal deformability as
\be
\label{eq:Taylor-del-pozzo}
\lambda (m) = \sum_k \frac{1}{k!} c_k \left( \frac{m-m_0}{M_\odot} \right)^k\,.
\ee
We checked that no universal relations exist among the dimensionless tidal coefficients $c_k$. This is consistent with the results of~\cite{Majumder:2015kfa}, in which we showed that the degree of universality depends strongly on how one normalizes the NS quantities in play. We repeat the Fisher analysis but choosing $(c_0,c_1)$ as tidal parameters instead of $(\bar \lambda_0^{(0)},\bar \lambda_0^{(1)})$. We choose the standard deviations of $\sigma_{c_0} = 5\times 10^{-23}$s$^5$ and $\sigma_{c_1} =2.5 \times  10^{-23}$s$^5$~\cite{Agathos:2015uaa} for their Gaussian priors. The fractional measurement accuracy of $c_0$ is shown as the blue dotted-dashed curve in Fig.~\ref{fig:Fisher-prime}. Interestingly, such an accuracy is better than that of $\bar \lambda_0^{(0)}$. This is because (i) the correlation between $c_0$ and $c_1$ is very weak thanks to the prior, and (ii) the $c_k$ parametrization has less degeneracy among non-tidal parameters compared to the $\bar \lambda_0^{(k)}$ parametrization. Therefore, the $\bar \lambda_0^{(0)}$ parametrization may not be better than the $c_0$ parameterization, though the former can be important for probing experimental relativity and cosmology, as we will discuss in Secs.~\ref{sec:gravitational} and~\ref{sec:cosmology} in more detail.

A word of caution is now needed. One cannot directly compare the fractional measurement accuracy of $\bar \lambda_s$, $\bar \Lambda$ and $\bar \lambda_1$ in Fig.~\ref{fig:Fisher-prior} with that of $\bar \lambda_0^{(0)}$ and $c_0$ in Fig.~\ref{fig:Fisher-prime}. This is because these two figures assume NS binaries with different masses. Moreover, the $(\bar \lambda_s,\bar \lambda_a)$ and $(\bar \Lambda, \delta \bar \Lambda)$ parameterizations have different properties compared to those of the $\bar \lambda_0^{(k)}$ parameterization. The former can be used to search for any NS binaries, while the latter is only suitable for those with NS masses that are close to some $m_0$, i.e.~with mass ratios sufficiently close to unity. Otherwise, the systematic error on $\bar \lambda_0^{(0)}$ due to mismodeling the tidal deformability for a given NS mass can dominate the statistical error if one does not include $\bar \lambda_0^{(k)}$ to sufficiently high order in the parameter set. If one does include enough terms in the Taylor expansion to minimize the systematic error, correlations among the different $\bar \lambda_0^{(k)}$ parameters may increase the statistical error on $\bar \lambda_0^{(0)}$.

\subsubsection{Systematic Errors}
\label{sec:sys}

One might wonder how much of a systematic error is introduced in the measurement of $\bar \lambda_s$ due to the fact that the $\bar \lambda_s$--$\bar \lambda_a$ relation is not perfectly EoS-universal (especially in the large mass and comparable mass ratio regime). We can roughly estimate this systematic error by making the following assumptions. First, since we are only interested in estimating the systematic error in the tidal parameter, we will neglect all statistical errors and any systematic errors in non-tidal parameters. Second, since we focus on an order of magnitude estimate, we will only retain the leading order term in the PN approximation to the tidal part of the GW  phase $\Psi^\mrm{tidal,5PN}$.

With these assumptions in hand, we can now estimate the systematic error as follows. Consider a NS binary GW signal with some injected tidal parameters $\bar{\lambda}_{s}^{(i)}$ and $\bar{\lambda}_{a}^{(i)}$, such that its leading PN order tidal phase, $\Psi^\mrm{tidal,5PN,injection}$, is 
\be
\label{eq:signal}
\Psi^\mrm{tidal,5PN,injection} \propto C_s \bar \lambda_s^{(i)} + C_a  \delta m \; \bar \lambda_a^{(i)}\,,
\ee 
where we have dropped an overall proportionality factor (including the frequency dependence of this term) that is common to both the injected and recovered tidal phase. The quantities $C_s$ and $C_a$ are functions of the mass ratio given by
\be
C_s \equiv 1+7 \eta - 31 \eta^2\,, \quad C_a \equiv 1+9 \eta - 11 \eta^2\,,
\ee
with $\delta m \equiv (m_2 - m_1)/M =  (1-q)/(1+q)$. Let us now recover this signal with a template that uses the $\bar \lambda_s$--$\bar \lambda_a$ relation, such that its leading PN order tidal phase $\Psi^\mrm{tidal,5PN,template}$ is
\be
\label{eq:template}
\Psi^\mrm{tidal,5PN,template} \propto C_s \bar \lambda_s + C_a  \delta m \; \bar \lambda_a \left( \bar \lambda_s \right)\,.
\ee
Notice that we use the $\bar \lambda_s$--$\bar \lambda_a$ relation in the second term. The recovered or \emph{best-fit} parameters are those which minimize the difference between the signal and the template (weighted by the spectral noise of the instrument). Therefore, a rough estimate of the recovered parameter $\bar \lambda_s^{(r)}$ is determined by equating Eq.~\eqref{eq:signal} to Eq.~\eqref{eq:template} with $\bar \lambda_s = \bar \lambda_s^{(r)}$, namely,
\begin{align}
 C_s \bar \lambda_s^{(i)} + C_a  \delta m \; \bar \lambda_a^{(i)} & =   C_s \bar \lambda_s^{(r)} + C_a  \delta m \; \bar \lambda_a \left( \bar \lambda_s^{(r)} \right) \nn \\
 & =  C_s \left( \bar \lambda_s^{(i)} + \Delta_\mrm{sys} \bar \lambda_s \right) + C_a  \delta m \; \bar \lambda_a \left( \bar \lambda_s^{(i)} + \Delta_\mrm{sys} \bar \lambda_s \right) \nn \\
 & \approx   C_s \left( \bar \lambda_s^{(i)} + \Delta_\mrm{sys} \bar \lambda_s \right)  + C_a  \delta m \; \left[ \bar \lambda_a \left( \bar \lambda_s^{(i)} \right) +\bar \lambda_a' \left( \bar \lambda_s^{(i)} \right) \Delta_\mrm{sys} \bar \lambda_s \right]\,,
 \label{eq:sys-error-est}
\end{align}
where on the second line we have assumed that the only difference between the recovered parameter and the injected one is the systematic error $ \Delta_\mrm{sys} \bar \lambda_s \equiv \bar \lambda_s^{(r)} - \bar \lambda_s^{(i)}$ on the parameter $\bar \lambda_s$. In the third line of Eq.~\eqref{eq:sys-error-est}, we expanded $\bar \lambda_a (\bar \lambda_s)$ assuming $ \Delta_\mrm{sys} \bar \lambda_s \ll \bar \lambda_s^{(i)}$. One can then solve the above equation for $\Delta_\mrm{sys} \bar \lambda_s$ to find
\be
\label{eq:sys-lambdas}
\frac{\Delta_\mrm{sys} \bar \lambda_s}{\bar \lambda_s^{(i)}} \approx\frac{C_a}{C_s + C_a \delta m \; \bar \lambda_a' \left( \bar \lambda_s^{(i)} \right)} \, \delta m \, \frac{\delta \bar \lambda_a}{\bar \lambda_s^{(i)} }\,,
\ee
with $\delta \bar \lambda_a \equiv \bar \lambda_a^{(i)} - \bar \lambda_a ( \bar \lambda_s^{(i)} )$.
Notice that the systematic error is proportional to $\delta m$.  In~\ref{app:sys} we rederive this equation with a different approach. In particular, we show explicitly that the correlation among parameters vanishes to the leading order.

\begin{figure}[htb]
\begin{center}
\includegraphics[width=7.cm,clip=true]{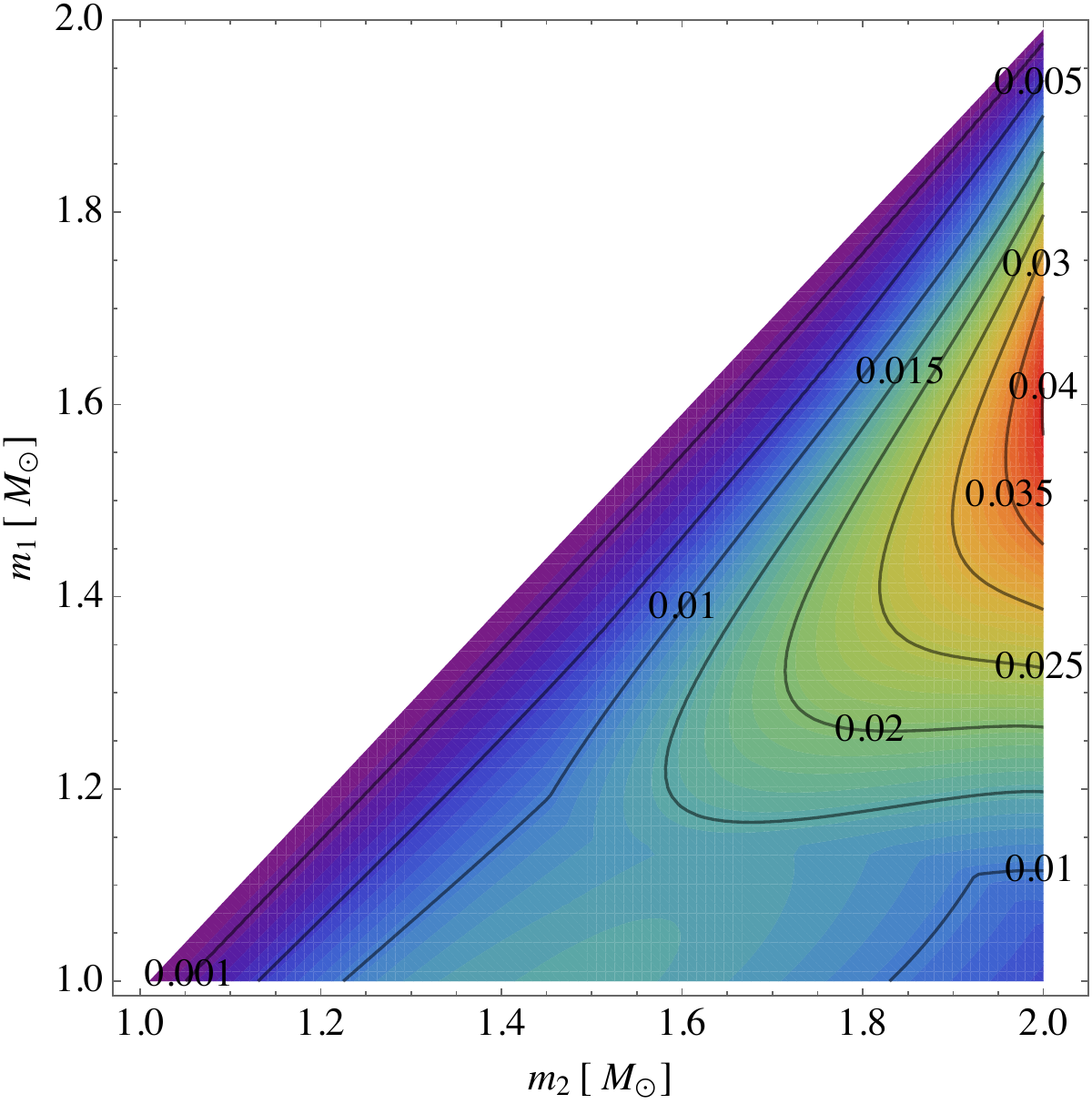}  
\caption{\label{fig:lambdas-sys} The maximum fractional systematic error on $\bar \lambda_s$ due to the EoS variation in the $\bar \lambda_s$--$\bar \lambda_a$ relation as a function of $m_1$ and $m_2$ (Eq.~\eqref{eq:sys-lambdas}). Observe that such an error is always smaller than $0.05$, much smaller than the fractional statistical error shown in Fig.~\ref{fig:Fisher-prior}.
}
\end{center}
\end{figure}

Figure~\ref{fig:lambdas-sys} shows the maximum fractional systematic error on $\bar \lambda_s$ as a function of $m_1$ and $m_2$ due to the EoS variation in the $\bar \lambda_s$--$\bar \lambda_a$ relation. At each $(m_1,m_2)$, we compute the fractional systematic error for each EoS and show the maximum error in Fig.~\ref{fig:lambdas-sys}. Comparing this with the fractional statistical error in Fig.~\ref{fig:Fisher-prior}, one finds that the systematic error is negligible. In fact, it is much smaller than the EoS variation of $\mathcal{O}(10 \%)$ in the $\bar \lambda_s$--$\bar \lambda_a$ relation. This is essentially because the tidal part of the GW phase that depends on $\bar \lambda_a$ is suppressed by $\delta m$, which is $\sim 0.05$ for $q = 0.9$. Thus, the $(m_{1},m_{2})$ regime where the binary Love relation is least EoS universal is also a part of the regime where the effect of $\bar \lambda_{a}$ on the GW phase is suppressed the most. In other words, the EoS variation becomes large in the large $m_1$ and $m_2$ region, while the suppression of systematic errors becomes smaller (or equivalently $\delta m$ becomes larger) in the small $m_1$ and large $m_2$ region. Therefore, the largest systematic error is realized for $m_1$ in the middle of its range and large $m_2$.  

\begin{figure*}[htb]
\begin{center}
\includegraphics[width=7.5cm,clip=true]{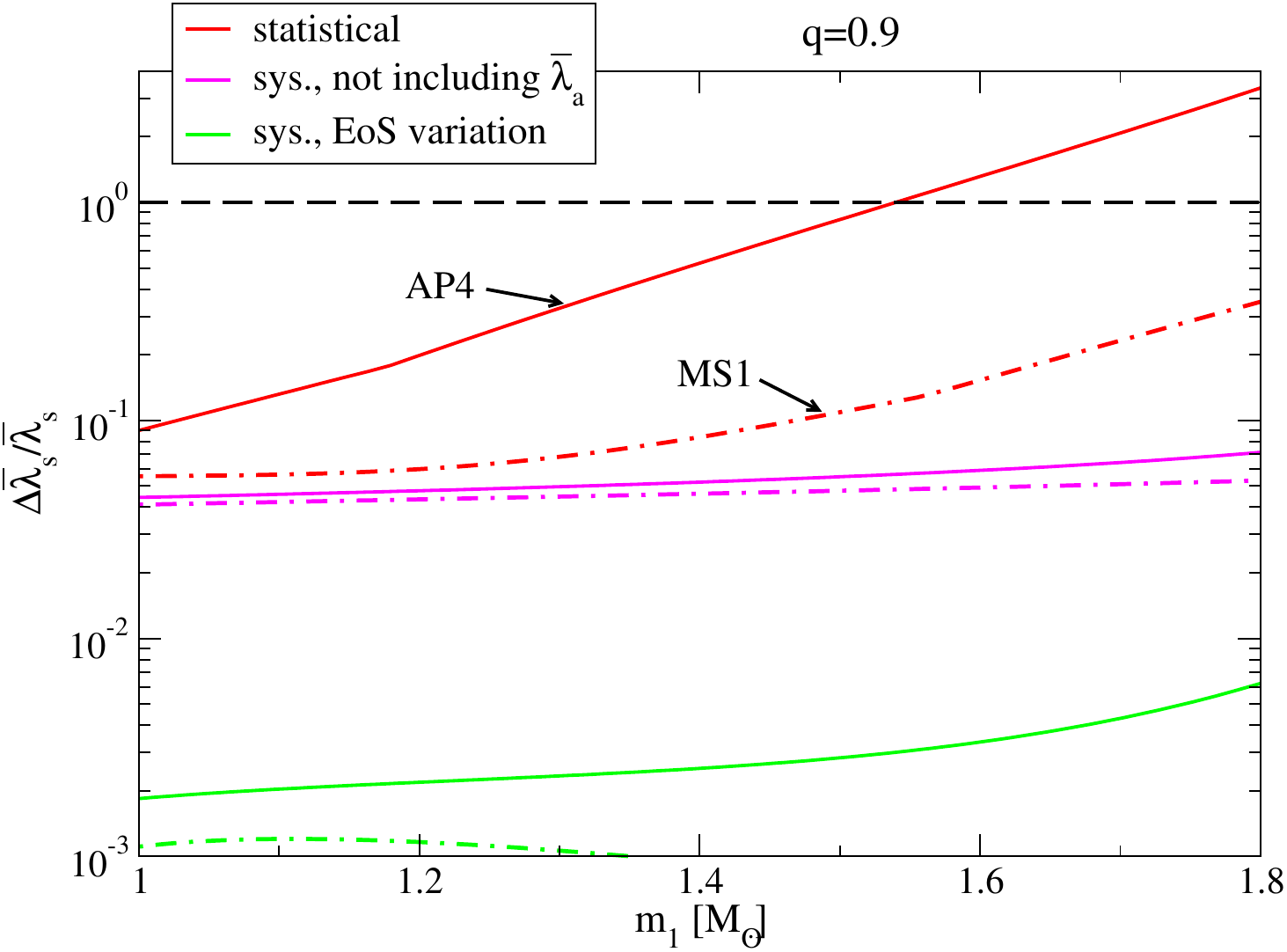}  
\includegraphics[width=7.5cm,clip=true]{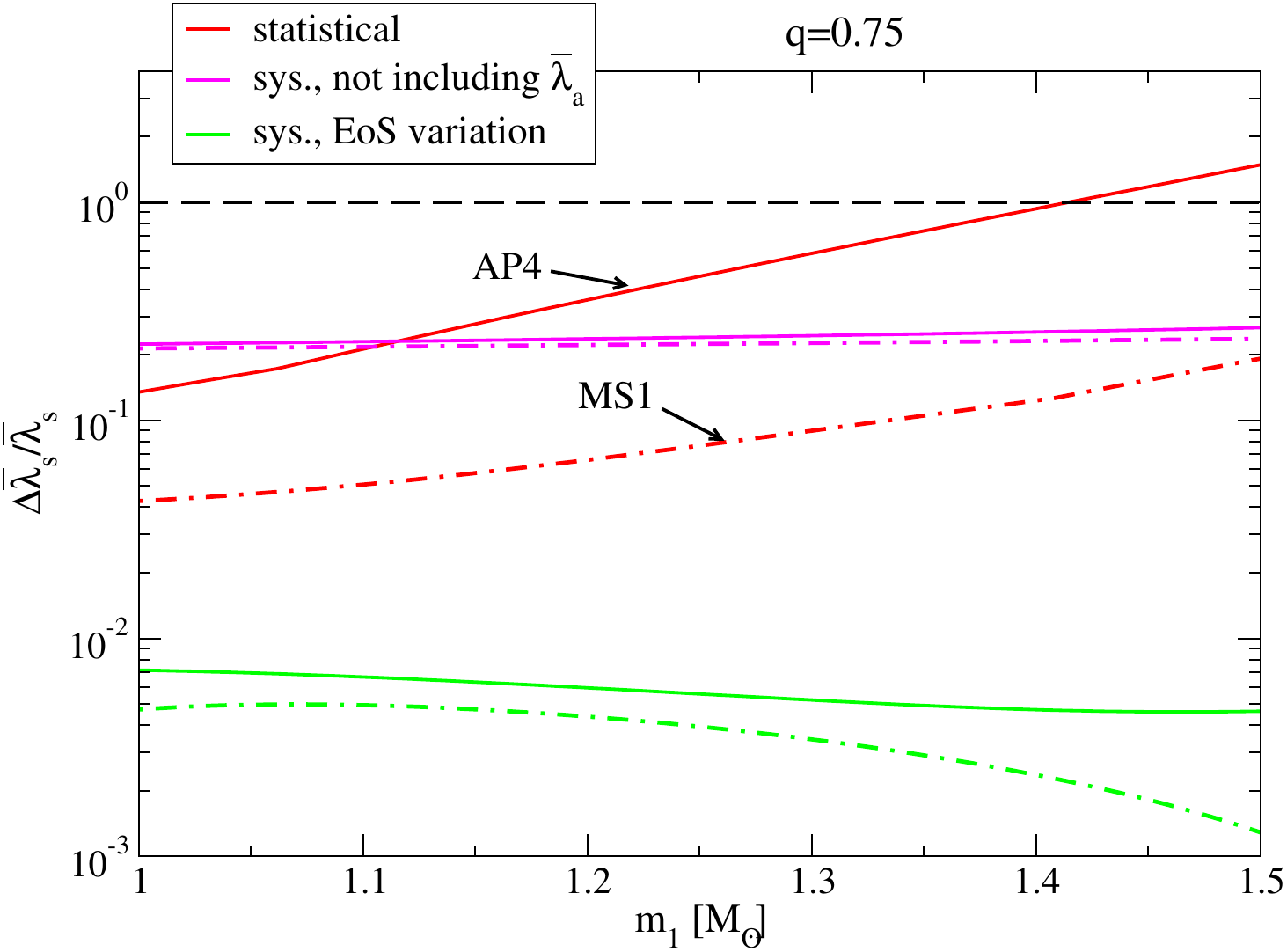}  
\caption{\label{fig:Fisher-sys-q}  Fractional measurement error on $\bar \lambda_s$ as a function of $m_1$ with $q=0.9$ (left) and $0.75$ (right) for two representative EoSs using Adv.~LIGO with SNR=30. We present statistical errors without including $\bar \lambda_a$ into the search parameter set (red), systematic errors due to not including $\bar \lambda_a$ given by Eq.~\eqref{eq:sys-lambdas2} (magenta) and due to the EoS variation in the $\bar \lambda_s$--$\bar \lambda_a$ relation given by Eq.~\eqref{eq:sys-lambdas} (green). Observe that the former systematics are important for small $q$, while the latter are always smaller than statistical errors. The former can also be important for relatively large $q$ if the SNR is large (in the case of third-generation interferometers).
}
\end{center}
\end{figure*}

Since the second term in Eq.~\eqref{eq:signal} is suppressed by $\delta m$, one might think that one does not need to include $\bar \lambda_a$ in the search parameter set in the first place. Whether this can be done depends on how large the systematic errors on $\bar \lambda_s$ are due to neglecting $\bar \lambda_a$ compared to the statistical errors on $\bar \lambda_s$. One can estimate the former by eliminating the second term on the right hand side of Eq.~\eqref{eq:sys-error-est}, which reduces to 
\be
\label{eq:sys-lambdas2}
\frac{\Delta_\mrm{sys} \bar \lambda_s}{\bar \lambda_s^{(i)}} \approx\frac{C_a}{C_s} \, \delta m \, \frac{\bar \lambda_a^{(i)}}{\bar \lambda_s^{(i)} }\,.
\ee
We present such systematic errors in Fig.~\ref{fig:Fisher-sys-q} for $q=0.9$ and 0.75 and the AP4 and MS1 EoSs. We also present statistical errors without including $\bar \lambda_a$ and systematic errors due to the EoS variation in the $\bar \lambda_s$--$\bar \lambda_a$ relation (Eq.~\eqref{eq:sys-lambdas}). Observe that if one does not take $\bar \lambda_a$ into account, systematic errors can dominate statistical errors for small $q$. This shows that using the binary Love relation is crucial to decrease the measurement error. Since statistical errors scale linearly with 1/SNR while systematic errors are independent of the SNR (see Eqs.~\eqref{eq:sys-lambdas} and~\eqref{eq:sys-lambdas2}), systematic errors can dominate the error budget even for large $q$ with third-generation GW interferometers such as ET.  

\section{Applications}
\label{sec:applications}

We now discuss whether the binary Love relations are useful in probing fundamental physics with gravitational wave observations. In particular, we will look at applications to astrophysics, nuclear physics, experimental relativity and cosmology.

\subsection{Astrophysics}

An improved measurement of the symmetric and antisymmetric tidal deformability also affects our ability to measure the individual tidal deformabilities, $\bar \lambda_{1,2}$. First, using propagation of error, we estimate the measurement accuracy of $\bar \lambda_a$ as 
\begin{align}
\label{eq:Delta-lambdaa}
\left( \Delta \bar \lambda_a \right)^2 &= \left( \Delta \bar \lambda_a^{(\mrm{fit})} \right){}^2 + \left(\frac{\partial \bar \lambda_a (\bar \lambda_s,q)}{\partial \bar \lambda_s}\right)^2 \left( \Gamma^{-1} \right)_{\bar \lambda_s \bar \lambda_s}  +  \left(\frac{\partial \bar \lambda_a (\bar \lambda_s,q)}{\partial q}\right)^2 \left( \frac{dq}{d\ln \eta} \right)^2 \left( \Gamma^{-1} \right)_{\ln \eta \ln \eta} \nn \\
& +  2 \frac{\partial \bar \lambda_a (\bar \lambda_s,q)}{\partial \bar \lambda_s}  \frac{\partial \bar \lambda_a (\bar \lambda_s,q)}{\partial q}  \frac{dq}{d\ln \eta}  \left( \Gamma^{-1} \right)_{\ln \eta\, \bar \lambda_s}\,, 
\end{align}
where $\Delta \bar \lambda_a^{(\mrm{fit})}$ corresponds to the EoS variation in the $\bar \lambda_s$--$\bar \lambda_a$ relation from the fit. Next, we estimate the error on $\bar \lambda_1$ and $\bar \lambda_2$ as 
\be
\label{eq:Delta-lambda12}
\Delta \bar \lambda_{1,2} = \sqrt{\left(\Delta \bar \lambda_s \right)^2 + \left(\Delta \bar \lambda_a \right)^2}\,.
\ee

The fractional measurement error on $\bar \lambda_1$ is shown in Fig.~\ref{fig:Fisher-prior}. Observe that such an error is smaller than that of the commonly-used chirp tidal deformability, $\bar \Lambda$, by a factor of 1.5. 
The fractional measurement error ($\Delta \bar \lambda_A / \bar \lambda_A$) of  $\bar \lambda_2$ is slightly larger than that of $\bar \lambda_1$, since $\Delta \bar \lambda_{1} = \Delta \bar \lambda_{2}$ but $\bar \lambda_1 > \bar \lambda_2$. Such a measurement of individual tidal deformabilities adds an important piece of astrophysical information on top of the masses and spins: via the I-Love-Q relations, they can be used to infer either the moment of inertia or the quadrupole moment of the stars.

\subsection{Nuclear Physics}
\label{sec:nuclear}

\begin{figure}[htb]
\begin{center}
\includegraphics[width=8.cm,clip=true]{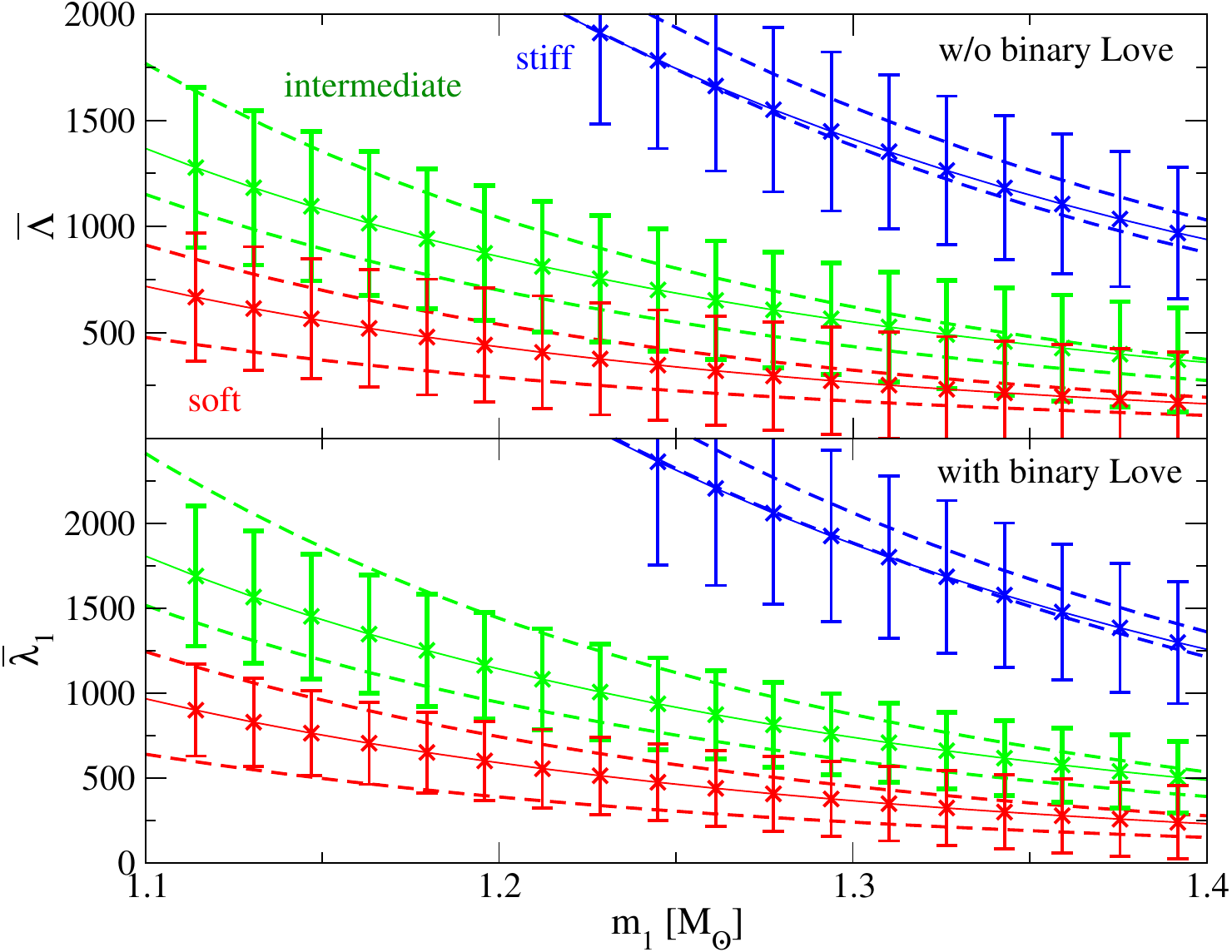}  
\caption{\label{fig:lambda-m-error} (Top) The relation between $\bar \Lambda$ and $m_1$ with $q=0.9$ for classes of soft (red dashed region), intermediate (green dashed region) and stiff (blue dashed region) EoSs. We also show 2-$\sigma$ error bars assuming one detects GWs emitted from NS binaries with a given $m_1$, $q=0.9$, an Adv.~LIGO SNR of 30 and EoSs WFF2, MPA1 and MS1b to represent the soft, intermediate and stiff classes respectively. Observe that it is relatively easy to distinguish the stiff class from the other two classes, while one may be able to distinguish between the soft and intermediate classes with $m_1 \lesssim 1.2 M_\odot$. (Bottom) Same as the top panel but for the relation between $\bar \lambda_1$ and $m_1$. The errors are calculated using the $\bar \lambda_s$--$\bar \lambda_a$ relation. Observe that in this case, one may be able to distinguish between the soft and intermediate classes with $m_1 \lesssim 1.3 M_\odot$.
}
\end{center}
\end{figure}

Let us now explain how the improved measurement accuracy of tidal parameters with the binary Love relations allows us to constrain the EoS. In particular, we study how well future GW observations can distinguish different classes of EoSs, namely soft, intermediate and stiff EoS classes. This classification can also be inferred by the clustering of the $\bar{\Lambda}$--$m$ curves for different EoS (see Fig.~\ref{fig:lambda-m-error}), which is the criterion we use in this paper. The eleven EoSs considered in Sec.~\ref{sec:binary-Love} can be classified as in Table~\ref{table:EoS}. 

The top panel of Fig.~\ref{fig:lambda-m-error} shows the range within each EoS class in Table~\ref{table:EoS} of the relation between $\bar \Lambda$ and $m_1$ for $q=0.9$. We also show error bars that correspond to 2-$\sigma$ measurement errors assuming Adv.~LIGO detects GWs emitted from NS binaries for each $m_1$ with $q=0.9$ and SNR=30. For the injections, we chose WFF2, MPA1 and MS1b EoSs as representative EoSs within the soft, intermediate and stiff classes respectively. We do not expect all of these measurements along an EoS to be realized, but rather, we would like to discuss how well one can distinguish each class from another if a single measurement is realized. Observe that it is relatively easy to distinguish the stiff class from the other two. On the other hand, one may be able to distinguish the soft and intermediate classes for $m_1 \lesssim 1.2 M_\odot$, since the edges of the error bars in one class do not overlap the other EoS region in this mass range.  

The bottom panel of Fig.~\ref{fig:lambda-m-error} is similar to the top panel, but it plots $\bar \lambda_1$ versus $m_1$ and it uses the $\bar \lambda_s$--$\bar \lambda_a$ relation in parameter estimation, as explained in the previous subsection. In this case, observe that the soft and intermediate classes may be distinguished for observations with $m_1 \lesssim 1.3 M_\odot$\footnote{In~\cite{Yagi:2015pkc}, we show a similar figure for $q=0.75$, in which one sees that the error bars for the stiff (soft) EoS classes decreases (increases) compared to those when $q=0.9$. Thus, distinguishing between the soft and intermediate EoS classes becomes more difficult as one decreases $q$.}, which is a slightly larger mass range than when using the $\bar \Lambda$ parameterization without the binary Love relations. This shows that an improvement in the measurement accuracy of tidal parameters using the binary Love relations may help us better distinguish between different classes of EoS.

\subsection{Experimental Relativity}
\label{sec:gravitational}

The improvement on the measurement accuracy of $\bar \lambda_0^{(0)}$ is also important from an experimental relativity standpoint. In~\cite{I-Love-Q-Science,I-Love-Q-PRD}, we suggested that the universal relation between the tidal deformability and the moment of inertia (the I-Love relation) would allow us to carry out extreme gravity tests of gravity without being affected by the uncertainties in the EoS. For example, one expects that future radio observations will measure the moment of inertia of the primary NS in the double binary pulsar J0737-3039 to $\sim 10\%$~\cite{lattimer-schutz,kramer-wex}, while the tidal deformability of a NS can be measured by Adv.~LIGO to a certain accuracy that depends on the Nature's EoS. Therefore, one can draw a measurement point in the I-Love plane with an error box that corresponds to the above measurements. In general, the I-Love relations in non-GR theories are different from that in GR. Therefore, one can constrain the coupling parameters in non-GR theories by requiring that non-GR I-Love relations be consistent with the error box in the I-Love plane. For example, in~\cite{I-Love-Q-Science,I-Love-Q-PRD} we considered dynamical Chern-Simons (dCS) gravity~\cite{jackiw,Smith:2007jm,CSreview}, a parity-violating theory of gravity motivated from heterotic superstring theory~\cite{Polchinski:1998rq,Polchinski:1998rr}, loop quantum gravity~\cite{Alexander:2004xd,Taveras:2008yf,Calcagni:2009xz} and effective field theories of inflation~\cite{Weinberg:2008hq}. We found that by combining future radio and GW observations, one can place a constraint on the theory that is six orders of magnitude stronger than the current bound from Solar System~\cite{alihaimoud-chen} and table-top~\cite{kent-CSBH} experiments.  

A practical difficulty of applying such a method to probe extreme gravity with two different observations is that the NS sources are different, and hence, one cannot, in principle, use the I-Love relation derived for a sequence of isolated NSs. Moreover, if one does not use the binary Love relations, it would be difficult to measure independent tidal deformabilities. In~\cite{I-Love-Q-Science,I-Love-Q-PRD}, we assumed that we could detect GWs emitted from an equal-mass (or nearly equal-mass) NS binary, whose NS masses were the same (or very close to the same) to that of the primary pulsar in J0737-3039 ($1.338M_\odot$). In such a situation, one has that $\bar \lambda_s = \bar \Lambda = \bar \lambda_1 = \bar \lambda_2$, and hence, one can measure individual tidal deformabilities by measuring $\bar \Lambda$. Moreover, since the NS masses are the same between the GW source and the primary pulsar, one can use the I-Love relation for isolated NSs. However, the event rate of detecting GWs emitted from such a binary or a similar binary might be rare. 

A more practical tidal parameter to probe extreme gravity is $\bar \lambda_0^{(0)}$. This quantity has at least three advantages: (i) one can use the relations for isolated NSs, if one chooses the fiducial mass $m_0$ to be the same as that of the NS used to measure the moment of inertia; (ii) one can measure $\bar \lambda_0^{(0)}$ from unequal-mass systems (provided their mass difference is not too large relative to $m_0$ that systematic errors due to the Taylor expansion dominate); (iii) one can improve the measurement accuracy by combining multiple GW detections. Thus, we see that the use of the $\bar \lambda_0^{(0)}$ parameterization allows us to relax many of the assumptions we made in~\cite{I-Love-Q-Science,I-Love-Q-PRD}, making this test of GR much more broadly applicable. 

\begin{figure}[htb]
\begin{center}
\includegraphics[width=8.cm,clip=true]{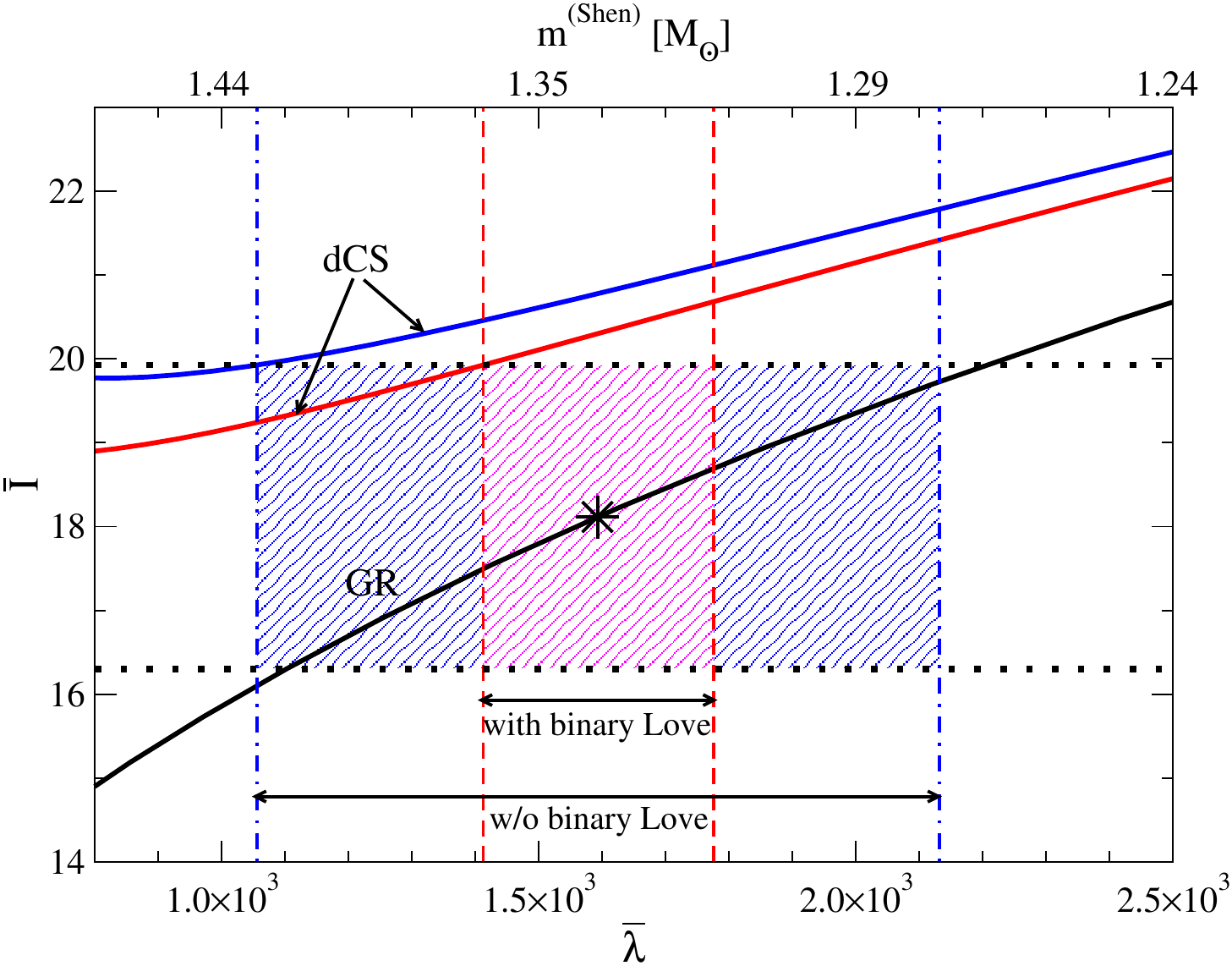}  
\caption{\label{fig:I-Love-Shen-error-CS} 
The universal I-Love relation for NSs in GR (black solid), together with the measurement error of $\bar I$ (black dotted) from future double binary pulsar observations and that of $\bar \lambda$ from future GW observations with (red dashed) and without (blue dotted-dashed) using the $\bar \lambda_0^{(0)}$--$\bar \lambda_0^{(1)}$ relation. We estimated the measurment accuracy of $\bar \lambda_0^{(0)}$ with $m_0 = 1.338M_\odot$, assuming Adv.~LIGO detects GWs emitted from a $(1.2,1.4)M_\odot$ NS binary with SNR=30 and Shen is the correct EoS. 
The top horizontal axis shows the mass corresponding to each $\bar \lambda$ for the Shen EoS in GR. Red and blue solid curves correspond to the I-Love relation in dCS with coupling constants that are marginally allowed from observational errors with and without using the $\bar \lambda_0^{(0)}$--$\bar \lambda_0^{(1)}$ relation respectively. Such a relation improves the projected constraint on dCS but only slightly.
}
\end{center}
\end{figure}

As shown in Sec.~\ref{sec:accuracy}, the $\bar \lambda_0^{(0)}$--$\bar \lambda_0^{(k)}$ relations further improves the measurement accuracy of $\bar \lambda_0^{(0)}$. Figure~\ref{fig:I-Love-Shen-error-CS} presents the I-Love relation in GR (black solid), together with errors corresponding to double binary pulsar and GW observations. Regarding the latter, we assumed Adv.~LIGO detects GWs emitted from a $(1.2,1.4)M_\odot$ NS binary with SNR=30. Following~\cite{I-Love-Q-Science,I-Love-Q-PRD}, we assumed that the Shen EoS (stiff) is the correct EoS, that gives a conservative constraint on dCS compared to softer EoSs. We estimated the measurement accuracy of $\bar \lambda_0^{(0)}$ with $m_0 = 1.338M_\odot$ with and without using the $\bar \lambda_0^{(0)}$--$\bar \lambda_0^{(1)}$ relation. Figure~\ref{fig:I-Love-Shen-error-CS} also presents the I-Love relation in dCS with the characteristic length scales of $\xi^{1/4} = 21.7$km (red solid) and $\xi^{1/4} = 22.8$km (blue solid)\footnote{The current strongest bound on dCS is $\xi^{1/4} \lesssim \mathcal{O}(10^8 \mrm{km})$~\cite{alihaimoud-chen,kent-CSBH}.} that are marginally allowed from the above measurements with and without using the binary Love relation respectively. This shows that the relation improves the bound on dCS, though only slightly (by $\sim 5\%$). Interesting future work includes studying the impact of the binary Love relations in testing theories other than dCS gravity.

\subsection{Cosmology}
\label{sec:cosmology}

Let us now discuss whether the binary Love relations can be used to improve the ability of probing cosmology with GW observations. Reference~\cite{messenger-read} proposed that if one knows the correct EoS, the measurement of tidal deformabilities can be used to break the degeneracy between the mass and the source redshift. This is because the tidal deformabilities depend on the intrinsic mass of the binary, while other parameters in the GW phase depend on the redshifted mass. Thus, if one can infer the intrinsic mass from the tidal deformabilities and one can measure the redshifted mass from the GW phase, one can then combine this information to infer the redshift. Moreover, the GW amplitude depends on the luminosity distance, which in turn encodes cosmological evolution information. Given a measurement of this distance and an inference of the redshift, one can then use NS binaries as \emph{standard sirens}~\cite{Schutz:1986gp} and extract cosmological parameters.  

The knowledge of how the tidal deformability depends on the intrinsic NS mass is crucial in such an analysis. This information can, in principle, be accurately measured from multiple GW signals from various NS binaries in local galaxies with third-generation interferometers, such as ET. The universal $\bar \lambda_s$--$\bar \lambda_a$ relation allows one to measure individual tidal deformabilities with an improved measurement accuracy compared to that of $\bar \Lambda$, which further helps one determine the intrinsic mass dependence of the tidal deformability.

Another route is to use the $\bar \lambda_0^{(0)}$--$\bar \lambda_0^{(k)}$ relation to infer this intrinsic mass dependence. As mentioned above, the critical aspect of performing GW cosmology through the tidal deformability measurement is to know the mass dependence of the latter. If one uses the Taylor expanded parametrization, one way of inferring the mass dependence is to change the fiducial mass $m_0$ from source to source and measure $\bar \lambda_0^{(0)}$ (or $c_0$) as a function of $m_{0}$ for each binary in local galaxies. Alternatively, one can take a different approach through the use of the $\bar \lambda_0^{(0)}$--$\bar \lambda_0^{(k)}$ relation. One can choose the same $m_0$ for various binaries that have masses close to $m_{0}$ and combine results from multiple detections to improve the measurement accuracy of $\bar \lambda_0^{(0)}$. One can then use the $\bar \lambda_0^{(0)}$--$\bar \lambda_0^{(k)}$ relation for various values of $k$ to extract the functional mass dependence of the tidal deformability. 

Let us investigate such an approach in more detail. Following~\cite{messenger-read}, we repeat the Fisher analysis performed in Sec.~\ref{sec:Fisher} but replacing the chirp mass with the redshifted one $\mathcal{M}_z = (1+z) \mathcal{M}$ and the tidal parameter set with ${\theta}_\mrm{tid}^{a} = (z)$, where $z$ represents the redshift. Regarding the relation between the mass and tidal deformability, we use the Taylor expanded expression in Eq.~\eqref{eq:Taylor} up to $k=1$. We use the noise curve of ET with the B configuration~\cite{mishra} and take the lower cutoff frequency to be $f_{\min} = 1$Hz. Following~\cite{DelPozzo:2015bna}, we choose the fiducial cosmological parameters as $(h,\Omega_m,\Omega_\Lambda,w) = (0.7,0.3,0.7,-1)$, where $h \equiv H_0/(100\mrm{km/s/Mpc})$ with $H_0$ representing the current Hubble constant, $\Omega_m$ and $\Omega_\Lambda$ corresponding to the dark matter and dark energy fractional densities respectively, and $w$ representing the cosmological EoS. These cosmological parameters are needed to calculate the luminosity distance for a given redshift, that enters in the amplitude of the gravitational waveform.

\begin{figure}[htb]
\begin{center}
\includegraphics[width=8.cm,clip=true]{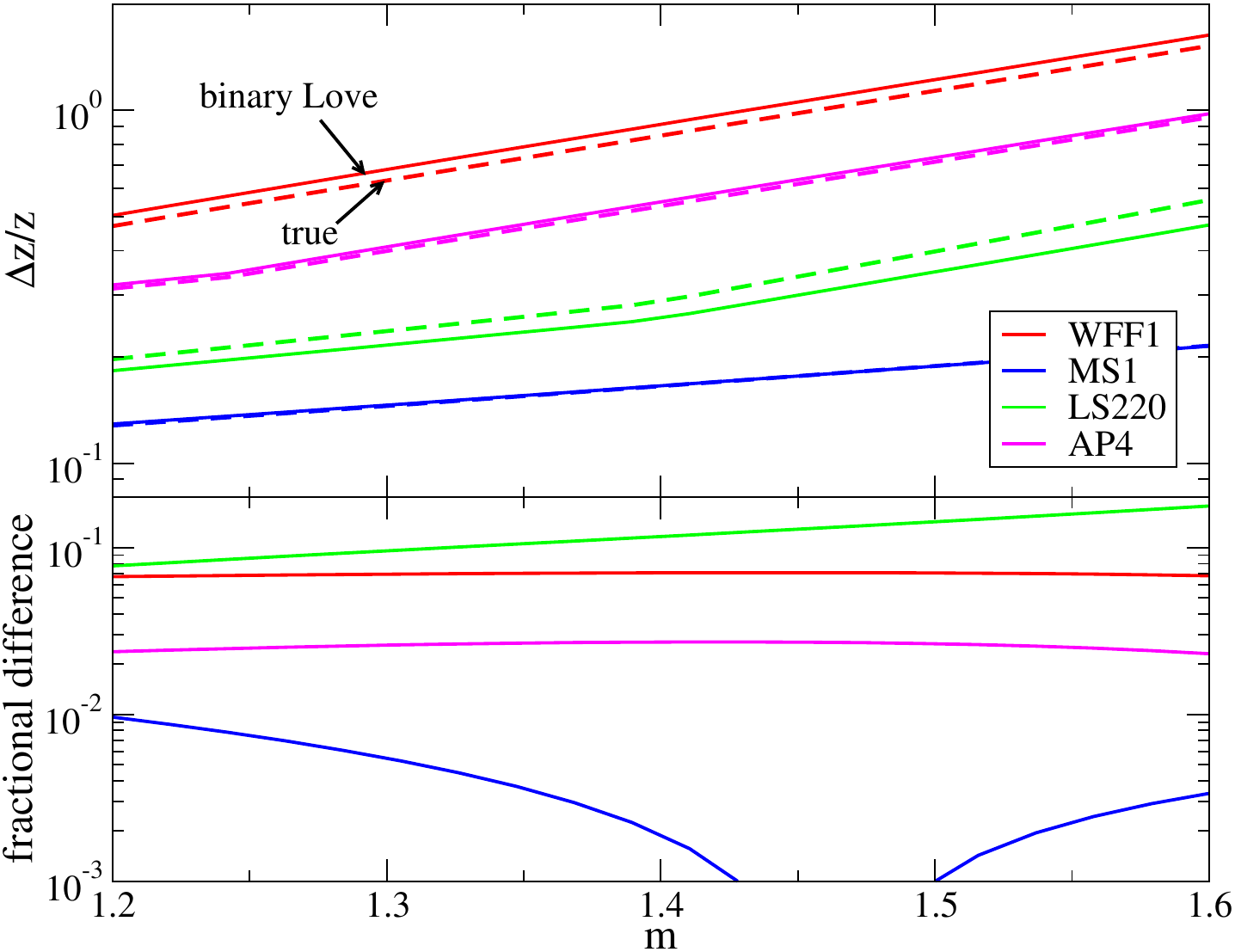}  
\caption{\label{fig:cosmology-frac-diff}  (Top) Fractional measurement accuracy of the redshift $z$ as a function of the individual NS mass $m$ for various EoSs assuming that we know the true $\bar \lambda_0^{(1)}$ \emph{a priori} (solid), and with $\bar \lambda_0^{(1)}$ obtained from the $\bar \lambda_0^{(0)}$--$\bar \lambda_0^{(1)}$ relation (dashed). In both cases, we assume that $\bar \lambda_0^{(0)}$ is known \emph{a priori}. We assume that one detects GWs emitted from equal-mass NS binaries with $m_1 = m_2 = m$ at $z=1$ using ET and we set $m_0 = m$.  (Bottom) Fractional difference between $\Delta z/z$ with true $\bar \lambda_0^{(1)}$and that obtained from the $\bar \lambda_0^{(0)}$--$\bar \lambda_0^{(1)}$ relation. Observe that the difference is $\sim 20\%$ at most for $m \in [1.2,1.6] M_\odot$. See the right panel of Fig.~3 in~\cite{Yagi:2015pkc} for a similar figure but as a function of $z$.
}
\end{center}
\end{figure}

The solid curves in the top panel of Fig.~\ref{fig:cosmology-frac-diff} represent fractional statistical errors on the redshift ($\Delta z/z$) against the individual NS mass $m$ for various EoSs and with $m_0=m$, assuming one detects GWs emitted from equal-mass NS binaries at $z=1$ with  which we know the true $\bar \lambda_0^{(0)}$ and $\bar \lambda_0^{(1)}$\footnote{The small difference in $\Delta z/z$ in Fig.~\ref{fig:cosmology-frac-diff} and in~\cite{messenger-read} is due to the fact that we are using conditions that are slightly different from those in~\cite{messenger-read}, such as the ET detector configuration, the high frequency cutoff, fiducial cosmological parameters and the inclusion of higher PN tidal terms in the waveform phase.}. We also show $\Delta z/z$ using the $\bar \lambda_0^{(0)}$--$\bar \lambda_0^{(1)}$ relation with dashed curves, assuming that $\bar \lambda_0^{(0)}$ is known \emph{a priori} from e.g.~GW measurements of NS binaries in local galaxies. The bottom panel shows the fractional difference between $\Delta z/z$ with and without using the $\bar \lambda_0^{(0)}$--$\bar \lambda_0^{(1)}$ relation. Observe that such a difference is always smaller than statistical errors (top panel) and should not significantly affect the measurement accuracy of cosmological parameters with GWs estimated in e.g.~\cite{DelPozzo:2015bna}. Such a result suggests that inferring $\bar \lambda_0^{(1)}$ from knowledge of $\bar \lambda_0^{(0)}$ using the $\bar \lambda_0^{(0)}$--$\bar \lambda_0^{(1)}$ relation should help probe cosmology with GWs.

Let us briefly comment on systematic errors on $z$ due to the EoS variation in the $\bar \lambda_0^{(0)}$--$\bar \lambda_0^{(1)}$ relation. Applying the expression for such errors given by Eq.~\eqref{eq:sys-Cutler-Vallisneri} that is valid to leading order in such errors, one can show that they are suppressed by $m-m_0$. Therefore, for equal-mass NS binaries ($m_1 = m_2 = m$) with $m_0 = m$, as we assumed in Fig.~\ref{fig:cosmology-frac-diff}, such systematic errors vanish.

\section{Discussion and Outlook}
\label{sec:conclusion}

Let us now discuss the possibility of applying the universal relations to GW observations of binaries with \emph{spinning} NSs. In such a case, one needs to include not only the individual spins of NSs, but also the individual quadrupole moments $\bar Q_A$, which also depend on the EoS. As we pointed out in~\cite{I-Love-Q-Science,I-Love-Q-PRD}, one can use the universal relations between $\bar Q_A$ and $\bar \lambda_A$ (the Q-Love relation) for an isolated NS to eliminate $\bar Q_A$ from the search parameters in the template family. Alternatively, as we show in~\ref{sec:binary-I-Love-Q}, one can easily combine the Q-Love and binary Love relations to express the two independent quadrupole moments in terms of a single tidal parameter, which also allows us to eliminate $\bar Q_A$ from the search parameters. One can also combine the universal relations between the moment of inertia and the tidal deformability (the I-Love relation) found in~\cite{I-Love-Q-Science,I-Love-Q-PRD} and the binary Love relations of this paper to express the independent moments of inertia in terms of a single tidal parameter.  The impact of all of this in parameter estimation will be analyzed elsewhere.

A possible avenue for future work includes improving the parameter estimation study by carrying out a Bayesian analysis~\cite{delpozzo,Wade:2014vqa,Lackey:2014fwa,Agathos:2015uaa} with a more natural choice of prior, such as a uniform one. Such an analysis is more reliable than the Fisher one done here, especially when the SNR is low, and allows us to easily estimate systematic errors on $\bar \lambda_0^{(0)}$ due to mismodeling the tidal deformability through a finite Taylor expansion. One can also improve the analysis by considering spinning NS binaries~\cite{Chatziioannou:2013dza,Chatziioannou:2014coa,Chatziioannou:2014bma}. For example, one can study how spin precession breaks degeneracies among tidal parameters and other template parameters, and how spin-corrections to the tidal deformability~\cite{Pani:2015nua} affects the measurement accuracy of the latter. Work along this line is currently in progress. Another avenue of possible future work includes finding similar universal relations among parameters in a binary that depend on higher multipole tidal deformabilities~\cite{Yagi:2013sva} and spin-induced multipole moments~\cite{Stein:2013ofa,Yagi:2014bxa,Chatziioannou:2014tha,Majumder:2015kfa}. 

\section*{Acknowledgments}
We would like to thank Katerina Chatziioannou for useful comments, suggestions and advice. 
We would like to also thank Benjamin Lackey for pointing out a few typos.
K.Y.~acknowledges support from JSPS Postdoctoral Fellowships for Research Abroad and NSF grant PHY-1305682.
N.Y. acknowledges support from the NSF CAREER Grant PHY-1250636. Some calculations used the computer algebra-systems MAPLE, in combination with the GRTENSORII package~\cite{grtensor}.

\appendix

\section{Effect of the Prior}
\label{app:prior}

\begin{figure}[htb]
\begin{center}
\includegraphics[width=8.cm,clip=true]{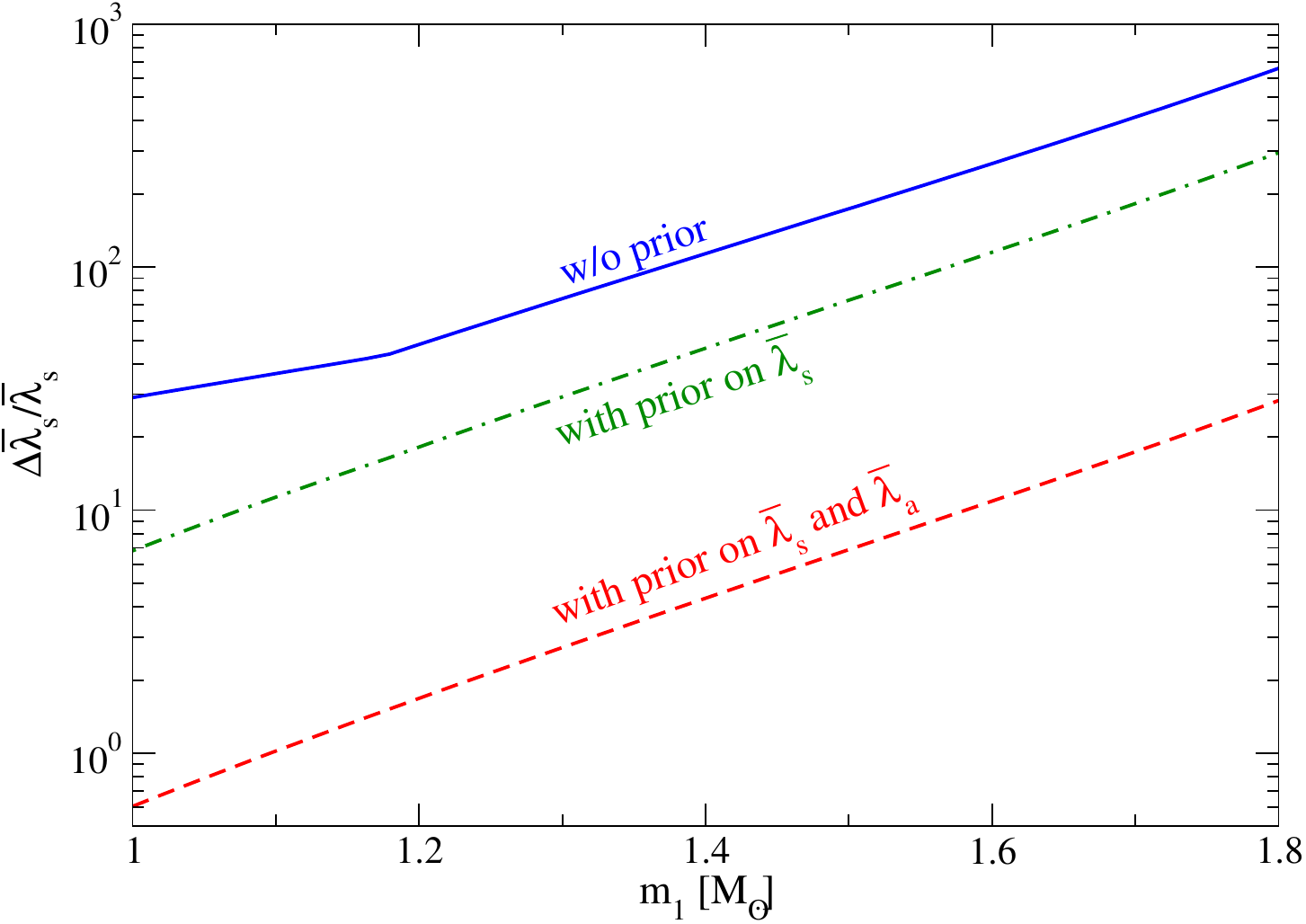}  
\caption{\label{fig:Fisher-prior-effect} The effect of the prior on the measurement accuracy of $\bar \lambda_s$ without using the $\bar \lambda_s$--$\bar \lambda_a$ relation. Each curve shows $\Delta \bar \lambda_s / \bar \lambda_s$ without the prior (blue solid), with only the prior on $\bar \lambda_s$ (green dotted-dashed) and with the prior on both $\bar \lambda_s$ and $\bar \lambda_a$ (red dashed). Observe that the prior on tidal parameters, in particular on $\bar \lambda_a$, affects the measurement accuracy of $\bar \lambda_s$ significantly.}
\end{center}
\end{figure}

Let us now discuss how the prior affects the measurement accuracy of $\bar \lambda_s$ when one does \emph{not} use the $\bar \lambda_s$--$\bar \lambda_a$ relation. First, when we include priors on both $\bar \lambda_s$ and $\bar \lambda_a$, we found that $\Delta \bar \lambda_a \sim \sigma_{\bar \lambda_a}$ irrespective of the injected mass. This shows that the measurement accuracy of $\bar \lambda_a$ is dominated by its prior, and indeed, if one does not impose such a prior, $\Delta \bar \lambda_a$ becomes much larger than $\sigma_{\bar \lambda_a}$. Since $\bar \lambda_s$ is strongly correlated to $\bar \lambda_a$, such a prior on $\bar \lambda_a$ also affects the measurement accuracy of $\bar \lambda_s$ significantly. To show this point explicitly, we present in Fig.~\ref{fig:Fisher-prior-effect} the fractional measurement accuracy of $\bar \lambda_s$ with (red dashed) and without (green dotted-dashed) imposing the prior on $\bar \lambda_a$, but in both cases imposing the prior on $\bar \lambda_s$. The former is the same as the red dashed curve in Fig.~\ref{fig:Fisher-prior}. Observe that the prior on $\bar \lambda_a$ improves the measurement accuracy of $\bar \lambda_s$ by approximately an order of magnitude. The green dotted-dashed curve actually corresponds to $\Delta \bar \lambda_s \sim \sigma_{\bar \lambda_s}$, and hence, the measurement accuracy is now dominated by the prior on $\bar \lambda_s$. 

If we further relax the prior on $\bar \lambda_s$, the measurement accuracy becomes worse, as shown by the solid blue curve in Fig.~\ref{fig:Fisher-prior-effect}. The slope changes at $m_1 \sim 1.2M_\odot$ because $f_{\max}$ changes from $f_\cont$ to $f_\ISCO$ at this critical mass. This figure shows that the prior on tidal parameters plays a crucial role in determining $\bar \lambda_s$ when one does not use the $\bar \lambda_s$--$\bar \lambda_a$ relation. On the other hand, we checked that the prior does not improve the measurement accuracy of $\bar \lambda_s$ at all if one uses the $\bar \lambda_s$--$\bar \lambda_a$ relation. This is because $\Delta \bar \lambda_s$ in this case is much smaller than the prior range.

\section{Systematic Errors on Tidal Parameters}
\label{app:sys}

In this appendix, we rederive the systematic error on $\bar \lambda_s$ in Eq.~\eqref{eq:sys-lambdas}  due to the EoS variation in the $\bar \lambda_s$--$\bar \lambda_a$ relation from a different approach. We follow~\cite{cutler-vallisneri} and estimate such an error as
\be
\label{eq:sys-Cutler-Vallisneri}
\Delta_\sys \bar \lambda_s = \left(  \Gamma^{-1} \right)^{\bar \lambda_s j} \left( [h_\inj - h_\temp] \bigg| \frac{\partial h_\temp}{\partial \theta^j} \right)\,,
\ee
where $h_\inj$ and $h_\temp$ are the injected and template gravitational waveforms. The Fisher matrix $\Gamma_{ij}$ is calculated with the template waveform and the right hand side of the above equation is evaluated at the injected parameters. As in Sec.~\ref{sec:sys}, we only consider the leading tidal part in the waveform phase. The only difference in $h_\inj$ and $h_\temp$ is that the latter evaluates $\bar \lambda_a$ through the $\bar \lambda_s$--$\bar \lambda_a$ relation. The Fourier transform of $h_\inj$ and $h_\temp$ are related by
\be
\tilde h_\inj = \tilde h_\temp \exp\left( i \delta \Psi \right)\,,
\ee
where
\be
\label{eq:delta-Psi}
\delta \Psi \propto C_a \delta m \left[ \bar \lambda_a^{(i)} - \bar \lambda_a \left( \bar \lambda_s^{(i)} \right) \right]\,. 
\ee
Therefore, the difference between $\tilde h_\inj$ and $\tilde h_\temp$ becomes
\be
\tilde h_\inj - \tilde h_\temp = \left[  \exp\left( i \delta \Psi \right) - 1 \right] \tilde h_\temp \approx   i \delta \Psi  \, \tilde h_\temp\,, 
\ee
where we neglected terms of $\mathcal{O} \left( \delta \Psi^2 \right)$. Combining the above equation with 
\be
\frac{\partial \tilde h_\temp}{\partial \bar \lambda_s} = i \frac{\partial \Psi}{\partial \bar \lambda_s} \tilde h_\temp\,,
\ee
with the tidal part of $\Psi$ given by Eq.~\eqref{eq:template}, one finds
\be
\tilde h_\inj - \tilde h_\temp = \delta \Psi \frac{\partial \tilde h_\temp}{\partial \bar \lambda_s} \left( \frac{\partial \Psi}{\partial \bar \lambda_s} \right)^{-1}\,.
\ee
Substituting this into Eq.~\eqref{eq:sys-Cutler-Vallisneri}, one finds the fractional systematic error as
\begin{align}
\frac{\Delta_\sys \bar \lambda_s}{\bar \lambda_s} &= \frac{1}{\bar \lambda_s}  \left(  \Gamma^{-1} \right)^{\bar \lambda_s j} \delta \Psi \left( \frac{\partial \Psi}{\partial \bar \lambda_s} \right)^{-1} \left( \frac{\partial h_\temp}{\partial \bar \lambda_s} \bigg| \frac{\partial h_\temp}{\partial \theta^j} \right) \nn \\
&= \frac{1}{\bar \lambda_s}  \left(  \Gamma^{-1} \right)^{\bar \lambda_s j} \delta \Psi \left( \frac{\partial \Psi}{\partial \bar \lambda_s} \right)^{-1} \Gamma_{\bar \lambda_s j} \nn \\
&= \frac{\delta \Psi}{\bar \lambda_s} \left( \frac{\partial \Psi}{\partial \bar \lambda_s} \right)^{-1}\,.
\label{eq:frac-sys-Cutler-Vallisneri}
\end{align}
Here, we pulled $\delta \Psi (\partial \Psi/\partial \bar \lambda_s)^{-1}$ out of the inner product as it is independent of the frequency.
Notice that the correlation among parameters encoded in the Fisher matrix vanishes automatically.
On the other hand, from Eq.~\eqref{eq:template}, one finds
\be
\label{eq:dPsi-dlambdas}
 \frac{\partial \Psi}{\partial \bar \lambda_s} \propto C_s + C_a \delta m \bar \lambda_a' \left( \bar \lambda_s \right)\,.
\ee
Substituting Eqs.~\eqref{eq:delta-Psi} and~\eqref{eq:dPsi-dlambdas} into Eq.~\eqref{eq:frac-sys-Cutler-Vallisneri} and evaluating the tidal parameters using the injected ones, one recovers Eq.~\eqref{eq:sys-lambdas}.

\section{Universal Binary I-Love-Q Relations}
\label{sec:binary-I-Love-Q}

In this section, we extend the binary Love relations presented in the previous section to include the moment of inertia $I_A$ and the rotation-induced, quadrupole moment $Q_A^\rot$. We work here in the slow-rotation limit and keep terms up to quadratic order in spin~\cite{hartle1967,Hartle:1968ht}. One can construct slowly-rotating NS solutions by treating spins as perturbations to a background non-spinning, spherically symmetric solution. One then solves the background and perturbed Einstein equations in the interior region order by order with a given EoS and imposing regularity at the center. This solution is then matched to an analytic, asymptotically flat exterior solution at the stellar surface at each order to determine integration constants. In particular, in Hartle-Thorne coordinates, the asymptotic behavior of the $(t,t)$ and $(t,\phi)$ components of the metric at spatial infinity is given by~\cite{hartle1967,Hartle:1968ht}
\begin{align}
g_{tt} &= -1 + \frac{2 (m + \delta m)}{r} + \frac{2 Q^\rot}{r^3} P_2(\cos\theta) + \mathcal{O}\left( \frac{m^4}{r^4} \right)\,,  \\
g_{t\phi} &= 2 \sin^2\theta \frac{I\, \Omega}{r}\,, 
\end{align}
where $m$ is the stellar mass of a non-rotating configuration, $\delta m$ is the spin correction to the mass at quadratic order in spin, $\Omega$ is the stellar angular velocity and the $P_\ell$'s are Legendre polynomials. We define the dimensionless moment of inertia $\bar I_A$ and dimensionless quadrupole moment $\bar Q_A$ as~\cite{I-Love-Q-Science,I-Love-Q-PRD} $\bar I_A \equiv I_A/m_A^3$ and $\bar Q_A \equiv - Q_A^\rot/(m_A^3\, \chi_A^2)$ respectively\footnote{Since we work in the slow-rotation limit, $m_A$ corresponds to the stellar mass of the $A$th body for a non-rotating configuration with a fixed central density.}, where $\chi_A \equiv S_A/m_A^2$ with $S_A$ representing the NS spin angular momentum. Next, following Sec.~\ref{sec:lambdas-lambdaa}, we define $\bar I_s$, $\bar I_a$, $\bar Q_s$ and $\bar Q_a$ as
\begin{align}
\bar I_s & \equiv  \frac{\bar I_1 + \bar I_2}{2}\,, \quad \bar I_a \equiv \frac{\bar I_1 - \bar I_2}{2}\,, \\
\bar Q_s & \equiv  \frac{\bar Q_1 + \bar Q_2}{2}\,, \quad \bar Q_a \equiv \frac{\bar Q_1 - \bar Q_2}{2}\,. 
\end{align}

With the universal I-Love-Q relations found in~\cite{I-Love-Q-Science,I-Love-Q-PRD} for isolated NSs and the binary Love relations discussed in Sec.~\ref{sec:lambdas-lambdaa}, one can easily derive similar EoS-insensitive relations that hold among $\bar \lambda_s$, $\bar \lambda_a$, $\bar I_s$, $\bar I_a$, $\bar Q_s$ and $\bar Q_a$. For example, let us look at the relation between $\bar \lambda_s$ and $\bar I_s$. First, from the definition of $\bar I_s$, $\bar \lambda_s$ and $\bar \lambda_a$ and the isolated I-Love relation $\bar I_A (\bar \lambda_A)$, one finds
\be
\bar I_s = \frac{\bar I_1 (\bar \lambda_1) + \bar I_2 (\bar \lambda_2)}{2} = \frac{\bar I_1 (\bar \lambda_s,\bar \lambda_a) + \bar I_2 (\bar \lambda_s,\bar \lambda_a)}{2}\,.
\ee
From the $\bar \lambda_s$--$\bar \lambda_a$ relation $\bar \lambda_a (\bar \lambda_s,q)$, one then finds
\be
\bar I_s (\bar \lambda_s, q) = \frac{\bar I_1 (\bar \lambda_s, q) + \bar I_2 (\bar \lambda_s, q)}{2}\,.
\ee
This shows that a $q$-dependent universal relation also exists between $\bar \lambda_s$ and $\bar I_s$. One can carry out similar calculations to find approximately universal relations among any two in the set $\{\bar \lambda_s$, $\bar \lambda_a$, $\bar I_s$, $\bar I_a$, $\bar Q_s$, $\bar Q_a\}$.

Let us provide another explicit example by considering the $\bar \lambda_s$--$\bar Q_s$ and the $\bar \lambda_s$--$\bar Q_a$ relations, which might be useful in future GW observations of unequal-mass NS binaries. First, let us look at the Newtonian limit, where $\bar Q_A$ has a compactness dependence given by~\cite{I-Love-Q-PRD}
\be
\bar Q_A = \frac{\alpha_n^{(\bar Q)}}{C_A}\,,
\ee
where we recall that $n$ is the polytropic index and
\be
\alpha_0^{(\bar Q)} = \frac{25}{8}\,, \quad \alpha_1^{(\bar Q)} = \frac{3 \pi^2 (15-\pi^2)}{4 (\pi^2 - 6)^2}\,.
\ee
Following a similar reasoning as that which led us to Eq.~\eqref{eq:lambdas-lambdaa-C-Newton}, one can derive the expression
\be
\label{eq:lambdas-Qsa-C-Newton}
\bar Q_{s,a} = \frac{\alpha_n^{(\bar Q)}}{2} \left( \frac{1}{C_1} \pm \frac{1}{C_2} \right) 
=  \frac{\alpha_n^{(\bar Q)}}{2 C_1} \left[1 \pm q^{2/(3-n)} \right]\,, 
\ee
where the $+$ and the $-$ correspond to $\bar Q_{s}$ and $\bar Q_{a}$ respectively. On the other hand, from Eq.~\eqref{eq:lambdas-lambdaa-C-Newton}, one can solve for $C_1$ in terms of $\bar \lambda_s$ as
\be
C_1 = \left\{ \frac{\alpha_n^{(\bar \lambda)}}{2 \bar \lambda_s} \left[ 1 + q^{10/(3-n)} \right] \right\}^{1/5}\,.
\ee
Substituting this equation into Eq.~\eqref{eq:lambdas-Qsa-C-Newton}, one finds the $\bar \lambda_s$--$\bar Q_s$ and $\bar \lambda_s$--$\bar Q_a$ relations in the Newtonian limit: 
\begin{align}
\label{eq:lambdas-Qsa-Newton1}
\bar Q_{s,a} &= F_n^{(\bar Q_{s,a})}(q) \ \bar \lambda_s^{1/5}\,, \\
\label{eq:lambdas-Qsa-Newton2}
F_n^{(\bar Q_{s,a})}(q) &\equiv \frac{\alpha_n^{(\bar Q)}}{2^{4/5} \left[ \alpha_n^{(\bar \lambda)} \right]^{1/5}} \frac{1 - q^{2/(3-n)}}{\left[1 + q^{10/(3-n)} \right]^{1/5}}\,. 
\end{align}

\begin{figure}[htb]
\begin{center}
\includegraphics[width=8.cm,clip=true]{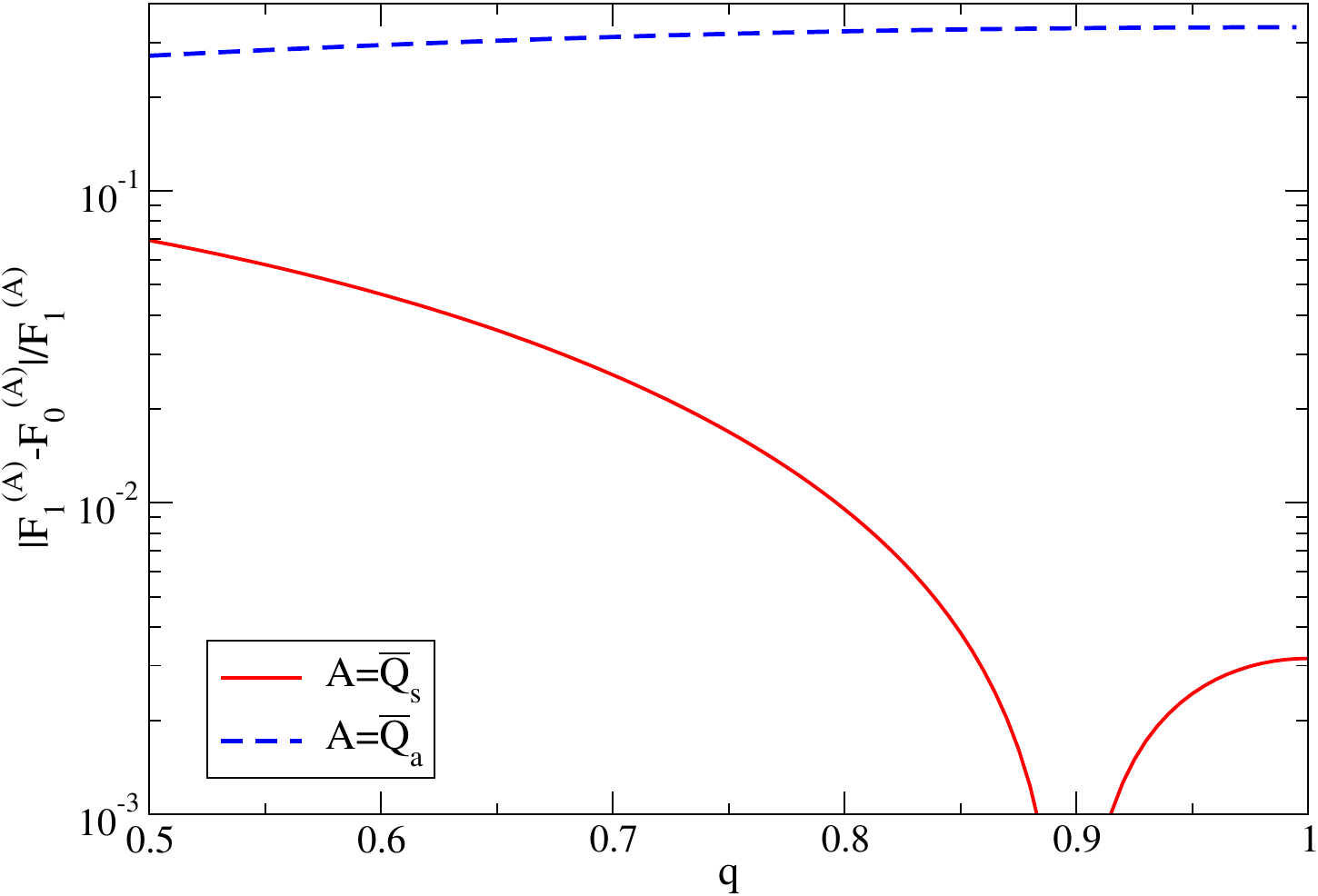}  
\caption{\label{fig:lambdas-Qsa-Newton} Fractional difference between the $n=0$ and $n=1$ Newtonian relations for the $\bar \lambda_s$--$\bar Q_s$ (red solid) and $\bar \lambda_s$--$\bar Q_a$ (blue dashed) relations given by Eqs.~\eqref{eq:lambdas-Qsa-Newton1} and~\eqref{eq:lambdas-Qsa-Newton2} respectively against $q$.
}
\end{center}
\end{figure}

Figure~\ref{fig:lambdas-Qsa-Newton} shows the fractional difference between the $n=0$ and $n=1$ Newtonian relations for the $\bar \lambda_s$--$\bar Q_s$ and $\bar \lambda_s$--$\bar Q_a$ relations. One sees that the EoS-variation is smaller than 7\% for the former, while it is $\sim 30\%$ for the latter. Interestingly, the fractional difference of the $\bar \lambda_s$--$\bar Q_s$ relation decreases as one increases $q$, a behavior opposite to that found for the $\bar \lambda_s$--$\bar \lambda_a$ and $\bar \lambda_s$--$\bar Q_a$ relations.

\begin{figure*}[htb]
\begin{center}
\includegraphics[width=7.5cm,clip=true]{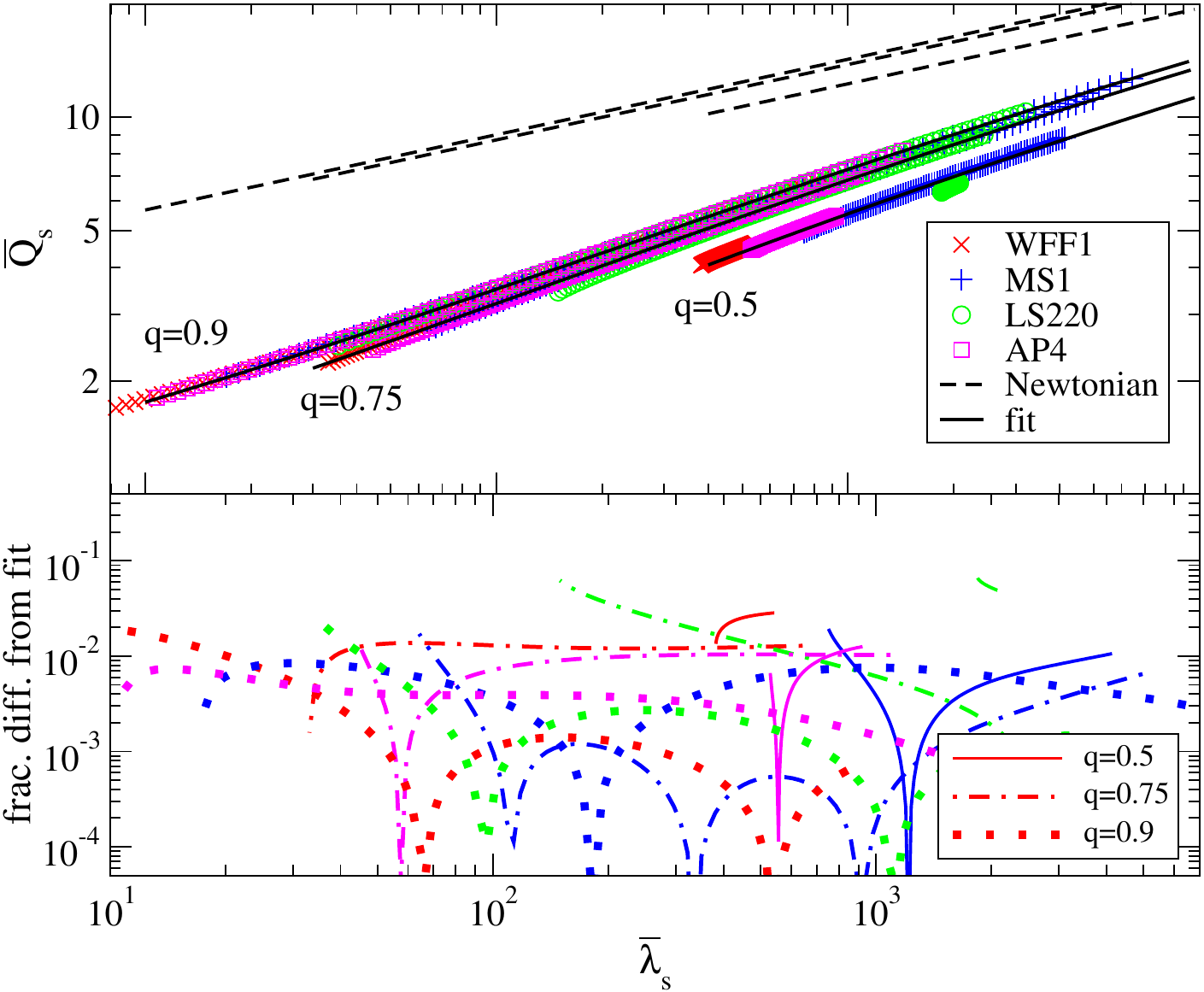}  
\includegraphics[width=7.5cm,clip=true]{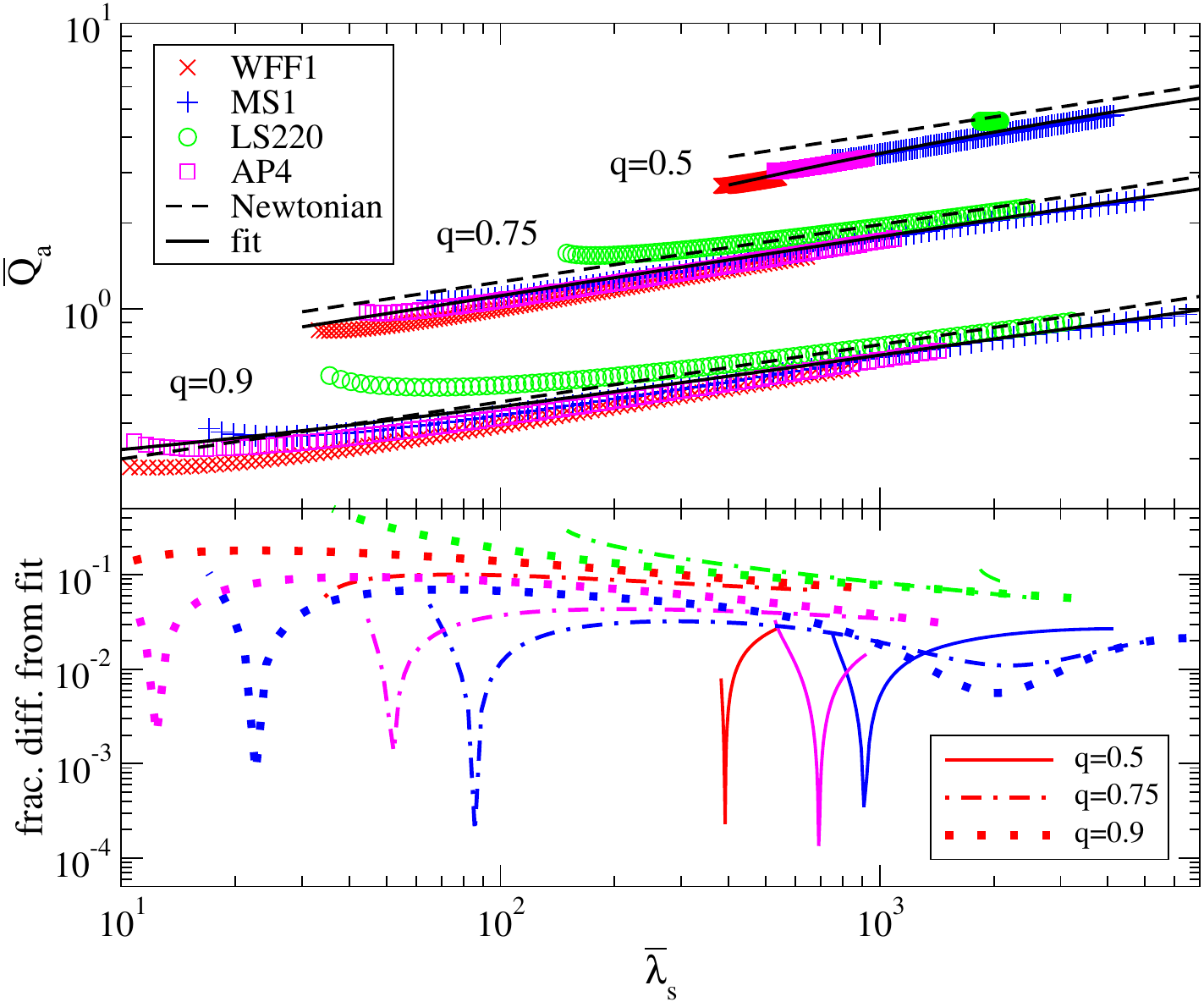}  
\caption{\label{fig:lambdas-Qsa} (Top) Universal $\bar \lambda_s$--$\bar Q_s$ (left) and $\bar \lambda_s$--$\bar Q_a$ (right) relations for four representative EoSs, with $q=0.5$, 0.75 and 0.9. We also present the Newtonian relations in Eq.~\eqref{eq:lambdas-Qsa-Newton1} and the fit in Eq.~\eqref{eq:fit} with the coefficients given in Table~\ref{table:coeff}. The former relations are with $n=1$ and not with $n=0.743$ because we do not have analytic relations for the latter. (Bottom) Same as the bottom left panel of Fig.~\ref{fig:lambdas-lambdaa} but for the $\bar \lambda_s$--$\bar Q_s$ and $\bar \lambda_s$--$\bar Q_a$ relations. Observe that the EoS variation of the $\bar \lambda_s$--$\bar Q_s$ relation is of $\mathcal{O}(1\%)$, which is of the same order as the I-Love-Q relations for isolated NSs.
}
\end{center}
\end{figure*}

The top panels of Fig.~\ref{fig:lambdas-Qsa} show numerically-obtained $\bar \lambda_s$--$\bar Q_s$ and $\bar \lambda_s$--$\bar Q_a$ relations for three different values of $q$, together with the Newtonian relations with $n=1$. The agreement between the  $\bar \lambda_s$--$\bar Q_s$ relation and the Newtonian one in the large $\bar \lambda_s$ region would have been much better if one would have used the Newtonian relation with an $n=3/4$ index, which can only be obtained numerically. We also show the fit given by Eq.~\eqref{eq:fit} with $x=1/\bar \lambda_s$, $y=\bar Q_s$ or $\bar Q_a$ and $n=1$, where the coefficients are given in Table~\ref{table:coeff}. For completeness, we also give the fitted coefficients for the $\bar \lambda_s$--$\bar I_s$ and $\bar \lambda_s$--$\bar I_a$ relations in Table~\ref{table:coeff}. 

The bottom panels of Fig.~\ref{fig:lambdas-Qsa} show the fractional difference between the numerical results and the fit for the $\bar \lambda_s$--$\bar Q_s$ and $\bar \lambda_s$--$\bar Q_a$ relations. Observe that the former relation is universal to $\mathcal{O}(1\%)$, which is the same accuracy as the isolated I-Love-Q relations~\cite{I-Love-Q-Science,I-Love-Q-PRD}. Observe also that the fractional difference decreases as one increases $q$, which is consistent with the Newtonian behavior of Fig.~\ref{fig:lambdas-Qsa-Newton}. On the other hand, the universality in the latter relation holds only to $\sim 10\%$. This is smaller than the maximum EoS-variation of 30\% in the Newtonian limit shown in Fig.~\ref{fig:lambdas-Qsa-Newton} because the realistic EoSs can be well described by a polytropic EoS with an index of $n=\{0.5,1\}$ and not $n=\{0,1\}$. One sees that the fractional difference increases as one increases $q$, which, again, is consistent with the Newtonian behavior of Fig.~\ref{fig:lambdas-Qsa-Newton}. This shows that the universal relations between symmetric quantities are better than those between symmetric and antisymmetric quantities.

\section*{References}

\bibliographystyle{iopart-num}
\bibliography{master}
\end{document}